\documentclass[nofootinbib]{revtex4}

\usepackage{bbold}
\usepackage{bbm}
\usepackage{graphicx}
\usepackage{color}
\usepackage{amsfonts}
\usepackage{amsmath}
\usepackage{slashed}
\def\ket#1{| #1 \ra }
\def\bra#1{\la #1 |}

\def\la{\langle}
\def\ra{\rangle}

\def\beq{\begin{equation}}
\def\eeq{\end{equation}}
\def\bea{\begin{eqnarray}}
\def\eea{\end{eqnarray}}
\def\barr{\begin{array}}
\def\earr{\end{array}}

\definecolor{MyDarkBlue}{rgb}{0,0.08,0.45} 
\definecolor{MyLightMagenta}{cmyk}{0.1,0.8,0,0.1} 
\definecolor{MLM}{cmyk}{0.1,0.8,0,0.1} 
\definecolor{MyDarkGreen}{rgb}{0,0.45,0.08} 
\definecolor{MDG}{rgb}{0,0.55,0.05}

\usepackage[textsize=footnotesize]{todonotes}

\begin{document}

\begin{titlepage}
\begin{center}
{\Large \bf Phase structure of the 1+1 dimensional massive Thirring model\\ from matrix product states} 
\vskip1cm {\large\bf
Mari~Carmen~Ba\~{n}uls$^{a,b}$,
Krzysztof~Cichy$^{c}$, Ying-Jer Kao$^{d}$,\\ \vspace{0.2cm}
C.-J.~David~Lin$^{e,f}$, Yu-Ping~Lin$^{g}$,
David~T.-L.~Tan$^{e}$}\\ \vspace{.5cm}
{\normalsize {\sl $^{a}$ Max-Planck Institut, f\"{u}r Quantenoptik,
    Garching 85748, Germany\\
    $^{b}$Munich Center for Quantum Science and Technology (MCQST), Schellingstr. 4, Munich 80799, Germany\\
$^{c}$ Faculty of Physics, Adam Mickiewicz University, Uniwersytetu Pozna\'nskiego 2, 61-614 Pozna\'{n}, Poland\\
$^{d}$ Department of Physics, National Taiwan University, Taipei 10617, Taiwan\\
$^{e}$ Institute of Physics, National Chiao-Tung University, Hsinchu
30010, Taiwan\\
$^{f}$ Centre for High Energy Physics, Chung-Yuan Christian
University, Chung-Li 32023, Taiwan\\
$^{g}$ Department of Physics, University of Colorado, Boulder,
Colorado 80309, USA
}}

\vskip1.0cm {\large\bf Abstract:\\[10pt]} \parbox[t]{\textwidth}{{
Employing matrix product states as an ansatz, we study the
non-thermal phase structure of the (1+1)-dimensional massive Thirring
model in the sector of vanishing total fermion number with staggered
regularization.     
In this paper, details of the implementation for this project are described.  
To depict the phase diagram of the model, we examine
the entanglement entropy, the fermion bilinear condensate and two
types of correlation functions.   Our investigation shows the existence of 
two phases, with one of them being critical and the other gapped.
An interesting feature of the phase structure is that the theory with
non-zero fermion mass can be conformal.
We also find clear numerical evidence that these phases are separated
by a transition of the Berezinskii-Kosterlitz-Thouless type.   
Results presented in this paper establish the possibility of using the
matrix product states for probing this type of phase transition in
quantum field theories.  
They can provide information for further
exploration of scaling behaviour, 
and serve as an important ingredient
for controlling the continuum extrapolation of the model.
}}
\end{center}
\end{titlepage}


\section{Introduction}
\label{sec:intro}
Many quantum field theories of interest cannot be studied with perturbative methods.
This concerns, most notably, quantum chromodynamics (QCD), but also seemingly simple models may have non-perturbative features.
Since the seminal idea of Wilson~\cite{Wilson1974}, the method of choice for such systems is usually to formulate them on the lattice.
In many cases, e.g.\ in QCD, this is the only way to obtain quantitative predictions directly from the Lagrangian.
The path integral corresponding to a discretized system is finite-dimensional, but usually this dimension is very large, implying no alternative to Monte Carlo sampling.
If such sampling is possible, the lattice provides an unambiguous way of reaching results with an arbitrary precision, with controllable total error.
This led to spectacular successes, particularly in the most important theory studied with lattice methods, i.e.\ QCD.
Several aspects of QCD were addressed with large-scale simulations, including hadron spectroscopy, hadron structure, thermal properties or fundamental parameters, such as the strong coupling constant and its scale dependence or the Cabibbo-Kobayashi-Maskawa matrix parameters. 
However, there are still areas in lattice field theories where the lattice has not offered quantitative or even qualitative insight due to an inherent shortcoming of the Monte Carlo approach.
This shortcoming is that not all theories admit a positive-definite probability measure, a necessary ingredient for stochastic sampling.
In the absence of positive-definite measure, a theory is said to have a sign problem, with a notable example of QCD at a non-zero chemical potential.
Even though techniques to alleviate this issue have been invented, no real solution readily available for QCD exists.
Another important example is real-time simulation, notoriously hard in quantum mechanics and quantum field theory in general.

Given these challenges, alternative approaches are being pursued.
Methods tested in the context of lattice gauge theories include e.g.\ Lefschetz thimbles \cite{Scorzato:2015qts,Bedaque:2017epw}, complex Langevin simulations \cite{Sexty:2014zya,Seiler:2017wvd} and density of states methods \cite{Langfeld:2016kty}.
In this paper,  we concentrate on the Hamiltonian approach in the framework of tensor networks (TN).
Tensor-network methods stem from quantum information theory and they have been heavily applied for description of, in particular, low-dimensional condensed matter systems.
Their main feature is that they can represent quantum states, taking correctly into account the relevant entanglement properties of a system.
In this way, they can define effective ansatzes for the wave function, which combined with efficient algorithms to utilize such ansatzes, can lead to a successful quasi-exact description of a broad class of physical theories.
A particularly efficient example of TN for one-dimensional systems is the so-called matrix product states (MPS) ansatz~\cite{aklt88,kluemper91,kluemper92,fannes92fcs,verstraete04dmrg,perez07mps}.  Generalizations to higher dimensions also exist, such as Projected Entangled Pair States (PEPS)~\cite{Verstraete:2004cf}.
For a pedagogical introduction to the TN approach, see e.g.\ Refs.~\cite{Verstraete2008,Eisert2010,Orus:2013kga,Silvi2019tns}.

The TN framework, mostly in combination with the Hamiltonian formulation, has also been used for solving lattice field theory models,
such as the $U(1)$ (the Schwinger model) and $SU(2)$ gauge theories, the $\phi^4$-theory and the $O(3)$ $\sigma$ model.
Among investigations that have been performed, one can name spectral computations~\cite{Banuls:2013jaa,Buyens:2013yza,Buyens:2014pga,Buyens:2015dkc,Buyens:2017crb,Zapp:2017fcr,Gillman:2017uir,Banuls:2017ena}, calculations of thermal properties~\cite{Banuls:2015sta,Banuls:2016lkq,Saito:2014bda,Saito:2015ryj,Buyens:2016ecr}, phase diagrams~\cite{Tagliacozzo:2012vg,Rico2013,Silvi2016,Zohar:2015eda,Zohar:2016wcf,Bruckmann:2018usp}, entanglement properties \cite{Pichler:2015yqa,Buyens:2015tea,Banuls:2017ena,Banuls:2017evv,Banuls:2018ckt}, non-zero chemical potential \cite{Banuls:2016hhv,Banuls:2016gid,Banuls:2016jws}, studies of the $\theta$-vacuum~\cite{Zache:2018cqq,Funcke:2019zna}, and real-time evolution~\cite{Buyens:2013yza,Kuhn:2015zqa,Pichler:2015yqa,Buyens:2015tea,Buyens:2016hhu,Gillman:2017ycq}.
The latter two are examples of computations hindered by a sign problem in traditional Monte Carlo simulations, which is, by construction, absent in the TN approach.
From a more formal perspective, the interplay of symmetries with TN has been a fertile research topic~\cite{Perez2008string,Sanz2009sym,Singh2010}
that has allowed, among others, a full classification of one dimensional gapped phases of matter~\cite{Chen2011phases,Schuch2011phases,Chen2011complete}
and the construction of topological states in two dimensional systems~\cite{Schuch2010peps,Chen2011twoDtopo,Williamson2016spt-mpo}.
In the particular case of gauge symmetries, it is possible to explicitly incorporate the local gauge invariance into the ansatz,
which can then be used to construct invariant models or perform numerical calculations~\cite{Tagliacozzo2014,Haegeman:2014maa,Zohar2015b,Silvi2014,Kull:2017gii,Zapp:2017fcr}. 
There are several aspects to these studies. 
From the computational perspective that occupies us in this work, their achievements are threefold.
Firstly, it is important to check whether physics of quantum field theoretical models can be well described in the TN language\footnote{A different TN approach, the tensor renormalization group method, in which one works with the partition function, has also been implemented for the investigation of several quantum field theories~\cite{Bazavov:2019qih,Shimizu:2014uva,Shimizu:2017onf,Takeda:2014vwa,Meurice:2019ddf,Denbleyker:2013bea,Kadoh:2018hqq,Kadoh:2018tis,Kawauchi:2016xng} and the formulation of quantum gravity~\cite{Asaduzzaman:2019mtx}.}.
Investigations by several groups led to unambiguously positive conclusions and the precision reached in these calculations 
in  (1+1)-dimensional cases
is in many cases unprecedentedly high.
Secondly, TN methods have offered some new insight about well-known models, in particular about their entanglement properties, behaviour at non-zero particle density and real-time evolution.  Finally, there is a close connection between TN formulations and proposals of quantum simulation for studying these theories in experiments with cold atoms, trapped ions or superconducting qubits~\cite{Wiese2013,Zohar2013,Hauke2013,Marcos2014,Dalmonte2016,Kuno:2016ipi}. Another related approach is the exploration of field theories employing full fledged quantum computers (for recent progress, see, e.g. Refs.~\cite{Klco:2018kyo,Klco:2018zqz,Alexandru:2019nsa}).  Indeed, the connections among (discretized formulations of) quantum field theories, condensed matter models and quantum information techniques are promoting a fruitful research ground, where the interdisciplinary effort can shed light on various research areas~\cite{Bermudez:2018eyh,Barros:2018mvf,Kuno:2018pcp}.

In this work, we concentrate on the investigation of the (1+1)-dimensional massive Thirring model\footnote{The massless model was proposed by W.~Thirring in 1958~\cite{Thirring:1958in} as an example of a solvable quantum field theory.}, with the Minkowski-space action
\begin{equation} 
\label{eq-action-thirring}
    S_{\mathrm{Th}}[\psi,\bar{\psi}] \
    = \int d^2x \left[ \bar{\psi}i \gamma^{\mu}\partial_{\mu}\psi \
      - m\,\bar{\psi}\psi \
      -\frac{g}{2} \left( \bar{\psi}\gamma_{\mu}\psi \right)\left( \bar{\psi}\gamma^{\mu}\psi \right) \right] \,,
\end{equation}
where $m$ denotes the fermion mass, and $g$ is the dimensionless four-fermion coupling constant.   The current paper summarizes our result for exploring the non-thermal (zero-temperature) phase structure of the above theory using the staggered-fermion regularization~\cite{Banks:1975gq, Susskind:1976jm}.  This is the first step in a research programme on studying the Thirring model with TN methods.  As described below in this section, the phase structure of the massive Thirring model can exhibit features such as infrared (IR) slavery, and the existence of a critical phase with a Berezinskii-Kosterlitz-Thouless (BKT) transition~\cite{Kosterlitz:1973xp}.  Understanding these features in lattice field theory using the TN approach 
is a worthy challenge. In particular, the connection between the numerical results and the continuum limit is a non-trivial aspect of the study, which
nevertheless is of 
fundamental importance for further applications of TN methods to other quantum field theories.
In addition, such exploration is essential for our future long-term work.  Our final goal in this research programme includes the investigation of the model with chemical potential, and various aspects of its real-time dynamics~\cite{Banuls:2020inprep}.  

The sector of vanishing total fermion number in the (1+1)-dimensional massive Thirring model is known to be S-dual to the sine-Gordon scalar field theory with zero total topological charge~\cite{Coleman:1974bu}.  The sine-Gordon theory is described by the classical action,
\begin{equation} 
\label{eq-action-SG}
    S_{\mathrm{SG}}[\phi] = \frac{1}{t}\int d^{2}x \left \{ \frac{1}{2} \left [\partial_{\mu}\phi(x) \right ]^2 \
    + \alpha \, \cos\phi(x) \right \} \,,
\end{equation}
with two couplings, $t$ and $\alpha$.  In Ref.~\cite{Coleman:1974bu}, Coleman obtained the duality relations 
\begin{eqnarray} 
\label{bosonization}
    g = \frac{4\pi^{2}}{t} - \pi  \, &,& \mbox{ } m = \frac{2\pi}{\Lambda} \left ( \frac{\alpha}{t} \right )\, , \nonumber\\
    \bar{\psi}\psi \leftrightarrow \frac{ \Lambda}{\pi}\mathrm{cos}\phi \, &,& \mbox{ } \bar{\psi} \gamma_{\mu} \psi  \leftrightarrow -\frac{1}{2\pi}\epsilon_{\mu\nu}\partial_{\nu}\phi \, ,
\end{eqnarray}
where $\Lambda$ is a cut-off scale.  In addition to the soliton solutions that can be related to fermions in the Thirring model~\cite{Mandelstam:1975hb}, the sine-Gordon theory exhibits interesting scaling behaviour and phase structure, which can be understood from studying its renormalization group (RG) equations~\cite{Amit:1979ab,Kaplan:2009kr,ZinnJustin:2000dr}.   Since this aspect of the theory is important for the current work, here we describe the scenario in slightly more detail.    

For convenience, let us define the following parameters,
\begin{equation}
\label{eq:sG_cov}
  \bar{t} \equiv \frac{1}{t} \, , \mbox{ } {\mathrm{and}} \mbox{ } z \equiv \frac{\alpha}{2t} \, .
\end{equation}
In terms of $\bar{t}$ and $z$, the RG equations (RGE's) of the sine-Gordon theory to ${\mathcal{O}} (z^{3})$ are~\cite{Kaplan:2009kr,ZinnJustin:2000dr}
\begin{eqnarray}
\label{eq:RGE_sG_for_t}
 \beta_{\bar{t}} &\equiv& \mu \frac{d \bar{t}}{d \mu} = -64 \pi \frac{z^{2}}{\Lambda^{4}} \, ,
 \\
 \label{eq:RGE_sG_for_z}
 \beta_{z} &\equiv& \mu \frac{dz}{d\mu} =  \frac{1 - 8 \pi \bar{t}}{4\pi \bar{t}} z - \frac{64\pi}{\bar{t}^{2}\Lambda^{4}} z^{3} \, ,
\end{eqnarray}
where $\mu$ is the renormalization scale.
From these RGE's, the crucial feature that can be seen immediately, in the limit $z/\Lambda^{2} \ll 1$, is the following scaling behaviour.
\begin{itemize}
 \item In the regime where $\bar{t} \gtrsim 1/8\pi$, i.e., $t \lesssim 8\pi$, the operator [cos $\phi (x)$] in the sine-Gordon theory is relevant, resulting in the existence of solitons in the model.
 \item On the contrary, at $\bar{t} \lesssim 1/8\pi$, i.e., $t  \gtrsim 8\pi$, the operator [cos $\phi (x)$] is irrelevant, and the model is a free bosonic theory at low energy.  In this case, from Eq.~(\ref{eq:RGE_sG_for_t}), the coupling $t$ will be scale-invariant in the IR.
\end{itemize}
These different scaling properties imply the existence of a phase transition at $t \sim 8\pi$.   To further understand the nature of this phase transition, we note that the sine-Gordon theory is known to be equivalent to the Coulomb-gas (vortex) sector of the two-dimensional O(2) $\sigma$~model (the XY model), and the coupling $\bar{t}$ can be interpreted as the temperature in the latter\footnote{See, for instance, Ref.~\cite{Huang:1990via} for the relation between the sine-Gordon kinks and the vortices in the XY model.}.   Therefore, the above scaling behaviour can be shown to be closely related to the BKT phase transition~\cite{Kosterlitz:1973xp}.   In fact, it can be easily verified that RGE's for the sine-Gordon model, Eqs.~(\ref{eq:RGE_sG_for_t}) and  (\ref{eq:RGE_sG_for_z}), bear the same characteristics as the Kosterlitz scaling equations~\cite{Kosterlitz:1974sm} for the XY model.  This is also in accordance with the Mermin-Wagner-Coleman no-go theorem~\cite{Mermin:1966fe,Coleman:1973ci}.   In a recent paper~\cite{Vanderstraeten:2019frg}, the MPS approach has been implemented for the XY model to study the BKT transition in statistical physics.
As we will demonstrate in this work, performing computations with the dual Thirring model using the MPS method enables us to probe this interesting phase structure.  In addition, it also allows for future investigation of real-time dynamics of this BKT phase transition~\cite{Banuls:2020inprep}.

\begin{figure}[!t]
  \centering
\includegraphics[width=10cm, height=6cm]{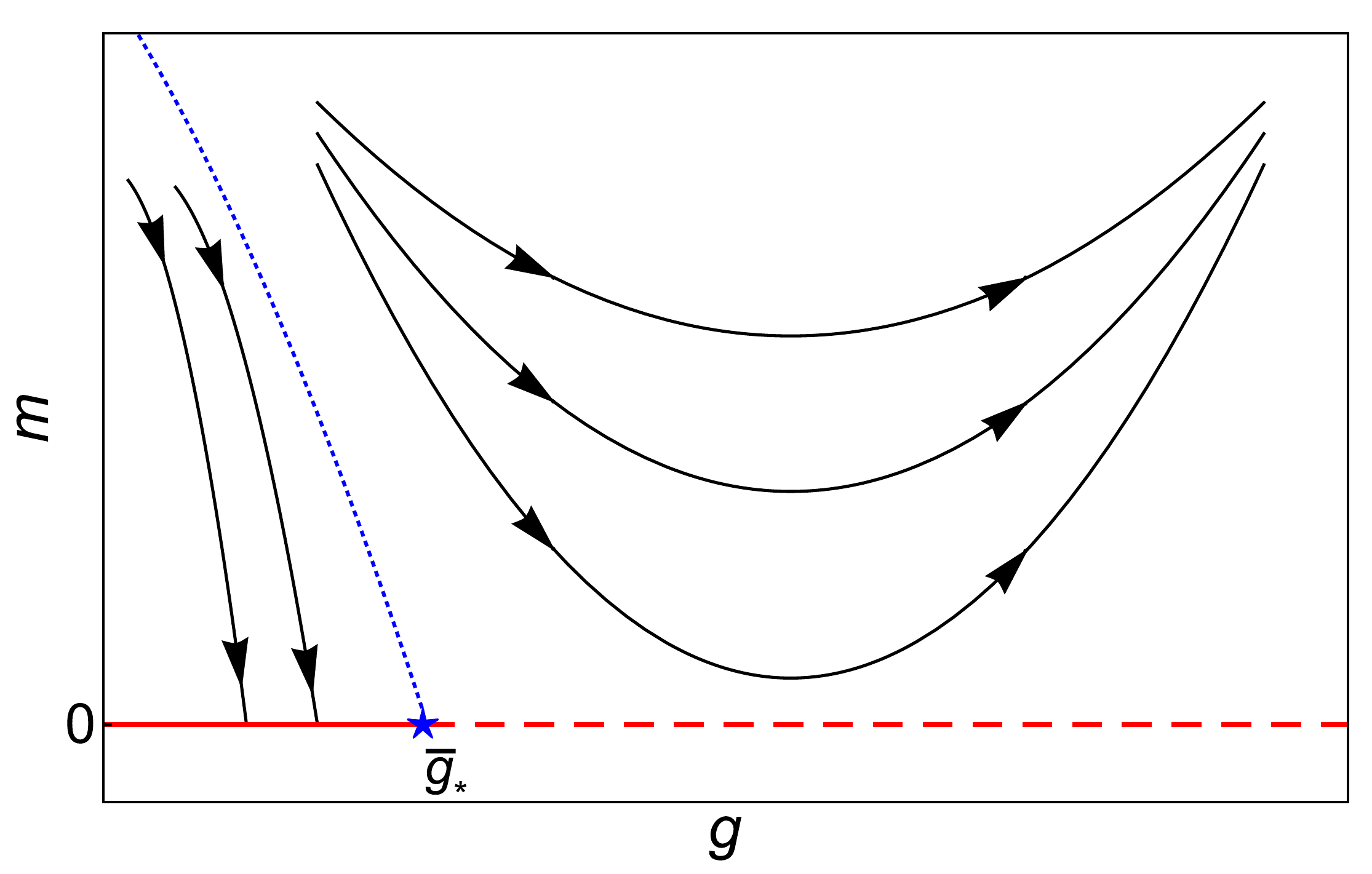}
\caption{Qualitative feature of RG flows of the massive Thirring model based on Eqs.~(\ref{eq:RGE_Thirring_for_g}) and (\ref{eq:RGE_Thirring_for_m}) in the regime where $-\pi < g$ and $m/\Lambda \lesssim 0.01$.  The arrows present the flows towards the IR limit.  The line $m=0$ is a fixed line under RG transformation.  It is separated into two sectors, with $g < \bar{g}_{\ast}$ being stable and $g > \bar{g}_{\ast}$ being unstable.}
  \label{fig:RG_flow}
\end{figure}

Employing the duality relations in Eq.~(\ref{bosonization}),
the sine-Gordon RGE's, Eqs.~(\ref{eq:RGE_sG_for_t}) and (\ref{eq:RGE_sG_for_z}), can be rendered into
\bea
\label{eq:RGE_Thirring_for_g}
 \beta_{g} &\equiv& \mu \frac{d g}{d \mu} = -64 \pi
 \left ( \frac{m}{\Lambda} \right )^{2} \, , \\
\label{eq:RGE_Thirring_for_m}
 \beta_{m} &\equiv& \mu \frac{d m}{d \mu} = m \left [ \frac{-2 (g +
   \frac{\pi}{2})}{g + \pi} - \frac{256 \pi^{3}}{(g+\pi)^{2}} \left (\frac{m}{\Lambda}\right )^{2} \right ]\, ,
\eea
with higher-order corrections in $(m/\Lambda)^{n}$.   These RGE's govern the scaling behaviour of the massive Thirring model in the limit $m/\Lambda \ll 1$.  
Equation~(\ref{eq:RGE_Thirring_for_g}) implies that $g$ always increases towards the IR limit as long as $m/\Lambda \not= 0$, although it does not signify asymptotic freedom.   It also concurs with the fact that the massless Thirring model is a conformal field theory~\cite{Thirring:1958in}.   To understand the implication of Eq.~(\ref{eq:RGE_Thirring_for_m}), we first notice that, 
for the duality relation in Eq.~(\ref{bosonization}) to be valid,
the four-fermion coupling is restricted to be in the range
\beq
\label{eq:g_lower_bound}
 g > -\pi \, .
\eeq
Since we rely on the duality to understand the RG flow of the model and thus how to correctly approach the continuum,
we need to restrict our exploration to this range of values.
With this constraint, the RGE in Eq.~(\ref{eq:RGE_Thirring_for_m}) leads to the expectation that a corresponding non-thermal phase transition occurs in the massive Thirring model at 
\beq
\label{eq:g_star}
 g = g_{\ast} \sim -\frac{\pi}{2} \, , 
\eeq
in the regime where $m/\Lambda \ll 1$,
with the exact value of $g_{\ast}$ being $(m/\Lambda)-$dependent\footnote{In Ref.~\cite{Coleman:1974bu}, it was pointed out that the spectrum of the massive Thirring model becomes unbounded at $g\sim -\pi/2$, and consequently the discussion of the duality was restricted to the regime $g \gtrsim -\pi/2$.}.   While the fermion mass, $m$, is a dimension-one coupling and remains relevant in the region $g > g_{\ast}$, it becomes irrelevant at $g < g_{\ast}$.  Obviously, equations~(\ref{eq:RGE_Thirring_for_g}) and (\ref{eq:RGE_Thirring_for_m}) predict
\beq
\label{eq:g_bar_star}
  \bar{g}_{\ast} = \lim_{\tiny{(m/\Lambda) \rightarrow 0 }} g_{\ast}(m/\Lambda) = -\frac{\pi}{2} \, .
\eeq
This means that the ``fixed line'' on the $g{-}m$ plane,  $m=0$ where both $\beta_{g}$ and $\beta_{m}$ vanish, is separated into two sectors, with $g < \bar{g}_{\ast}$ being stable and $g > \bar{g}_{\ast}$ being unstable.
It is also straightforward to show that generically, from Eq.~(\ref{eq:RGE_Thirring_for_m}), $g_{\ast}(m/\Lambda)$ decreases with growing $m/\Lambda$ under the condition $m/\Lambda \ll 1$.   Figure~\ref{fig:RG_flow} shows qualitatively the above features for the RG flows of the Thirring model in the regime where $-\pi < g$ and $m/\Lambda \lesssim 0.01$.
Furthermore, Eqs.~(\ref{eq:RGE_Thirring_for_g}) and (\ref{eq:RGE_Thirring_for_m}) predict that the $\bar{\psi}\psi$ operator becomes irrelevant in the regime $g < g_{\ast}(m/\Lambda)$, although its classical dimension is one.  This indicates that a large anomalous dimension is generated in the Thirring model.  As will be discussed in Sec.~\ref{sec:scaling_and_continuum_limit}, these features can guide the study of the continuum limit of the model.  

It should be stressed that the above discussion is based on an expansion in terms of $m/\Lambda$.  In this project, we carry out non-perturbative study for the non-thermal phase structure of the massive Thirring model through lattice simulations, employing the method of MPS.   Our investigation can shed light on the scaling behaviour of the theory beyond perturbation theory.  
As mentioned earlier in this section, results
  reported in this paper are the first step in a research programme
  where we will further explore other aspects of this model, such as
  its properties out of equilibrium and relevant real-time dynamics.
  Although the massive
  Thirring model has been extensively examined in the past, our study
  presented in this article does reproduce known properties of the
  theory with novel ingredients.  In addition to implementing a new
  strategy for research on lattice-regularized Thirring model, to our knowledge, the discussion
  in Sec.~\ref{sec:entanglement_entropy} is the first time that entanglement entropy is
  used for probing the zero-temperature phase
  structure of this quantum field theory.

The rest of this paper is organized in the following way.  Section~\ref{sec:latt_spin} contains the formalism of the theory that we simulate, and Sec.~\ref{sec:strategy} gives details of the numerical implementation.  In Sec.~\ref{sec:results}, we present main numerical computations in this project.  The outcome of these computations is used in Sec.~\ref{sec:discussion} for addressing the phase structure, the scaling behaviour, as well as the continuum limit of the massive Thirring model in 1+1 dimensions.   We then conclude in Sec.~\ref{sec:conclusion}.  Preliminary results of this work were presented in our contributions to the proceedings for the Lattice conferences in Refs.~\cite{Banuls:2017evv, Banuls:2018ckt}.

\section{Lattice formulation and the corresponding spin model}
\label{sec:latt_spin}
In this section, we first describe subtleties in constructing the Hamiltonian at the operator level in the continuum, then discuss the latticization procedure of the system and the comparison with the quantum spin-chain model.   In our numerical implementation, we use the XXZ-model Hamiltonian in Sec.~\ref{sec:XXZ}.

\subsection{The Hamiltonian operator at quantum level and the staggered regularization}
\label{eq:latt_hamiltonian}
To perform lattice simulations using the MPS approach, we first have to obtain the corresponding Hamiltonian of the classical action in Eq.~(\ref{eq-action-thirring}).   At the quantum level, the Hamiltonian operator cannot be related to the Lagrangian through a straightforward Legendre transform.   The main subtlety arises from quantum effects that modify the current conservation laws, leading to an ambiguity in defining the vector current that appears in the four-fermion operator in Eq.~(\ref{eq-action-thirring})\footnote{In 1+1 dimensions, the vector and the axial-vector currents are related to each other.}.    In the path integral formalism, these effects are easily understood via analysing anomalies that result from the fermionic measure in a field-redefinition procedure~\cite{Das:1985ci}.    When working with the operator formalism, this ambiguity can be accounted for by employing a non-local definition of the currents~\cite{Schwinger:1962tp, hagen1967new}.   As explained in the Introduction, we are interested in studying the sector of zero total fermion number in the Thirring model.  To ensure that fermion number is conserved, we choose to maintain vector-current conservation in this work.   Incorporating quantum effects in the energy-momentum tensor at the operator level, one derives the Hamiltonian~\cite{hagen1967new},
\begin{equation} 
\label{eq-H_Th-continuum}
    H_{{\mathrm{Th}}} = \int dx \left[ -i Z_{\psi}(g)\bar{\psi}\gamma^{1}\partial_{1}\psi
    + m_{0}\bar{\psi}\psi
    + \frac{g}{4} \left(\bar{\psi}\gamma^{0}\psi\right)^2 
    - \frac{g}{4}\left(1+\frac{2g}{\pi}\right)^{-1} \left(\bar{\psi}\gamma^{1}\psi\right)^2 \right] \,,
\end{equation}
where $Z_{\psi}(g)$ is the wavefunction renormalization constant that
can be found in Refs.~\cite{Mueller:1972md, Gomes:1972yb}.  Notice
that the structure of the four-fermion operators and their couplings
in $H_{{\mathrm{Th}}}$ are different from that in the action of
Eq.~(\ref{eq-action-thirring}).  This originates from the effects of quantum
anomalies discussed above.  
Furthermore, in
Eq.~(\ref{eq-H_Th-continuum}) we denote the fermion mass as $m_{0}$,
to emphasize that in the simulations the input mass is bare.  As discussed below in Sec.~\ref{sec:XXZ}, the input $g$
can be related to the spectrum of a one-dimensional spin
chain~\cite{Luther:1976mt, Bergknoff:1978bm, Alcaraz:1995JPhysA}.

In this project, we work with the standard representation of the Dirac matrices in two dimensions,   
\begin{equation}
\label{eq:gamma_matrices_std_rep}
 \gamma^{0} = \sigma^{z}, \mbox{ } \gamma^{1} = i \sigma^{y}, \mbox{ } \gamma^{5} = \gamma^{0} \gamma^{1} = \sigma^{x}\, ,
\end{equation}
where $\sigma^{i}$'s are the Pauli matrices.  In this representation, the Hamiltonian can be expressed in terms of the two components of the Dirac spinor, $\psi_{1}$ and $\psi_{2}$,
\begin{equation} 
\label{eq-H_Th-continuum-component}
    H_{{\mathrm{Th}}} = \int dx \Big\{ -i Z_{\psi} (g)\left( \psi_{2}^{*}\partial_{x}\psi_{1} \
        + \psi_{1}^{*}\partial_{x}\psi_{2}\right) \
        + m_{0}\left(\psi_{1}^{*}\psi_{1} \
        - \psi_{2}^{*}\psi_{2} \right) 
        + \tilde{g}(g) \, \psi_{1}^{*}\psi_{1}\psi_{2}^{*}\psi_{2} \Big\} \, ,
\end{equation}
where
\begin{equation} 
\label{eq-current-renor-const-continuum}
    \tilde{g}(g) = \frac{g}{2}\left[1+\left(1+\frac{2g}{\pi}\right)^{-1}\right] \,.
\end{equation} 

Having obtained the Hamiltonian operator in Eq.~(\ref{eq-H_Th-continuum-component}), we now proceed to discretize it, following the strategy as detailed in Refs.~\cite{Banks:1975gq, Susskind:1976jm}.  Upon latticizing the one-dimensional space, we implement the staggered regularization by keeping only $\psi_{1}$ on the even site, and $\psi_{2}$ on the odd site.  That is, 
\begin{eqnarray}
\label{eq-kogut-susskind-fermion}
    \psi_{1}(x) \rightarrow \frac{1}{\sqrt{a}}\,c_{2n} \, ,  \mbox{ }
    \psi_{2}(x) \rightarrow \frac{1}{\sqrt{a}}\,c_{2n+1} \, , 
\end{eqnarray}
where $a$ is the lattice spacing, and $c_{i}$ is the resulting dimensionless one-component fermion field on the $i$-th spatial lattice site.  This procedure leads to the discretized theory with
\begin{equation} 
\label{eq-H_latt}
    H_{{\mathrm{Th}}}^{{\mathrm{(latt)}}} 
        = -\frac{i}{2a} Z_{\psi}(g) \sum_{n=0}^{N-2} 
            \Big( c^{\dagger}_{n}c_{n+1} - c^{\dagger}_{n+1}c_{n} \Big)
            + m_{0}\sum_{n=0}^{N-1} \left(-1\right)^{n} c^{\dagger}_{n}c_{n} 
        + \frac{\tilde{g}(g)}{2a} \sum_{n=0}^{N-2} \
            c^{\dagger}_{n}c_{n} c^{\dagger}_{n+1}c_{n+1} \, , 
\end{equation}
with $N$ being the total number of sites.   Notice that since we are only latticizing the spatial direction in 1+1 dimensions, the staggered regularization completely removes the fermion doubling problem.  That is, $H_{{\mathrm{Th}}}^{{\mathrm{(latt)}}}$ describes only one ``taste'' of fermion with the effective lattice spacing being $2a$.

\subsection{The XXZ spin chain}
\label{sec:XXZ}

The Hamiltonian in Eq.~(\ref{eq-H_latt}) contains fermionic degrees of freedom,
which TN can in general handle without causing a loss in efficiency~\cite{Kraus2010,Corboz2010,Pineda2010}. But in the case of one spatial dimension, 
 it is more convenient to map the fermions to spin matrices by applying a Jordan-Wigner transformation,
\begin{equation}
\label{eq:JW_trans}
     S_{n}^{-} = \exp\left[ -\pi i \sum_{j=1}^{n-1} \left(c_{j}^{\dagger}c_{j}+\frac{1}{2}\right) \right] c_{n} \,,  \mbox{ }
     S_{n}^{+} =  c_{n}^{\dagger} \, \exp\left[ \pi i \sum_{j=1}^{n-1} \left(c_{j}^{\dagger}c_{j}+\frac{1}{2}\right) \right] \, , 
\end{equation}
where $S_{n}^{\pm} = S_{n}^{x} \pm i S_{n}^{y}$, and $S^{\alpha}_{n} = \sigma^{\alpha}/2$ ($\alpha=x,\,y,\,z$) with $[S^{\alpha}_{n}, S^{\beta}_{m}] = 0$ when $n \not= m$.   Application of this transformation on $H_{{\mathrm{Th}}}^{{\mathrm{(latt)}}}$ results in the quantum spin chain,
\begin{equation}
\label{eq-H_spin}
    H_{{\mathrm{Th}}}^{{\mathrm{(spin)}}} 
        = -\frac{Z_{\psi}(g)}{2a} \sum_{n=0}^{N-2} \left( S_{n}^{+}S_{n+1}^{-} 
            + S_{n+1}^{+}S_{n}^{-} \right)
            + m_{0} \sum_{n=0}^{N-1} \left(-1\right)^{n} \left( S_{n}^{z}+\frac{1}{2} \right) 
       + \frac{\tilde{g}(g)}{2a} \sum_{n=0}^{N-1} \left( S_{n}^{z}+\frac{1}{2} \right) \ 
            \left( S_{n+1}^{z}+\frac{1}{2} \right) \,.
\end{equation}
In this work, we adopt open boundary conditions. This 
avoids the appearance of boundary operators that make the Jordan-Wigner transformation more complicated than Eq.~(\ref{eq:JW_trans})
and, most importantly for our simulations, it allows us to use the most efficient variants of the MPS algorithms
\footnote{Notice, nevertheless, that periodic boundary conditions can also be treated by MPS~\cite{Pippan2010pbc}.}.


The spin-chain Hamiltonian in Eq.~(\ref{eq-H_spin}) is precisely that
of the XXZ model coupled to staggered  and uniform external fields,
in a regime in which the latter is equal to the anisotropy.  
But using exactly the form in Eq.~\eqref{eq-H_spin} turns out not to be the adequate strategy,
as the choice of parameters in terms of $g$ does not account for the lattice effects,
an issue that has been studied in the literature of condensed matter physics~\cite{Luther:1976mt, Luscher:1976pb}.
Based on a matching of the soliton bond-state spectrum 
between the exactly solvable field theory at $m_0=0$ and the Bethe-ansatz solution of the XXZ chain,
the correct choice was found to be
\begin{equation}
Z_{\psi}(g) \to \nu(\gamma),\quad \tilde{g}(g) \to \tilde{\Delta}(\gamma),
\end{equation}
with
\begin{equation}
\label{eq:nu_and_Delta_tilde}
    \nu(\gamma) = \frac{2 \gamma}{\pi\sin (\gamma )} \, , \mbox{ }
    \tilde{\Delta}(\gamma) = \frac{4 \gamma}{\pi}\cot (\gamma ) \, , 
\end{equation}
where the parameter $\gamma$ can be related to the four-fermion
coupling, $g$, in the Thirring model by~\cite{Luther:1976mt, Alcaraz:1995JPhysA},
\begin{equation}
\label{eq:gamma_and_g}
  \gamma = \frac{\pi - g}{2} .
\end{equation}

%
Notice that the functions $\tilde{\Delta}(g)$ and $\tilde{g}(g)$
(in Eq.~(\ref{eq-current-renor-const-continuum})) agree to
${\mathcal{O}}(g^{2})$.   The parameter $\nu(\gamma)$ accounts for the
wavefunction renormalization.

For convenience,  we define 
\begin{equation}
\label{eq:H_XXZ__to_sim}
 H_{{\mathrm{XXZ}}} = \frac{\nu (g)}{a} \bar{H}_{{\mathrm{sim}}} ,
\end{equation}
where
\begin{equation}
\label{eq:H_sim}
 \bar{H}_{{\mathrm{sim}}} = -\frac{1}{2} \sum_{n}^{N-2} \left( S_{n}^{+}S_{n+1}^{-} 
            + S_{n+1}^{+}S_{n}^{-} \right)
            + a \tilde{m}_{0} \sum_{n}^{N-1} \left(-1\right)^{n} \left(
              S_{n}^{z}+\frac{1}{2} \right)  
        + \Delta (g) \sum_{n}^{N-1} \left( S_{n}^{z}+\frac{1}{2} \right) \ 
            \left( S_{n+1}^{z}+\frac{1}{2} \right) \,,
\end{equation}
with 
\begin{equation}
\label{eq:m_tilde_0_and_Delta}
 \tilde{m}_{0} = \frac{m_{0}}{\nu} \, , \mbox{ } \Delta (g) =\cos \left (
   \frac{\pi - g}{2} 
 \right ) \, .
\end{equation}
This Hamiltonian, $\bar{H}_{{\mathrm{sim}}}$, is what we actually
implement for the simulations in this work.  That is, the exploration
of the phase structure of the (1+1)-dimensional massive Thirring model
is performed via scanning the two spin-chain parameters, $a\tilde{m}_{0}$
and $\Delta(g)$ in Eq.~(\ref{eq:H_sim}), that can be translated back
to the two couplings in Eq.~(\ref{eq-H_Th-continuum}). It is
  obvious from Eq.~(\ref{eq:m_tilde_0_and_Delta}) that we only explore
the regime $-1 < \Delta(g) \le 1$.  This corresponds to $-\pi < g \le \pi$,
which encompasses the expected phase-transition point, $g\sim -\pi/2$,
discussed in Sec.~\ref{sec:intro}, and keeps $\Delta (g)$ single-valued.  

The XXZ Hamiltonian with a uniform magnetic field has been profusely studied in the literature.
The simultaneous presence of a staggered component in the same direction affects the phase diagram,
but has been much less explored, beyond the initial work by Alcaraz et al.~\cite{Alcaraz:1995JPhysA}, 
and some recent results for particular values of the parameters~\cite{Moradmard2014}.
Furthermore, as discussed in the next section, our
  simulation restricts the system to be in the sector of vanishing total
  $z-$component of the spin.  This is necessary in order to employ
  duality properties to understand numerical results.

\section{Strategy for numerical simulations}
\label{sec:strategy}
\subsection{Matrix product states}
In this paper, we are interested in the phase structure of the Thirring model at zero temperature.
Hence, we need to search for the ground state of the system for different sets of parameters.
For small system sizes, it is possible to find the ground state numerically, either by diagonalizing the full Hamiltonian or by targeting only the very lowest state, e.g.\ with the Lanczos algorithm.
However, the main difficulty in this approach is that when one enlarges the system size, $N$, the dimension of the Hilbert space increases exponentially with $N$.
This exponential explosion can sometimes be evaded by using tensor network methods~\cite{Verstraete2008,Cirac2009rg,Schollwoeck2011,Huckle2013tns,Orus:2013kga,Silvi2019tns}. 
In particular, the MPS ansatz~\cite{fannes92fcs,Oestlund1995,Vidal2003,Verstraete2004,perez07mps} is especially adequate for one dimensional quantum many-body systems.
For a one-dimensional lattice system with $N$ sites, each of them with
a $d-$dimensional local Hilbert space generated 
by $\{\ket{\sigma_i}\}_{\sigma_i=1}^d$, 
a generic quantum state can be written
\begin{equation}
\label{eq:state}
    |\Psi\rangle = \sum_{\sigma_{1},\cdots,\sigma_{N}} c_{\sigma_{1}\cdots\sigma_{N}}\,|\sigma_{1}\cdots\sigma_{N}\rangle \,.  
\end{equation}
In a MPS, the basis coefficients  $c_{\sigma_{1}\cdots\sigma_{N}}$ adopt the particular form,
\begin{equation}
\label{eq:MPS}
    |\Psi\rangle = \sum_{\sigma_{1},\cdots,\sigma_{N}} \mathrm{tr}[M_1^{\sigma_{1}}M_2^{\sigma_{2}} \cdots M_{N-1}^{\sigma_{N-1}}M_N^{\sigma_{N}}]\,|\sigma_{1}\cdots\sigma_{N}\rangle \,,
\end{equation}
where each $M_k^{\sigma}$ is a $D\times D$ matrix. With open boundaries, $M_1^{\sigma}$ and $M_N^{\sigma}$ are instead $D-$dimensional vectors, and the trace reduces to the simple product of matrices. 

The parameter $D$ is called the bond dimension and determines the maximum amount of entanglement in the state. 
With respect to a bipartition of the chain in two regions $A|B$, the entanglement entropy is defined as the von Neumann entropy of 
the reduced state,
\begin{equation}
\label{eq:entanglement_entropy_generic}
    S_{A|B} \equiv S(\rho_A)= -\mathrm{tr}( \rho_A \log \rho_A)=-\mathrm{tr}( \rho_B \log \rho_B),
\end{equation}
where $\rho_A=\mathrm{tr}_{B}\ket{\Psi}\bra{\Psi}$ is the reduced density matrix for subsystem $A$ (and analogously for $B$, $\rho_B=\mathrm{tr}_{A}\ket{\Psi}\bra{\Psi}$)
\footnote{For a bipartition of a pure state $\ket{\Psi}$, both $\rho_A$ and $\rho_B$ have the same spectrum, determined by the Schmidt decomposition of $\ket{\Psi}$ with respect to the bipartition,
and the entropy of both reduced states is the same, $S(\rho_A)=S(\rho_B)$.}
. In the generic case, the entropy 
of $A$ can be as large as the number of sites it contains (its volume). If the state is a MPS and $A$ is a subchain, it is easy to see that
$S_{A|B} \leq 2 \log D $, i.e. it  is bounded from above by a constant independent of the system size. 
The MPS ansatz hence satisfies by construction an area law~\cite{Eisert2010}.

Any state $\ket{\Psi}$ can be exactly written in the MPS form with bond dimension $D\leq d^{\lfloor N/2 \rfloor}$.
Alternatively, a MPS approximation with a maximum bond dimension $D_{\mathrm{cut}}$ can in principle be found by 
performing a sequence of singular value decompositions (SVD)\footnote{Notice that for a general state this still has an exponential cost, although approximation schemes have been proposed~\cite{Savostyanov2011approx}.}
across each bond of the chain, and retaining only the largest $D_{\mathrm{cut}}$ singular values 
for each of them~\cite{Vidal2003,perez07mps}.
The distance to the true state can be bounded by the discarded weights, which thus give an estimation of the truncation error.

For the purpose of practical calculations, the representation is useful only if $D$ is relatively small.
This is in fact the case for states which satisfy certain entropic area law~\cite{Schuch2008}, and, in particular
for ground states of local Hamiltonians with a gap~\cite{Verstraete2006,Hastings2007,Hastings2007a}.
Even in the gapless case, where the entropy presents logarithmic corrections to the area law~ \cite{Calabrese:2004eu}, a MPS approximation is possible with bond dimension that scales
only polynomially (not exponentially) with the system size~\cite{Verstraete2006}.

These properties, together with the existence of several efficient algorithms for finding MPS approximations to the
ground states of one-dimensional problems~\cite{Verstraete2008,Schollwoeck2011,Huckle2013tns,Orus:2013kga,Silvi2019tns},
make the MPS ansatz one of the most precise numerical methods for strongly correlated models in one spatial dimension.

Here we use a variational search to optimize the MPS that minimizes the energy with respect to the Hamiltonian (\ref{eq:H_sim}).
The algorithm, closely related to the density matrix renormalization group\footnote{To our knowledge, there has not been extensive
  exploration of quantum field theories, as well as the XXZ spin model
  in Eq.~(\ref{eq:H_sim})
  with the presence of a staggered magnetic field, using the original DMRG
  approach.} (DMRG)~\cite{white92dmrg,schollwoeck05dmrg},
has a computational cost that scales as $N D^3$ (for open boundaries)\footnote{Variants of the algorithm exist also for periodic boundary conditions, in which case the cost scales with $N D^5$~\cite{Pippan2010pbc,Pirvu2011pbc}.} . 
The method proceeds by optimizing the $d\times D \times D$ dimensional tensors $M$ one by one, in a sequential manner, sweeping iteratively over the chain back and forth until the desired level of convergence has been reached in the energy. Namely, we choose to stop the iteration when
 the relative change in energy is smaller than the tolerance $\epsilon=10^{-7}$.
Because the optimization of a local tensor can only decrease the energy, the algorithm is guaranteed to converge (although it could be to a local minimum).
This provides an approximation to the ground state of the model, which can be systematically improved by increasing the bond dimension.

\subsection{Matrix product operator}
\label{sec:MPO}
The form of the Hamiltonian most suitable for numerical calculations is given in Eq.~(\ref{eq:H_sim}), representing a spin-1/2 system with a local Hilbert space of dimension two.
Since the duality between the massive Thirring model and the sine-Gordon model is only valid in the zero-charge sector, the simulations should be performed in this sector, corresponding in the language of the spin model to $S^{z}_{tot} = \sum_{n}S^{z}_{n}=0$. 
Within the MPS framework, it is possible to construct states that satisfy this constraint explicitly, by imposing the proper structure in the tensors $M$. In this work, instead, we use generic tensors, and introduce a penalty term in the Hamiltonian to ensure that the ground state is in the desired $S^{z}_{tot} $ sector, namely
\begin{equation} \label{eq-penalty-term}
    \bar{H}_{{\mathrm{sim}}}^{{\mathrm{penalty}}} = \bar{H}_{{\mathrm{sim}}} + \lambda \left( \sum_{n=0}^{N-1} S_{n}^{z} - S_{{\mathrm{target}}} \right)^2 \,,
\end{equation}
where the magnitude of $\lambda$ should be chosen large enough to ensure that all the undesired states have energy above the lowest state of the targeted sector. 

For efficient contraction of MPSs with an operator, such as the Hamiltonian, the operator has to be expressed also in the tensor network language, i.e.\ as the so-called Matrix Product Operator (MPO) \cite{pirvu10mpo,McCulloch2007}.
We find that the MPO representation of Eq.~(\ref{eq-penalty-term}) is given by
\begin{subequations}
\begin{align}
    & W^{[0]}=
    \begin{pmatrix}
         \mathbb{1}  & -\frac{1}{2} S^{+} & -\frac{1}{2} S^{-} & 2\lambda S^{z} & \Delta S^{z} & \beta_{n}S^{z}+\alpha\mathbbm{1} \\
    \end{pmatrix} \,,\\
    & W^{[N-1]}=
    \begin{pmatrix}
        \beta_{n}S^{z}+\alpha\mathbbm{1} \\ S^{-} \\ S^{+} \\ S^{z} \\ S^{z} \\ \mathbbm{1}\\
    \end{pmatrix} 
\end{align}    
\end{subequations}
for the tensors at the boundaries and
\begin{equation}
    W^{[n]}=
    \begin{pmatrix}
        \mathbbm{1} & -\frac{1}{2} S^{+} & -\frac{1}{2} S^{-} & 2\lambda S^{z} & \Delta S^{z} & \beta_{n}S^{z}+\alpha\mathbbm{1} \\
        0 & 0 & 0 & 0 & 0 & S^{-} \\
        0 & 0 & 0 & 0 & 0 & S^{+} \\
        0 & 0 & 0 & \mathbbm{1} & 0 & S^{z} \\
        0 & 0 & 0 & 0 & 0 & S^{z} \\
        0 & 0 & 0 & 0 & 0 & \mathbbm{1}
    \end{pmatrix} \,,
\end{equation}
in the bulk, where
\begin{equation}
    \beta_{n} = \Delta + \left(-1\right)^{n} \tilde{m}_{0}a - 2\lambda \, S_{target} \,,\, \alpha = \lambda \left( \frac{1}{4} + \frac{S_{target}^{2}}{N} \right) + \frac{\Delta}{4} \,.
\end{equation}

\subsection{Simulation details}
\label{sec:simulation}
We describe now our simulation strategy.
The variational search begins with a randomly-initialized MPS with bond dimension $D=50$. 
To have reliable results, we aim to observe convergence in $D$.
With several different values of $D$, one can investigate the truncation error systematically, and extrapolate the physical quantities to the infinite-$D$ limit (see Sec.~\ref{sec:fermion_bilinear_condensate}).
Having results with $D=50$, we gradually increase the bond dimension to 100, 200, 300, 400, 500 and finally 600.
To do so, the size of the optimized tensors is increased to the desired $D$ value, and the additional components are
initialized to zero or a small random number. 
This MPS is used as initial guess for the variational procedure, which is run again until convergence.  
This is repeated, successively increasing the bond dimension, until our final $D=600$ is reached.

Similarly, we study finite size effects, using four system sizes, $N = 400, 600, 800, 1000$, which allows us to perform an infinite volume extrapolation.
We cover a wide parameter range to study the phase structure. 
The coupling $\Delta(g)$ is chosen from the range $-0.9 \leq \Delta \leq 1.0$, with five different masses, $\tilde{m}_0 a = 0.0, 0.1, 0.2, 0.3, 0.4$. 
To study the mass dependence in more detail, for some values of the coupling, we simulate also additional masses, $\tilde{m}_0 a=0.005,0.01,0.02,0.03,0.04,0.06,0.08,0.13,0.16$.
We set the parameter of the penalty term, $\lambda$, to 100, and target the zero-charge sector ($S_{target} = 0$).

\begin{figure}[!t]
  \centering
\includegraphics[width=8.8cm]{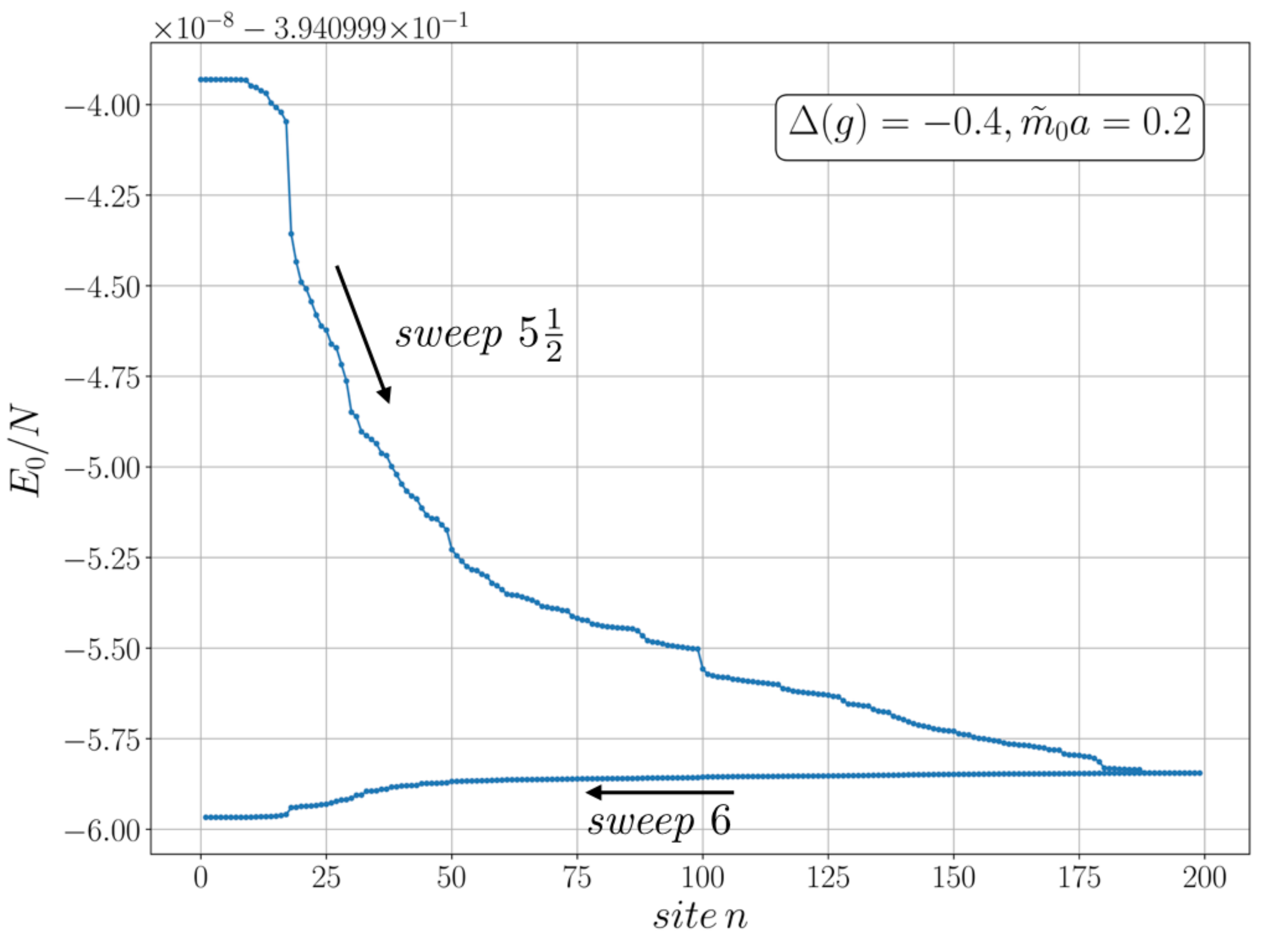}
\hspace{0.0cm}
\includegraphics[width=8.8cm]{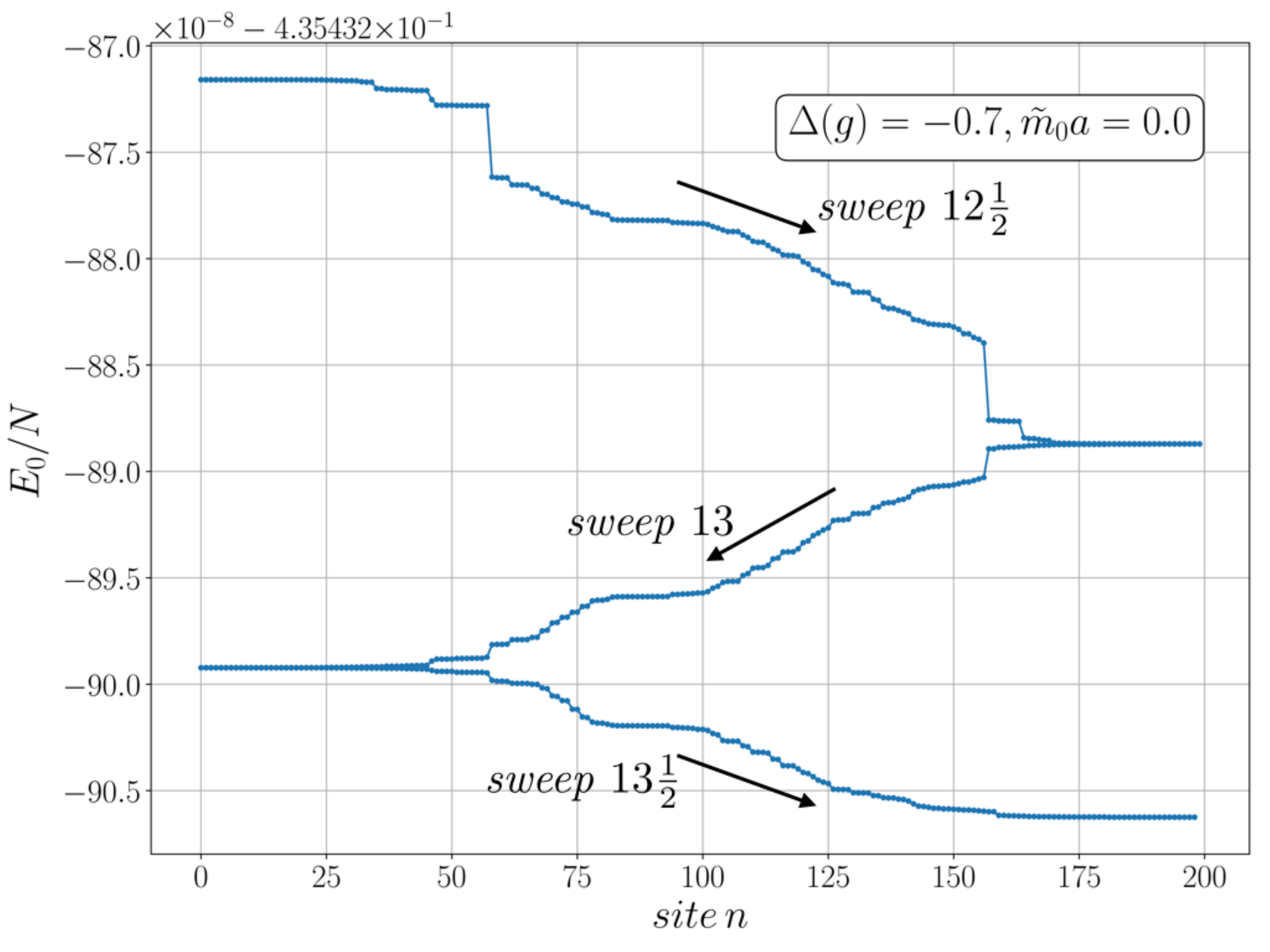}
  \caption{Fast (left) and slow (right) convergence of the variational
    algorithm in our simulations.}
  \label{fig:DMRG_convergence}
\end{figure}

In performing the search of the ground state using the variational method,
we observe that the convergence of the algorithm is slower 
in some regions of the parameter space, namely for $m_0 a=0$ and, in the massive case, for large negative $\Delta(g)$.
This is consistent with a regime where the theory may become critical. 
Figure~\ref{fig:DMRG_convergence} shows examples of these fast- and slow-convergence cases.   
For the slow cases, not only it takes more sweeps for the algorithm to converge, but also iterations of the Jacobi-Davidson solver 
used to solve for the local tensors
are also significantly more time-consuming.

\section{Numerical results for probing the phase structure}
\label{sec:results}
This section describes numerical computations for quantities that can be employed to probe the non-thermal
phase structure of the Thirring model.   As indicated by the
perturbative RGE's, Eq.~(\ref{eq:RGE_Thirring_for_g}) and
(\ref{eq:RGE_Thirring_for_m}) in Sec.~\ref{sec:intro}, it is expected that
there are at least two phases in the massive Thirring model, with the
$\bar{\psi}\psi$ operator in Eq.~(\ref{eq-action-thirring}) being
relevant in one of them and irrelevant in the other.   Since this
$\bar{\psi}\psi$ operator is dual to the cos$(\phi)$ term in the
sine-Gordon theory in Eq.~(\ref{eq-action-SG}), one envisages that in
the regime where $\bar{\psi}\psi$ is irrelevant in the Thirring model, the
corresponding bosonic theory is free.   Furthermore, since we are investigating
two-dimensional systems, and the sine-Gordon model is closely related
to the XY model~\cite{Huang:1990via}, it is foreseen that the phase
transition in the Thirring model is of BKT type.  Below, we show that these
features in the phase structure can be observed using the
ground state obtained with the technique of MPS described in Sec.~\ref{sec:strategy}.

In the following, we demonstrate calculations for the entanglement
entropy (Sec.~\ref{sec:entanglement_entropy}), the fermion bilinear
condensate, $\la \bar{\psi} \psi \ra$ (Sec.~\ref{sec:fermion_bilinear_condensate}),
as well as density-density and fermion-antifermion correlators
(Sec.~\ref{sec:correlators}).  Our results of these four objects can
be employed to obtain knowledge of the phase structure that we
summarize in the next section.


\subsection{Entanglement entropy}
\label{sec:entanglement_entropy}
To investigate the non-thermal phase structure of the massive Thirring model
in 1+1 dimensions, we first study the von Neumann entanglement entropy
for the ground state of the XXZ spin chain with finite spatial extent of size $N$ (cf. Eq.~(\ref{eq:entanglement_entropy_generic})).
Cutting the  $n$-th link of the chain divides the chain in two parts, containing $n$, and $N-n$
sites,  for $n = 0, 1, 2, \ldots , N-1 $.
If the state is a MPS, the corresponding Schmidt decomposition gives at most $D$ non-zero values, $\{s_{\alpha}(n)\}$, 
which can be  easily recovered from the MPS representation~\cite{perez07mps}.
The entanglement entropy with respect to this cut can thus be computed as 
\beq
\label{eq:finite_size_entanglement_entropy}
 S_{N} (n) = - \sum_{\alpha=1}^{D} s_{\alpha}(n)^2 \  \mathrm{ln} \ s_{\alpha}(n)^2.
\eeq
Since the XXZ spin chain studied in this work is
equivalent to the massive Thirring model, this entanglement entropy,
computed in the spin model, is
a useful tool to probe properties of the ground state in the latter.
In particular, it can be employed to identify critical points in the
field theory.
As demonstrated by Calabrese and Cardy~\cite{Calabrese:2004eu,
Calabrese:2009qy}, at criticality, $S_{N}(n)$ exhibits the scaling behaviour,
\begin{equation} 
\label{eq-entropy-cardy-finite-sys}
    S_{N}(n) = \frac{c}{6}\mbox{ } {\mathrm{ln}}\left[\frac{N}{\pi}\sin\left(\frac{\pi n}{N}\right)\right] + k  \,,
\end{equation}
where $c$ is the central charge, and $k$ is a constant that can be
argued to reflect the boundary effects.

\begin{figure}[!t]
  \centering
\includegraphics[width=6.3cm]{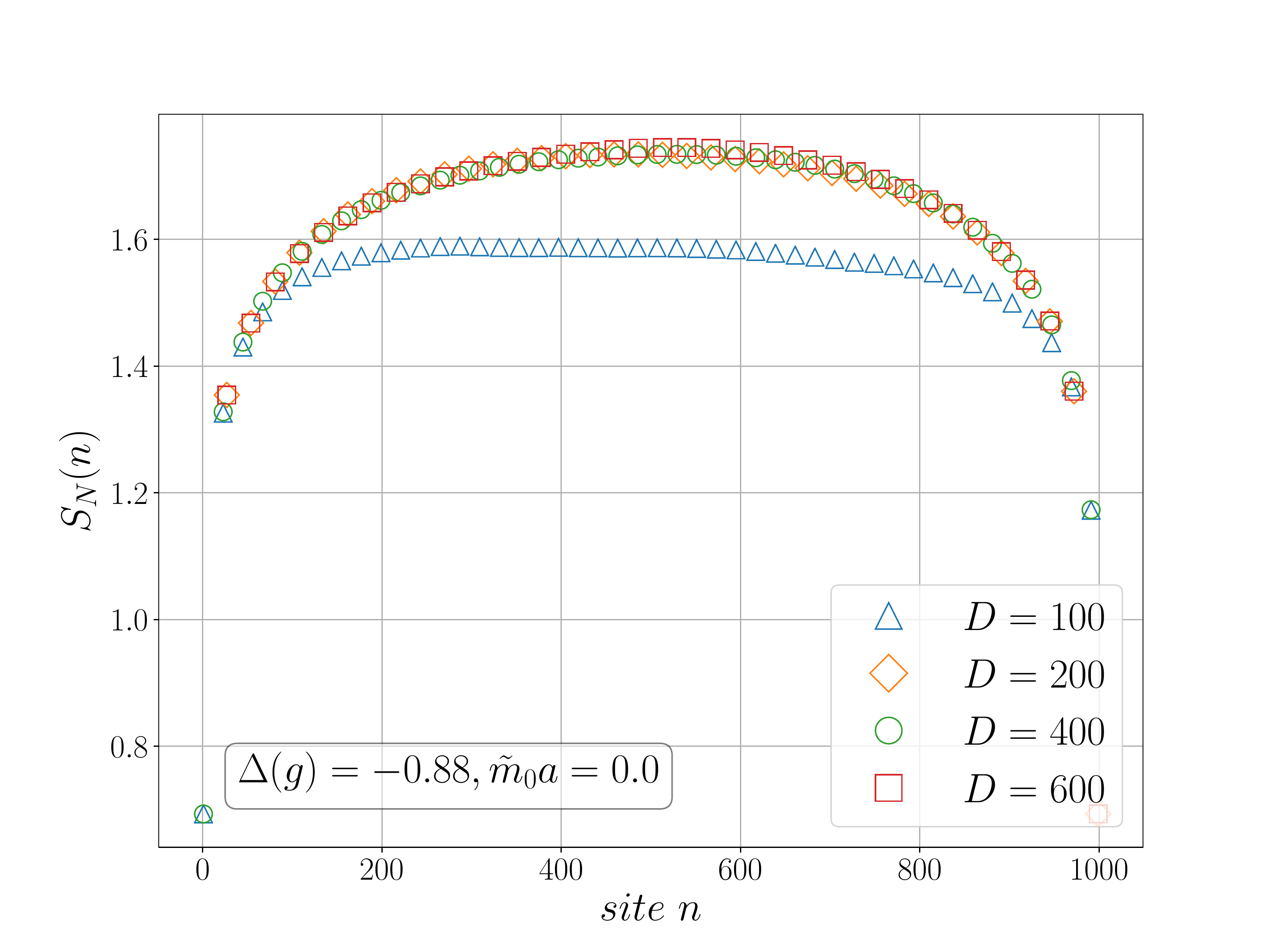}
\hspace{1.0cm}
\includegraphics[width=6.3cm]{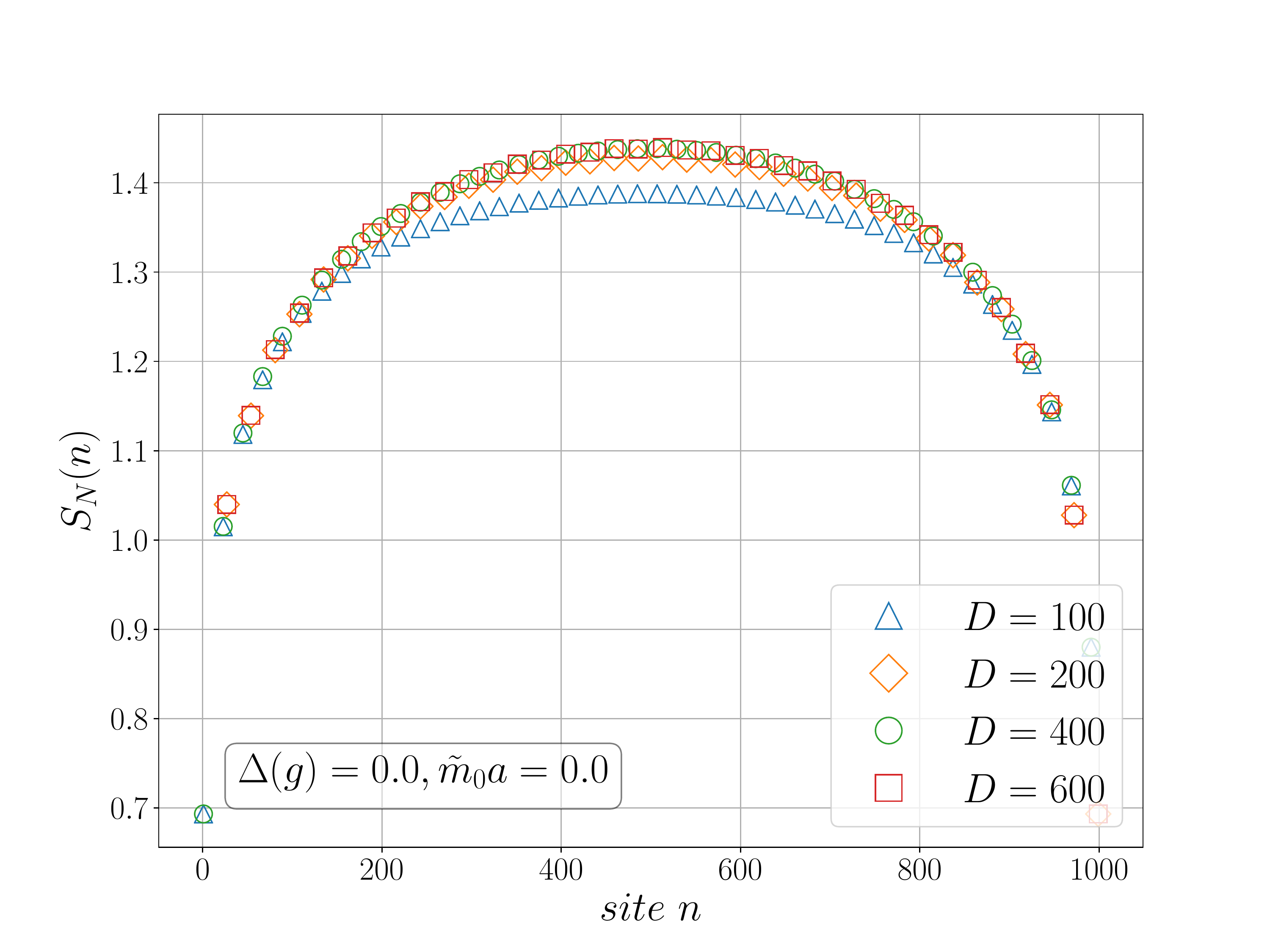}\\
  \caption{Entanglement entropy, $S_{N}(n)$, for $\tilde{m}_{0} a = 0.0$, at
  $\Delta (g) = -0.88$ (left) and $\Delta (g) = 0.0$ (right).}
  \label{fig:entropy_am_0}
\end{figure}

\begin{figure}[!t]
  \centering
\includegraphics[width=6.3cm]{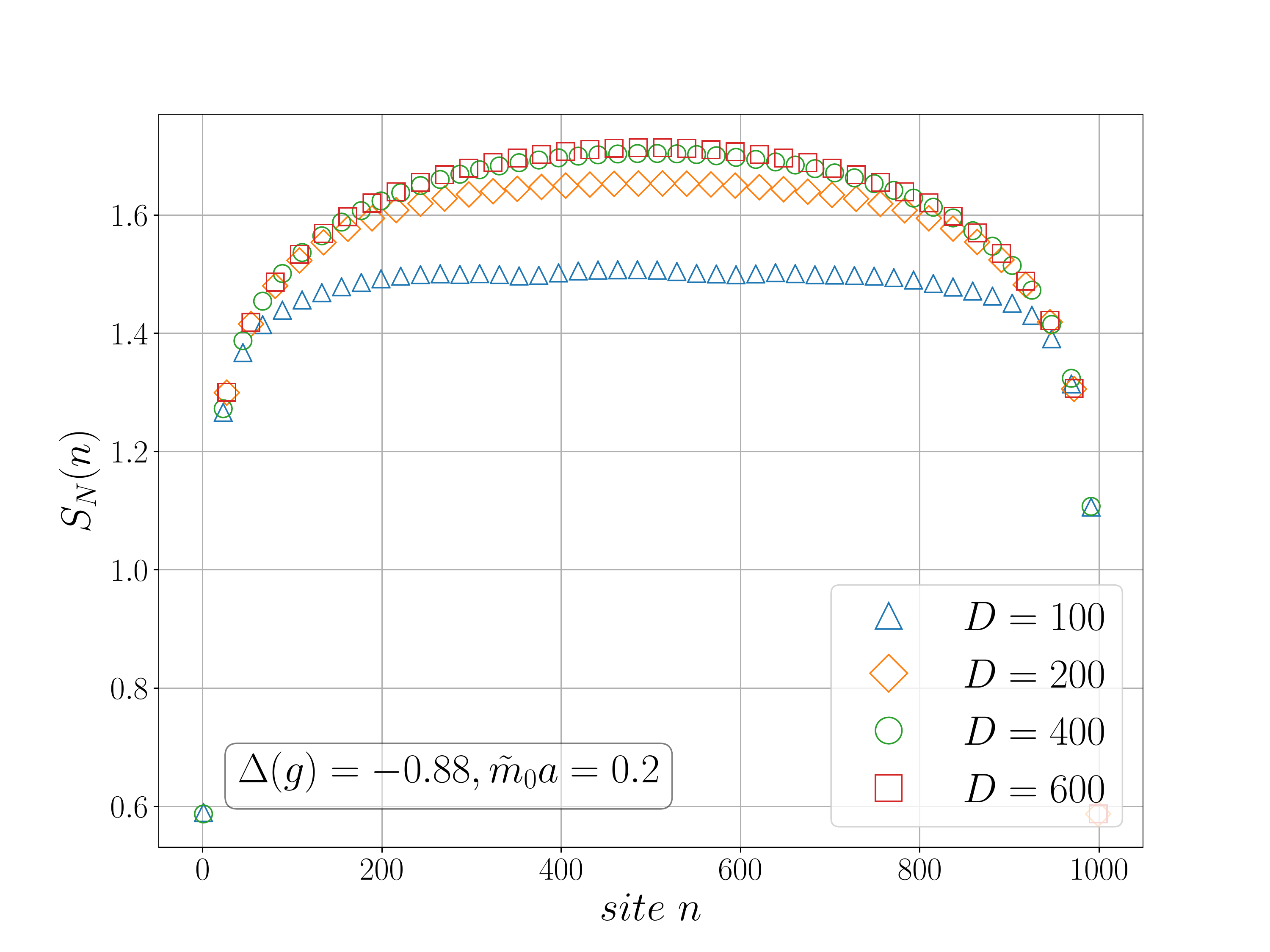}
\hspace{-0.7cm}
\includegraphics[width=6.3cm]{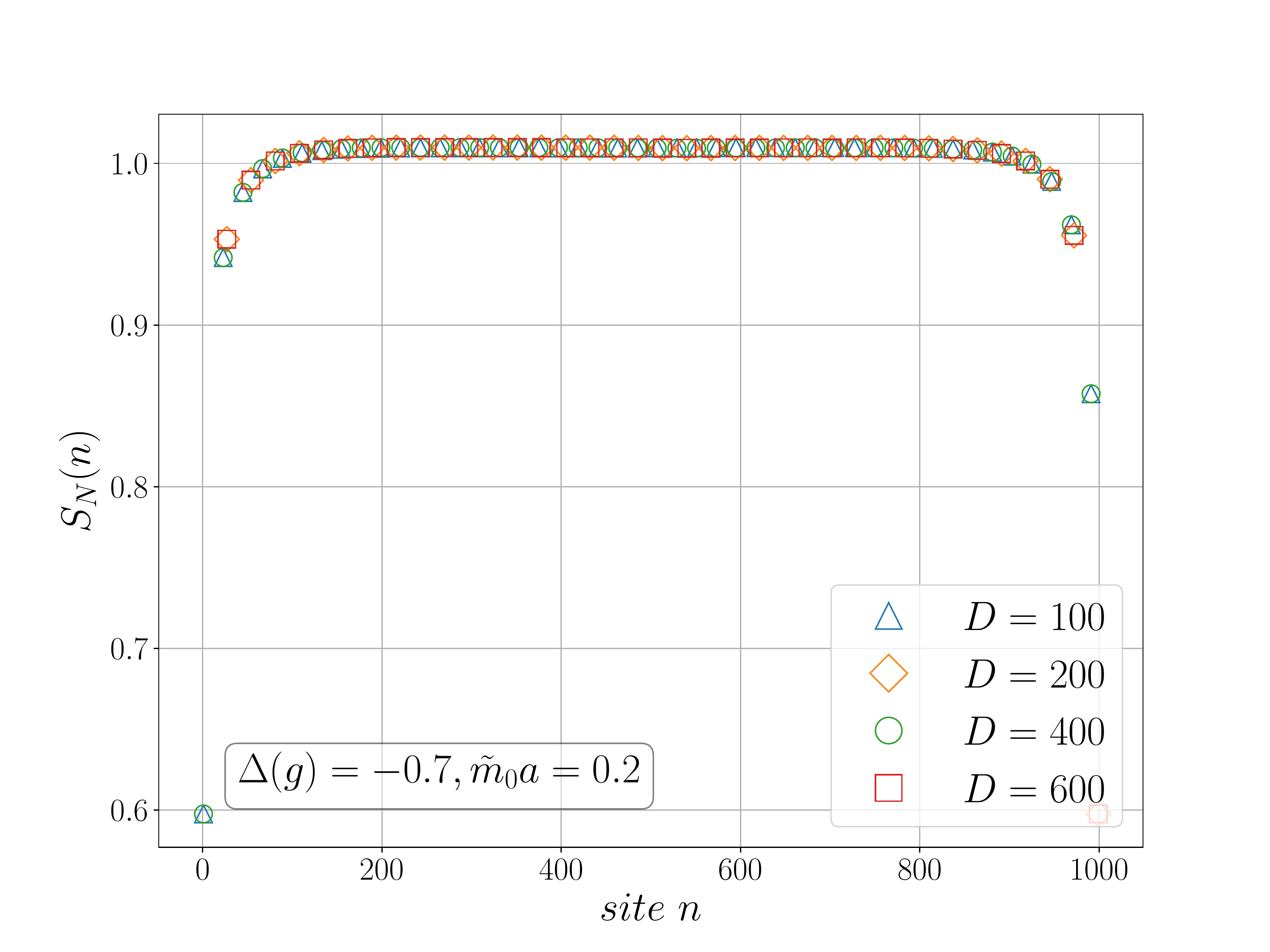}
\hspace{-0.7cm}
\includegraphics[width=6.3cm]{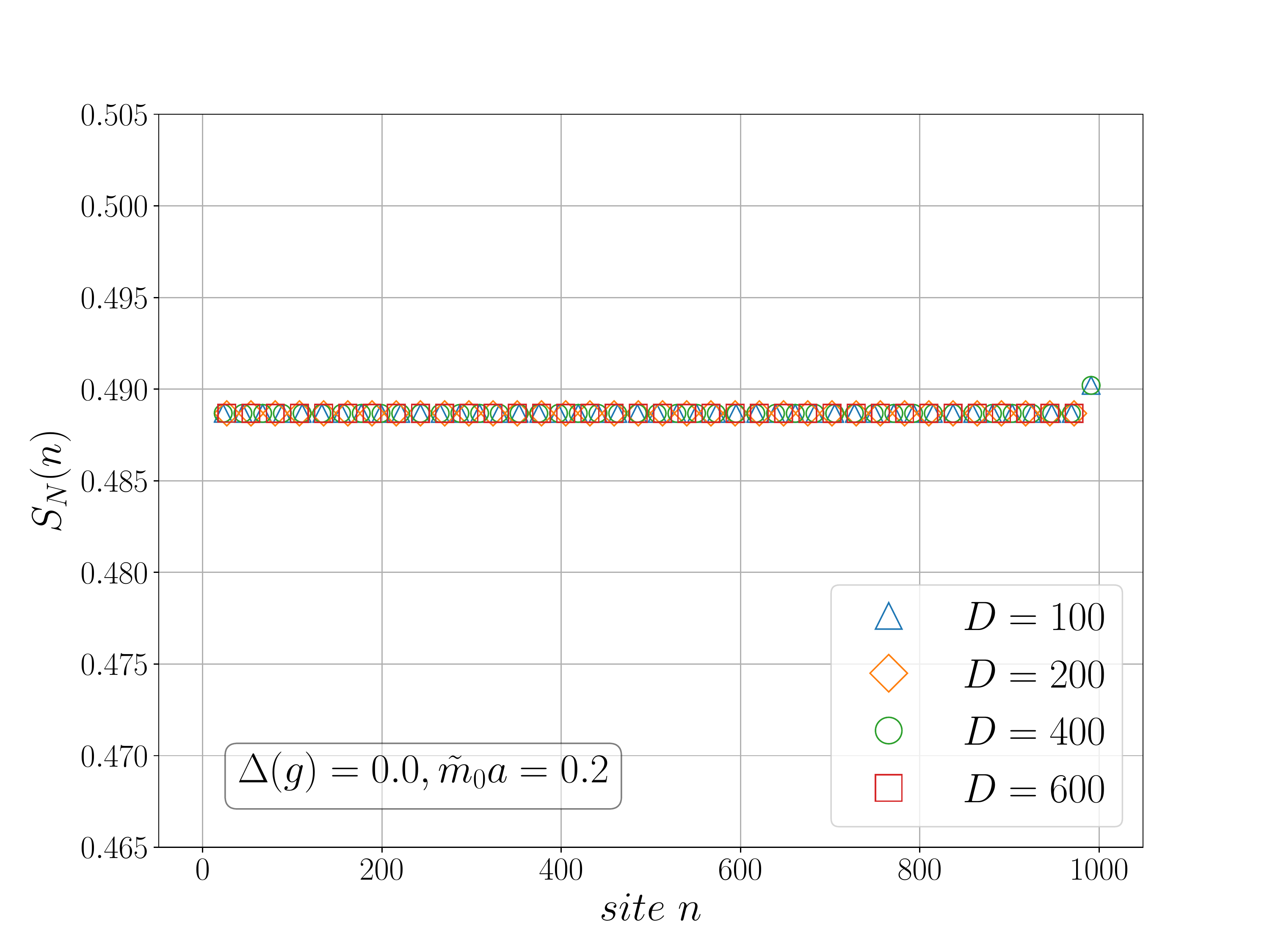}
  \caption{Entanglement entropy, $S_{N}(n)$, for $\tilde{m}_{0} a = 0.2$, at
  $\Delta (g) = -0.88$ (left),  $\Delta (g) = -0.7$ (middle) , and $\Delta (g) = 0.0$ (right).}
  \label{fig:entropy_am_point2}
\end{figure}

Figure~\ref{fig:entropy_am_0} shows examples of the
$S_{N}(n)$ from our analysis at $\tilde{m}_{0}a = 0$.  
In both plots displayed in this figure, the critical Calabrese-Cardy scaling
behaviour is manifest when the bond dimension is large enough. This property is present at all
values of $\Delta (g)$ for vanishing $\tilde{m}_{0}a = 0$, signalling
that the system is at criticality.  In
Fig.~\ref{fig:entropy_am_point2}, we exhibit results from performing the same scaling test for 
$\tilde{m}_{0}a = 0.2$ at three choices of $\Delta (g)$. 
In this case, as in results from all other  simulations with non-vanishing
$\tilde{m}_{0}a$, the Calabrese-Cardy scaling is observed
only for $\Delta (g)$
smaller than a certain value,
$\Delta_{\ast}$, which depends on $\tilde{m}_{0}a$ 
$[\Delta_{\ast} = \Delta_{\ast} (\tilde{m}_{0}a)]$.  This is indicated
by the left panel of Fig.~\ref{fig:entropy_am_point2}.  In the middle
panel of the same figure, we show a case where $\Delta (g) \gtrsim
\Delta_{\ast}$.  It is clear that the scaling behaviour is no longer
present.  Finally, when $\Delta (g) \gg
\Delta_{\ast}$ (right panel of Fig.~\ref{fig:entropy_am_point2}), the
entanglement entropy, $S_{N}(n)$, is almost independent of $n$ and its
value is small.  This gives strong hints that the theory is in a
gapped phase
at $\Delta > \Delta_{\ast}$ when $\tilde{m}_{0}a \not= 0$\footnote{If one forces a fit of data presented in
  the right panel of Fig.~\ref{fig:entropy_am_point2} to the scaling
  formula, Eq.~(\ref{eq-entropy-cardy-finite-sys}), the resultant
  central charge is zero.  
  This brings tension with the C-theorem.   
  It will also be shown in Sec.~\ref{sec:correlators}
  that in this regime, the study of fermion correlators confirms that
  the theory is not in the conformal phase, 
  and the critical scaling predicted by Calabrese and Cardy does not apply.}.
   Furthermore, our data show
a clear trend that $\Delta_{\ast}$ decreases with increasing
$\tilde{m}_{0}a$.   In the regime of very small $\tilde{m}_{0}a$
($\lesssim 0.04$), we see that $\Delta_{\ast} \sim -1/\sqrt{2}$.  This
is what one expects from the small-mass RGE's,
Eqs.~(\ref{eq:RGE_Thirring_for_g}) and (\ref{eq:RGE_Thirring_for_m}),
as discussed in Sec.~\ref{sec:intro}.   The above features of the
phase structure are confirmed by our analysis of fermion correlators,
which will be discussed in Sec.~\ref{sec:correlators}.

\begin{figure}[!t]
  \centering
\includegraphics[width=9.0cm]{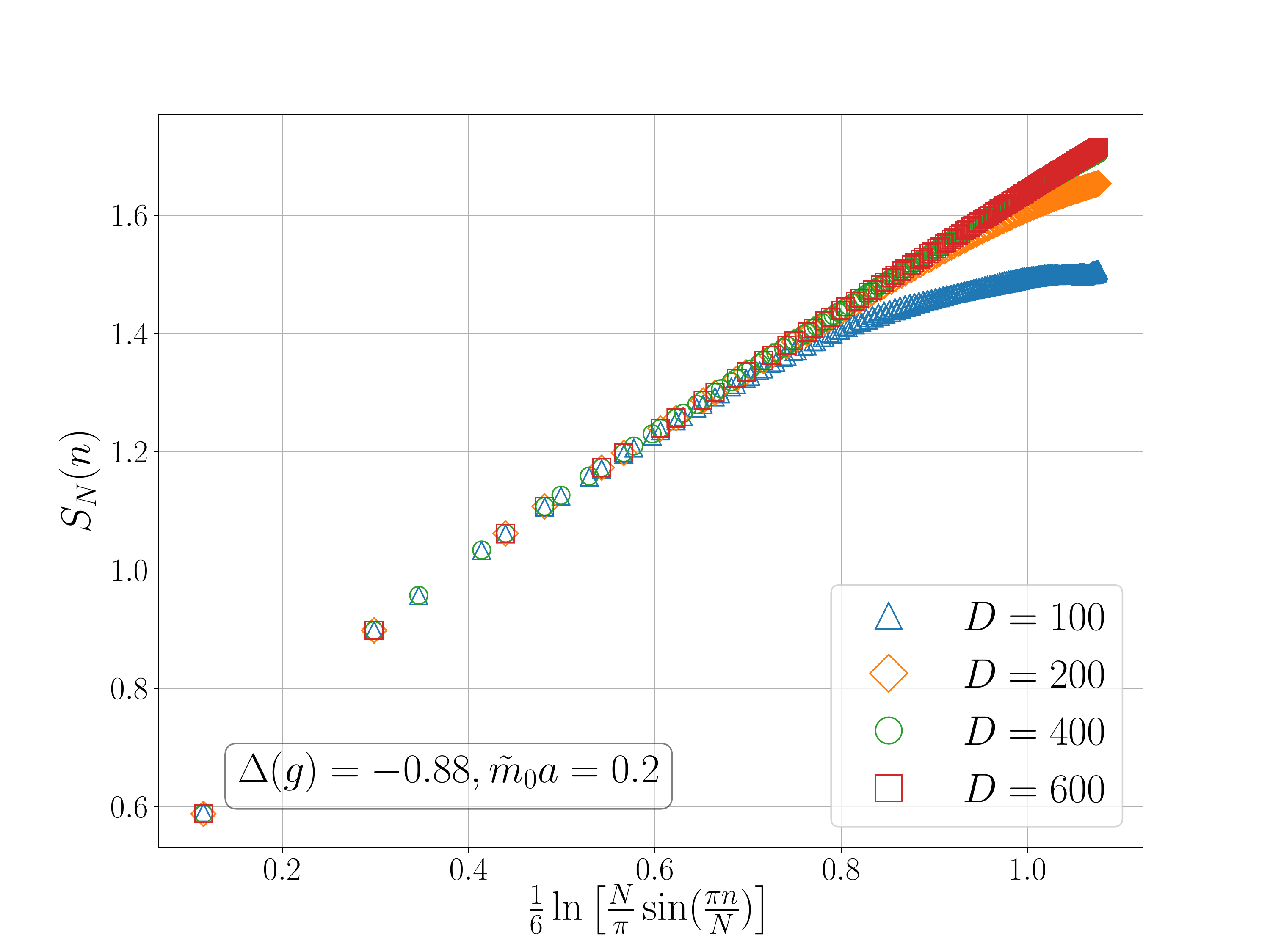}
  \caption{Finite-size entanglement entropy, $S_{N}(n)$, plotted
    against $X$ defined in Eq.~(\ref{eq:def_X}).}
  \label{fig:entropy_log}
\end{figure}

In the cases where the Calabrese-Cardy scaling behaviour is observed,
one can use Eq.~(\ref{eq-entropy-cardy-finite-sys}) to determine the central charge, $c$.  For this purpose, we study
the entanglement entropy, $S_{N}(n)$, as a function of $X$, where
\beq
\label{eq:def_X}
 X = \frac{1}{6}\mbox{ }
 {\mathrm{ln}}\left[\frac{N}{\pi}\sin\left(\frac{\pi
       n}{N}\right)\right] \, .
\eeq
In Fig.~\ref{fig:entropy_log}, we display such an example for $\Delta
(g) = -0.88$ and $\tilde{m}_{0}a = 0.2$, where the Calabrese-Cardy scaling is manifest.
From this plot, it is obvious that at large enough bond dimension, one
can read off the central charge and the result is $c \sim 1$.
Throughout the entire regime where the system is found to be critical,
we extract the central charge by 
fitting $S_{N}(n)$ computed at the largest bond
dimension ($D=600$) in this work to
Eq.~(\ref{eq-entropy-cardy-finite-sys}).  Such analysis
finds that $c$ is always at most $\sim 1\%$ away from one.  
This is consistent with a free bosonic field theory at criticality,
and is in accordance with the prediction of the RGE's,
Eq.~(\ref{eq:RGE_sG_for_t}) and (\ref{eq:RGE_sG_for_z}).  That is,
there exists a phase where the cos$[\phi(x)]$ operator in the the dual
sine-Gordon model Eq.~(\ref{eq-action-SG}) is irrelevant, so is the
$\bar{\psi}\psi$ operator in the dual massive Thirring model.


To summarize, we find strong evidence for the existence of
  two phases in the massive Thirring model.  In one of the phases the
  Calabrese-Cardy scaling is valid, while in the other this scaling
  behaviour is absent.  Through the detailed studies of the
  correlators, presented in Sec.~\ref{sec:correlators}, it will be
  established that the theory is at criticality in the former, and the
  latter phase is gapped.

\subsection{Fermion bilinear condensate}
\label{sec:fermion_bilinear_condensate}
In order to obtain further information for the nature of the 
phase transition discussed in Sec.~\ref{sec:entanglement_entropy}, we investigate the chiral condensate,
\beq
\label{eq:chiral_condensate_def}
 \hat{\chi} =a \chi  = a  \left | \left \la  \bar{\psi} \psi 
 \right \ra \right |   \,  .
\eeq
Under the Jordan-Wigner transformation,
\beq
\label{eq:condensate_JW}
 \hat{\chi}  \xrightarrow{{\mathrm{JW}}\mbox{
   }{\mathrm{transformation}}}  \frac{1}{N} \left | \left \la \sum_{n=0}^{N-1}
 (-1)^{n} S^{z}_{n} \right \ra \right | \, .
\eeq
That is, the chiral condensate in the (1+1)-dimensional Thirring model corresponds to
the staggered magnetization in the XXZ spin chain.  
In the spin chain model, the magnitude of anisotropy never exceeds one
according to Eq.~(\ref{eq:m_tilde_0_and_Delta}).  This implies that a N\'{e}el state can only arise
when the staggered magnetic field, $a\tilde{m}_0\not= 0$, is applied.
For the corresponding quantum field theory, the Thirring model, this
means that the chiral condensate is expected to be zero in the
massless limit.  Such a feature is consistent with the fact that the massless
Thirring model in 1+1 dimensions is a conformal field theory. 
Furthermore, due to the presence of a uniform magnetic field at
non-zero $\Delta (g)$ in the spin model,  it is obvious that the value
of the staggered magnetization (chiral condensate in the Thirring model)
will exhibit $\Delta-$dependence.

It is worth noticing that the Mermin-Wagner-Coleman
theorem~\cite{Mermin:1966fe,Coleman:1973ci} is applicable for the
theory under investigation.  As also elaborated in
Ref.~\cite{Witten:1978qu}, this means that the
condensate defined in Eq.~(\ref{eq:chiral_condensate_def}) is not a
viable order parameter for phase transitions in 1+1 dimensions, and
can be non-vanishing when the theory is in the critical phase.
We compute the chiral condensate at all values of $\Delta
(g)$ and $a\tilde{m}_{0}$ where numerical calculations are performed.
At every choice of $[ \Delta(g), a\tilde{m}_{0} ]$, we first estimate
$\hat{\chi}$ in the limit of infinite bond
dimension.   Although there are no errors on the ``raw data'' for
$\hat{\chi}$ at finite $D$, we employ the following procedure to
assign errors on infinite$-D$ results of the condensate.
\begin{enumerate}
\item Using the raw data of $\hat{\chi}$ at $D = 500$ and 600,
  $\hat{\chi}_{500}$ and $\hat{\chi}_{600}$,
  a ``linear extrapolation'' to
  the $D\rightarrow\infty$ limit is first performed.   Result for the chiral
  condensate from this step is denoted by
  $\hat{\chi}_{\infty}^{({\mathrm{temp}})}$.
\item Estimate the central value of the infinite$-D$ condensate, $\hat{\chi}_{\infty}$, by computing
\beq
\label{eq:infinite_D_chi_central_value}
 \hat{\chi}_{\infty} = \frac{1}{2} \left [
   \hat{\chi}_{\infty}^{({\mathrm{temp}})} + \hat{\chi}_{600} \right ] \, .
\eeq
\item The (symmetric) error on $\hat{\chi}$ in the limit of infinite bond
  dimension is evaluated by
\beq
\label{eq:infinite_D_chi_error}
 \delta\hat{\chi}_{\infty} = \frac{1}{2} \left |
   \hat{\chi}_{\infty} - \hat{\chi}_{600} \right | \, .
\eeq
\item If the error, $\delta\hat{\chi}_{\infty}$, obtained in the
  previous step is smaller than the chosen precision of the variational algorithm in this
work, $\epsilon = 10^{-7}$, we replace the numerical value of
$\delta\hat{\chi}_{\infty}$ by $\epsilon$.
\end{enumerate}
It is worth noticing that the above procedure is a conservative
approach for assigning errors to $\hat{\chi}$ in the infinite$-D$
limit, $\hat{\chi}_{\infty}$.   This is
because our data show that the chiral condensate converges very well
at $D \ge 400$.

\begin{figure}[!t]
  \centering
\includegraphics[width=8.8cm]{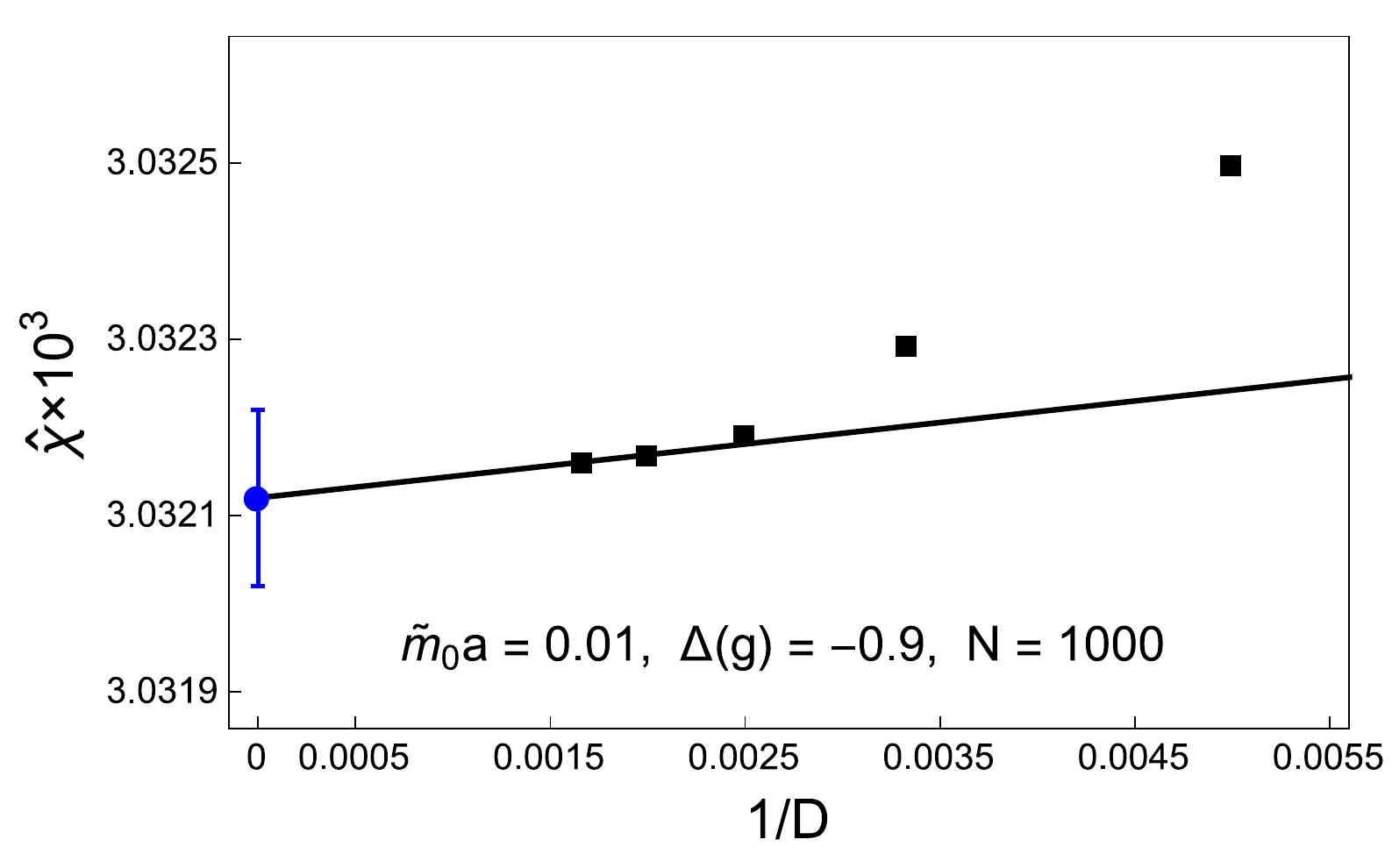}
\hspace{-0.0cm}
\includegraphics[width=8.8cm]{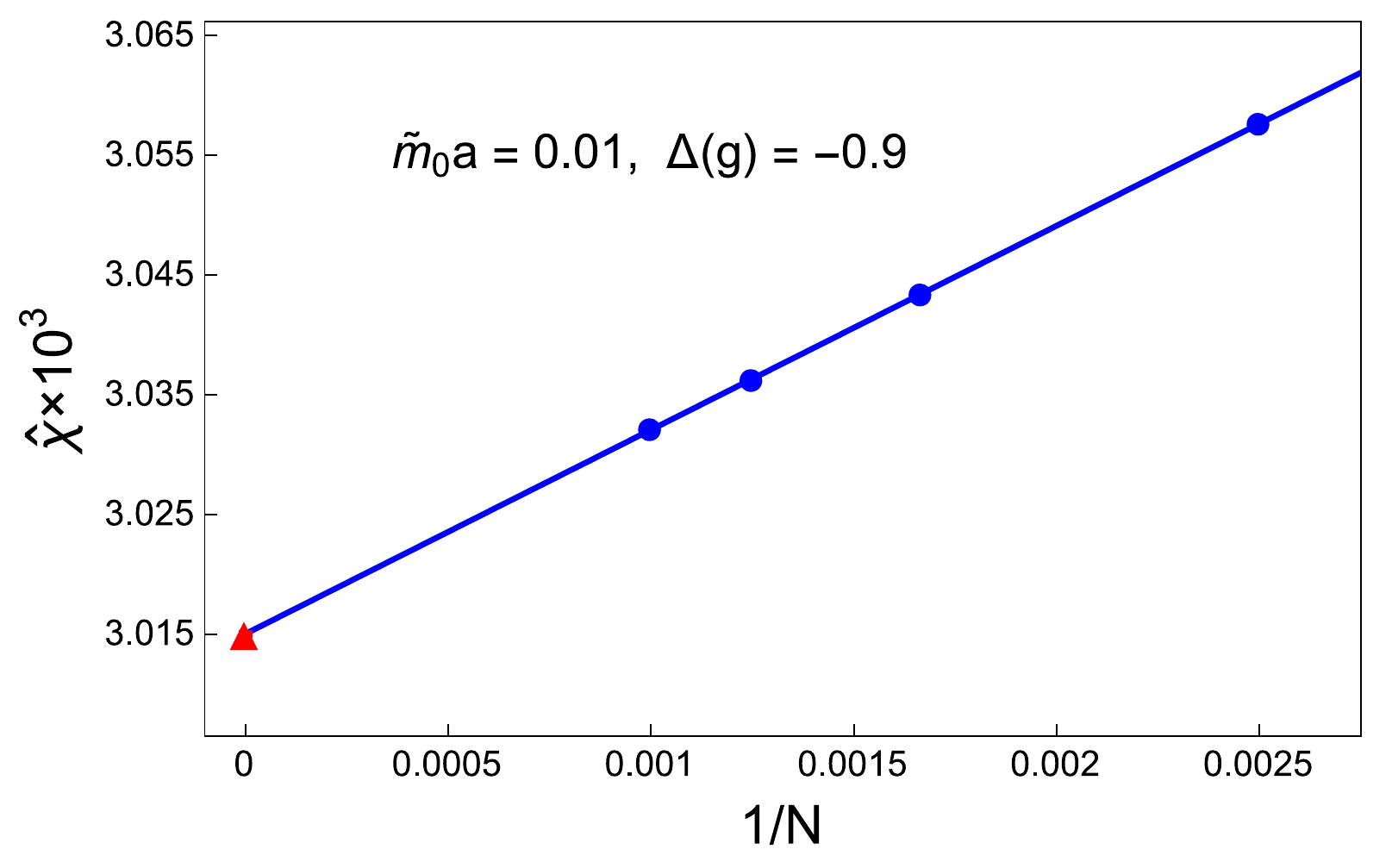}
  \caption{Extrapolations of $\hat{\chi}$ to the $D \rightarrow \infty$ at $N=1000$
    (left) and to the $N \rightarrow \infty$ limits (right) at $[ \Delta(g), a\tilde{m}_{0} ] =
  [-0.9, 0.01]$, employing the procedure described in the texts.   In
  the left panel, the central value of point at $1/D = 0$ is
  $\hat{\chi}_{\infty}$ computed using
  Eq.~(\ref{eq:infinite_D_chi_central_value}).  The error of this
  point is $\delta
  \hat{\chi}_{\infty} = \epsilon = 10^{-7}$, since the method of
  extracting it from Eq.~(\ref{eq:infinite_D_chi_error}) results in a
  value smaller than $\epsilon$. The straight line in this plot is the extrapolation
  for determining $\hat{\chi}_{\infty}^{({\mathrm{temp}})}$. Notice that errors on the data points and the
  extrapolated result for the right panel are too small to be
  discernible on the plot.}
  \label{fig:condensate_N_extrap}
\end{figure}

Results of $\hat{\chi}$ in the infinite$-D$ limit are then used for the procedure
of taking the thermodynamic limit, $N \rightarrow \infty$.  
For this extrapolation in $N$, we use data points at $N = 400$, 600,
800 and 1000.
Figure~\ref{fig:condensate_N_extrap} shows examples of 
extrapolations in $D$ and $N$ for $[ \Delta(g), a\tilde{m}_{0} ] = [-0.9, 0.01]$.
In this figure, results of the condensate obtained at $D=200, 300, 400, 500, 600$ and $N=400, 600,
800, 1000$ are exhibited.  From these plots, one can easily observe that $\hat{\chi}$ depends
very mildly 
on $D$ for $D \ge 400$, and it also extrapolates smoothly to
$N\rightarrow\infty$ using data at $N \ge 400$.   Such a feature is in fact
observed for all choices of $[ \Delta(g), a\tilde{m}_{0} ]$ in this
work.  As mentioned above, this means that our method of assigning
errors to $\hat{\chi}_{\infty}$ is a conservative approach.

We also remark that finite volume effects (FVE) in the Thirring model were investigated a long time ago in Ref.~\cite{Hochberg:1984td}.
The authors compared an exact solution for the infinite-volume mass spectrum with results of Monte Carlo simulations at finite volume and concluded that FVE are suppressed to the level of a few percent when $N\gtrsim50$.
In our case, despite using different observables, we observe $\mathcal{O}(1\%)$ FVE at $N\in[400,1000]$, which is qualitatively consistent with the results from Ref.~\cite{Hochberg:1984td}.

\begin{figure}[!t]
  \centering
\includegraphics[width=8.8cm]{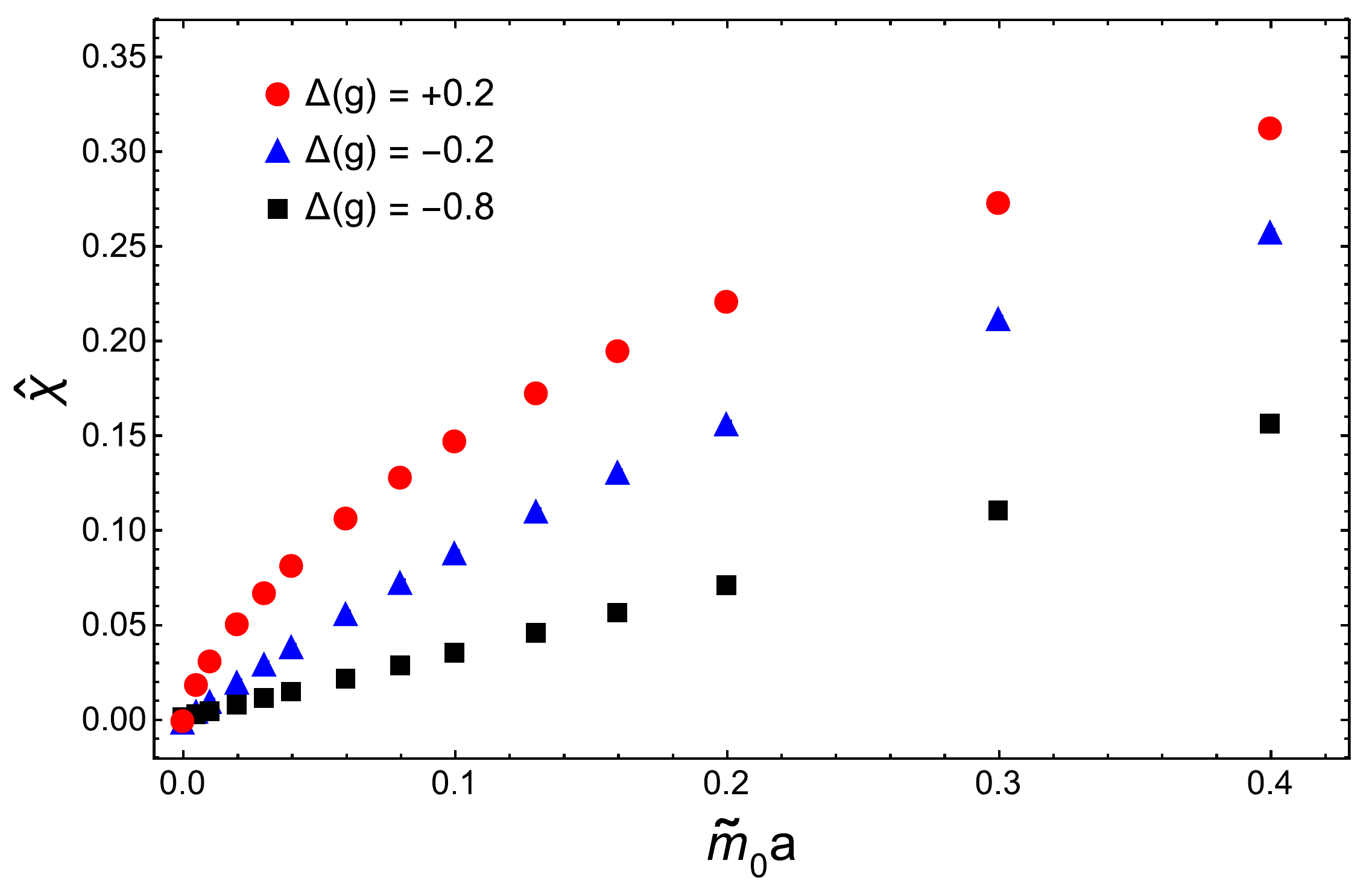}
\hspace{-0.0cm}
\includegraphics[width=8.8cm]{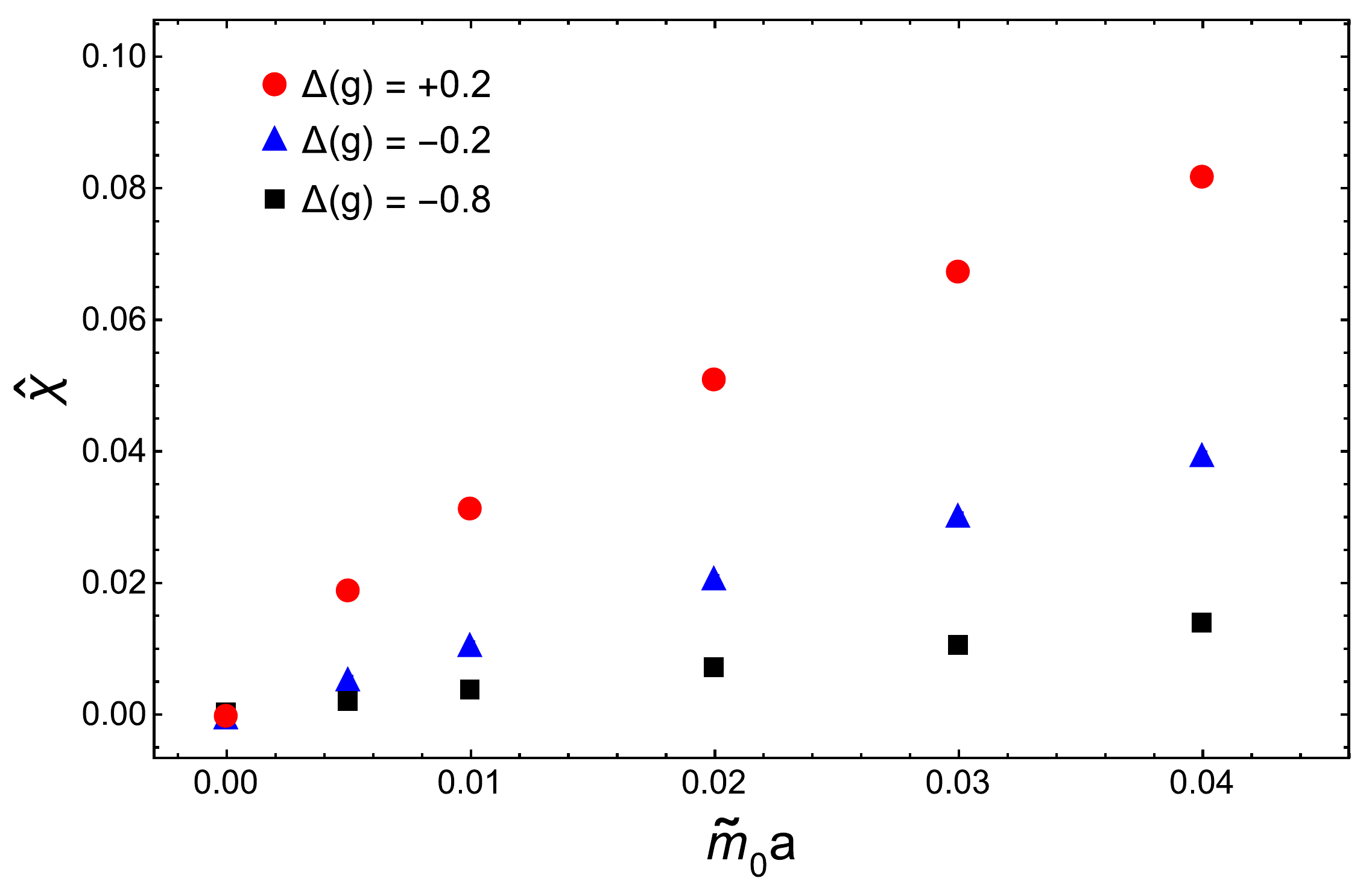}
  \caption{The dependence on $a\tilde{m}_{0}$ in the chiral condensate
    at $\Delta (g) = 0.2$, $-0.2$ and $-0.8$ in the limit $D\rightarrow
    \infty$ and $N \rightarrow \infty$.  The left panel shows
    results at all values of $a\tilde{m}_{0}$ in this work, while the
    right panel displays only those at $a\tilde{m}_{0} \le 0.04$.  Notice that errors on the data points are too small to be
  discernible on the plots.}
  \label{fig:condensate_mass_dep}
\end{figure}

Figure~\ref{fig:condensate_mass_dep} demonstrates representative results for the 
chiral condensate
in the limit of infinite bond dimension and system size.  In this figure,
we only show $\hat{\chi}$ at three
values of the four-fermion coupling constant, corresponding to $\Delta
(g) = 0.2$, $-0.2$ and $-0.8$.     The condensate computed at other
choices of $\Delta (g)$ exhibits the same feature as those in this figure.  
According to results of the entanglement entropy
discussed in Sec.~\ref{sec:entanglement_entropy}, the theory is in the gapped (massive) phase at
$a\tilde{m}_{0}\not= 0$ for $\Delta > \Delta_{\ast}$ where
$\Delta_{\ast}$ is mass-dependent, while it can be in the
critical phase at $\Delta < \Delta_{\ast}$.   This feature will be
further confirmed by the investigation of fermion correlators,
presented in Sec.~\ref{sec:correlators}.
We notice that, in
Fig.~\ref{fig:condensate_mass_dep}, $\hat{\chi}$ is
non-vanishing at $\Delta(g) = -0.8$ for all data points with 
$a\tilde{m}_{0} \not= 0$.  Through the study of the entanglement entropy
reported in Sec.~\ref{sec:entanglement_entropy}, we also know that for
$\Delta(g) = -0.8$, the theory can be in two different phases
separated at $a\tilde{m}_{0} \sim 0.2$.   In other words, the chiral
condensate can be non-zero in both phases.
This means that $\chi$ is not an order
parameter for the observed phase transition, providing further evidence
that this transition can be of the BKT-type~\cite{Witten:1978qu}.

In Fig.~\ref{fig:condensate_mass_dep}, it can be seen that
$\hat{\chi}$ extrapolates smoothly to zero at vanishing
$a\tilde{m}_{0}$.  As mentioned above, this is in accordance with the
fact that the massless Thirring model in 1+1 dimensions is a
conformal field theory.   Furthermore, we find that the chiral
condensate computed directly at $a\tilde{m}_{0} = 0$ is zero for
all values of $\Delta (g)$.   Given that all simulations that lead to
results in Fig.~\ref{fig:condensate_mass_dep} are
performed at finite system sizes, we carry out checks for
$a\tilde{m}_{0} = 0$ calculations with infinite-size simulations by
employing the variational uniform MPS (VuMPS) 
method~\cite{Zauner-Stauber:2018kxg}.   These checks confirm that
$\hat{\chi}$ obtained from simulations at $a\tilde{m}_{0}=0$ indeed
vanishes.  Results of the VuMPS approach will be
published in a separate article where we will report our study of
real-time dynamics associate with the BKT phase transition in the
massive Thirring model~\cite{Banuls:2020inprep}.

\subsection{Correlation functions}
\label{sec:correlators}
%

We now proceed to present results for the correlation functions in the Thirring model.
We will consider two types of correlators: density-density 
(which in the continuum is $\langle\bar\psi(x_1)\psi(x_1)\bar\psi(x_2)\psi(x_2)\rangle$, and in terms of the discrete fermions corresponds to $\langle c_n^\dagger c_n c_m^\dagger c_m\rangle/a^2$) 
and fermion-antifermion 
($\langle \bar\psi(x_1)\psi(x_2)\rangle$ or $\langle c_n^\dagger c_m\rangle/a^2$).

The connected part of the former can be expressed in the spin language as:
%
\begin{equation}
C_{\rm zz}(x)=\frac{1}{N_x}\sum_n \left[ \langle S^z(n) S^z(n+x) \rangle  - \langle S^z(n)\rangle\langle S^z(n+x) \rangle  \right ].
\end{equation}
The sum over $n$ indicates that, for a given distance $x$, we average over all pairs of spins within the 200-site subchain in the middle of the lattice, to avoid boundary effects.
Despite using open boundary conditions, translational invariance is realized almost ideally in this region.
We take $x$ to be odd, i.e.\ we look at the correlator between even-odd or odd-even sites.
The number of pairs corresponding to a given distance $x=1,\ldots,199$ is denoted by $N_x$ and is equal to $200-x$.
To calculate the connected part of the correlator, we subtract the product of single-site expectation values.

The fermion-antifermion correlation function is expressed in terms of the spin operators as the following string correlator:
\begin{equation}
\label{eq:Cstr}
C_{\rm string}(x)=\frac{1}{N_x} \sum_n \langle S^+(n) S^z(n+1)\ldots S^z(n+x-1) S^-(n+x) \rangle ,
\end{equation}
with the same subchain averaging procedure as for the distance-dependent part of $C_{\rm zz}(x)$.  This correlator can be shown to correspond to the soliton-soliton correlation function in the dual sine-Gordon theory~\cite{Mandelstam:1975hb}.

We first concentrate on the short-distance behaviour of both correlation functions.
The decay of $C_{\rm zz}(x)$ and $C_{\rm string}(x)$ is expected to be power-law in the critical phase and power-exponential in the gapped phase.
To test this behaviour, for all parameter values and for both types of correlators, we perform fits using the power-law ansatz:
\begin{equation}
\label{eq:pow}
C^{\rm pow}(x)=\beta x^\alpha+C,
\end{equation}
with the fitting parameters $\alpha$, $\beta$, and the constant $C$ allowed only for $C_{\rm string}(x)$.
Moreover, we fit three types of exponential ansatzes:
\begin{equation}
\label{eq:powexp}
C^{\rm pow-exp}(x)=B x^{\eta}A^x+C,
\end{equation}
\begin{equation}
\label{eq:2exp}
C^{\rm 2exp}(x)=B_1A_1^x + B_2A_2^x+C,
\end{equation}
\begin{equation}
\label{eq:3exp}
C^{\rm 3exp}(x)=B_1A_1^x + B_2A_2^x + B_3A_3^x+C,
\end{equation}
with fitting parameters $A$, $B$, $\eta$ or $A_i$, $B_i$ (we assume $A_1>A_2>...$), and the constant $C$ allowed in the case of the string correlator.
Even if the true behaviour is power-law or mixed power-exponential decay, a MPS with finite bond dimension can only approximate it 
by a finite sum of exponentially decaying terms. Their decay rate, in a translationally invariant case, is determined by the 
eigenvalues  $\{\lambda_i\}$ of the transfer matrix $\mathbb{E}\equiv \sum_{\sigma} {M^{\sigma}}^* \otimes M^{\sigma}$, more concretely
by the ratios $\lambda_i/\lambda_1$, where $\lambda_1$ is the largest one.
Only in the limit of infinite $D$ would it be able to reproduce a true power law at arbitrarily long distance. 
Thus, we also test how well a multi-exponential ansatz can describe the observed data.
The parameters $A$ or $A_i$ will be related to the aforementioned ratios of transfer matrix eigenvalues -- in particular $A$ is the ratio of the second largest to the largest eigenvalue, $\lambda_2/\lambda_1$.
Thus, in the critical phase (strictly conformal in the massless case and in the continuum), $A=1$ and the power-exponential fitting ansatz becomes the power-law one.
A discussion is in place also for the constant $C$ in the fermion-antifermion correlator.
Such a constant cannot emerge in the continuum theory.
However, on the lattice, discretization effects may cause the constant to appear.
In the critical phase, all lattice effects should vanish in the infinite volume limit, since the theory is conformal in this regime (the mass operator is irrelevant, as discussed above in the context of RG equations).
Thus, the constant should be zero, which we check explicitly.
In the gapped phase, however, the mass operator becomes relevant and an energy scale appears.  This breaks the conformality, and the continuum limit requires $\tilde{m}_{0}a\rightarrow0$.
In such a situation, a constant may be generated in the string correlator.
This constant is expected to increase the further away from the continuum limit one simulates -- i.e.\ when the fermion mass or $\Delta(g)$ is increased.
At large mass and deep in the gapped phase, a string order configuration may be favoured as an artefact of the staggered discretization.
In spin language, the system is subject to a staggered magnetic field, leading to a N\'eel-type phase when this field is strong.
Inspecting the definition of this correlator, Eq.\ (\ref{eq:Cstr}), it becomes clear that this renders a constant at large distances.

\begin{figure}[!t]
  \centering
\includegraphics[width=9.5cm]{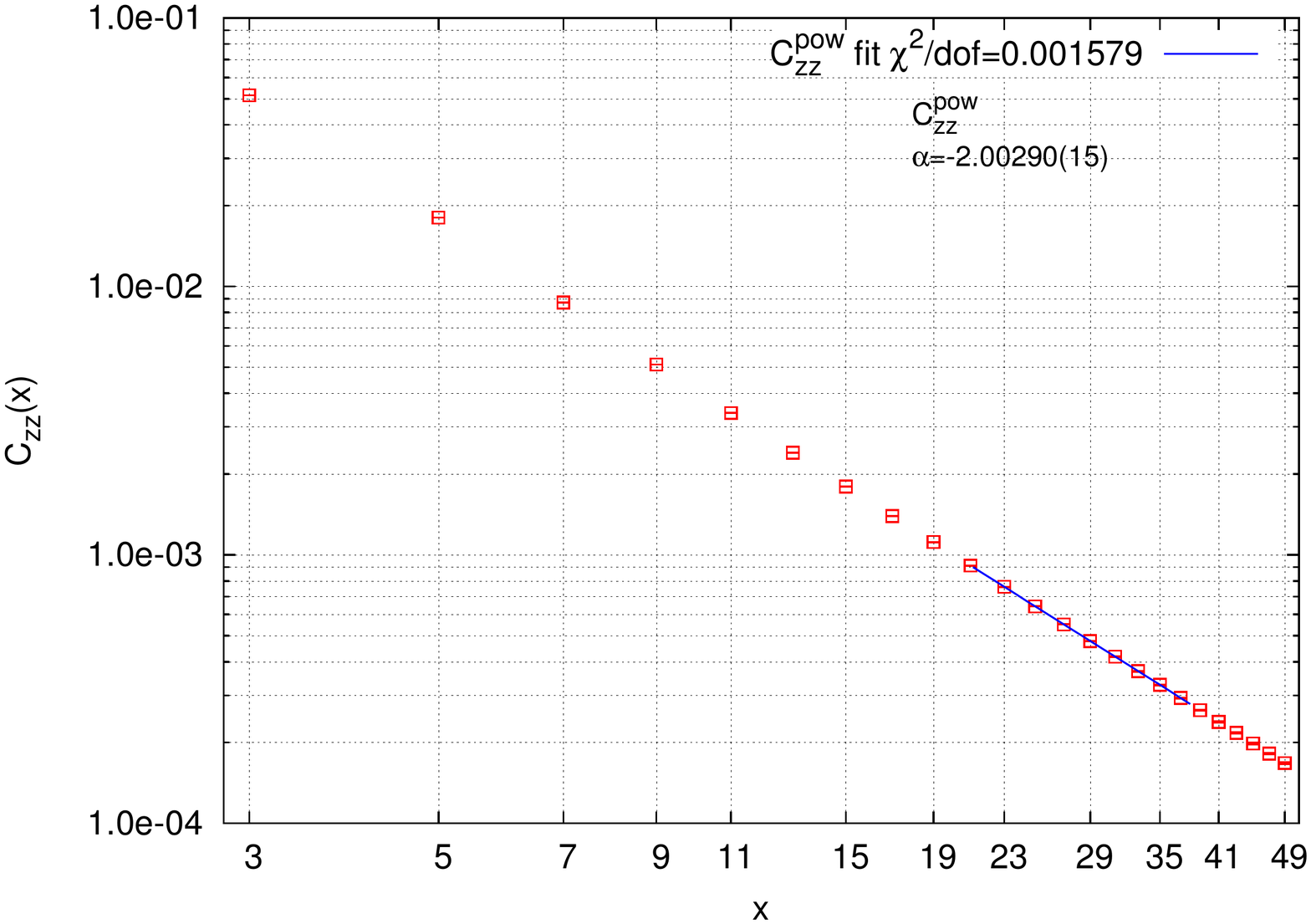}
\hspace{-1.3cm}
\includegraphics[width=9.5cm]{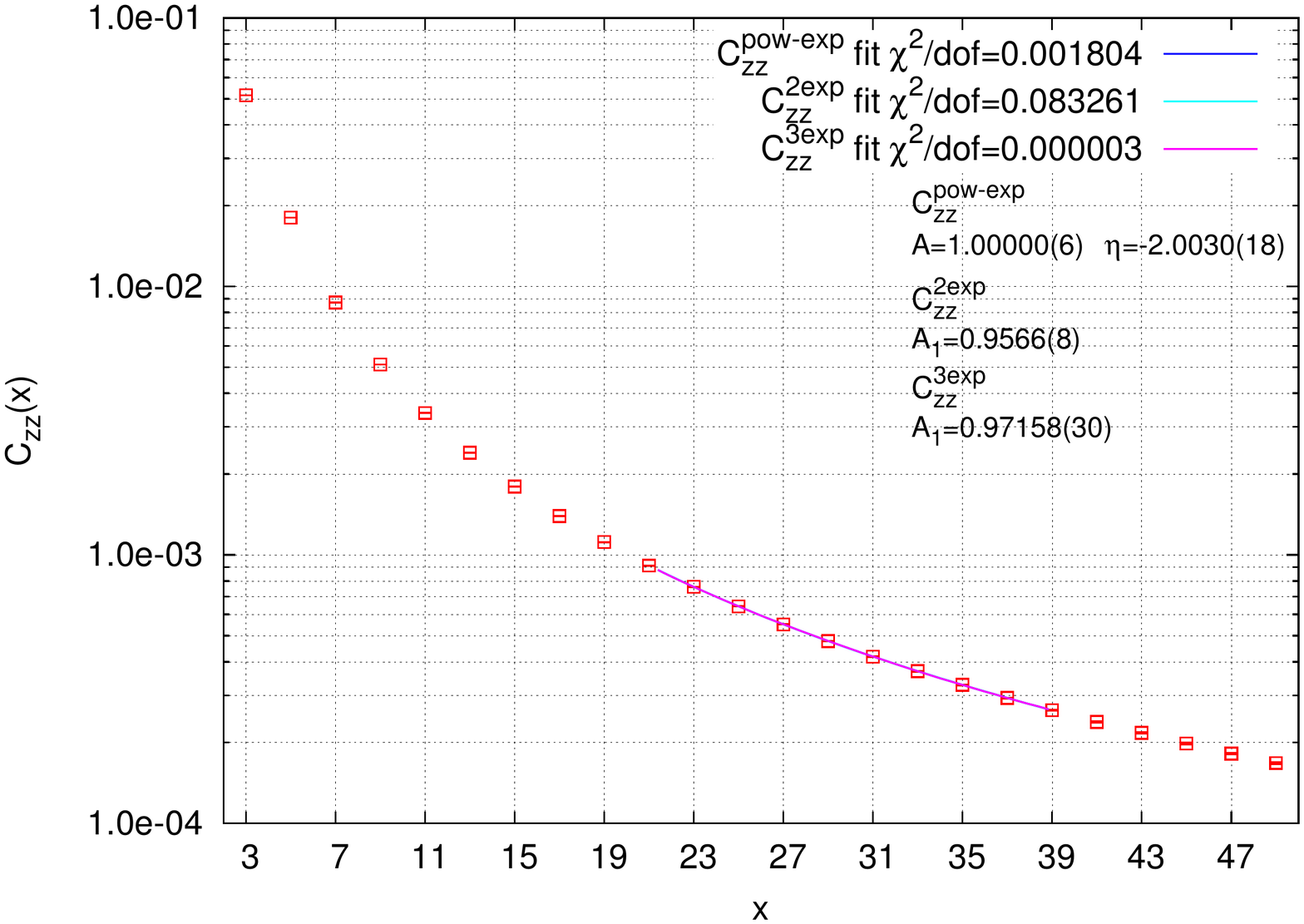}
  \caption{Density-density correlator $C_{\rm zz}(x)$ for $N=1000$, $\tilde{m}_{0}a = 0.02$, $\Delta(g)=-0.7$. In the left panel, we show an example fit of the power-law fitting ansatz (\ref{eq:pow}) in the interval $x\in[21,39]$. In the right panel, the fits are for three types of exponential ansatzes (\ref{eq:powexp})-(\ref{eq:3exp}), in the same interval.}
  \label{fig:Czz_g-07}
\end{figure}

In Fig.\ \ref{fig:Czz_g-07}, we show example fits for the density-density correlator, for a system size of 1000 sites, coupling $\Delta(g)=-0.7$ and a small fermion mass $\tilde{m}_{0}a = 0.02$. 
All of the results of this subsection are obtained in the limit of infinite bond dimension, using a procedure analogous to the one described in Sec.~\ref{sec:fermion_bilinear_condensate}.
The left panel presents the power-law fit of Eq.\ (\ref{eq:pow}) and the right panel the three exponential-type fits, Eqs.\ (\ref{eq:powexp})-(\ref{eq:3exp}).
In both cases, the fitting range is distances between $x=21$ and $x=39$, i.e.\ 10 consecutive data points.
For the case depicted in these plots, we expect that the system is close to the point of the BKT transition and hence $C_{\rm zz}$ should be well described by the power-law fitting ansatz.
Indeed, as the left panel of Fig.\ \ref{fig:Czz_g-07} demonstrates, such a fit provides an excellent description of the correlator decay, following a straight line on a log-log plot.
Obviously, the power-exponential fitting ansatz works equally well, since it boils down to the power-law functional form when $A=1$, as reproduced by the fit.
A single exponential is not enough to describe the data, as clear from visual inspection of the right panel -- there is clear curvature visible with only the $y$-axis in logarithmic scale.
However, two-exponential and three-exponential fits provide good description of the data.
Nevertheless, they do not reproduce the expected power-law behaviour of the critical phase, i.e.\ the fitting parameter $A_1$ is different from one (although it increases towards one when more exponentials are included).
The rate of the power decay of the correlator in the critical phase is described by the parameter $\alpha$, equal to $\eta$ for $A=1$.
The expected value of $\alpha$ is $-2$, well reproduced by the fit \cite{Rams2018corr,Dugave2015}.

\begin{figure}[!t]
  \centering
\includegraphics[width=9.5cm]{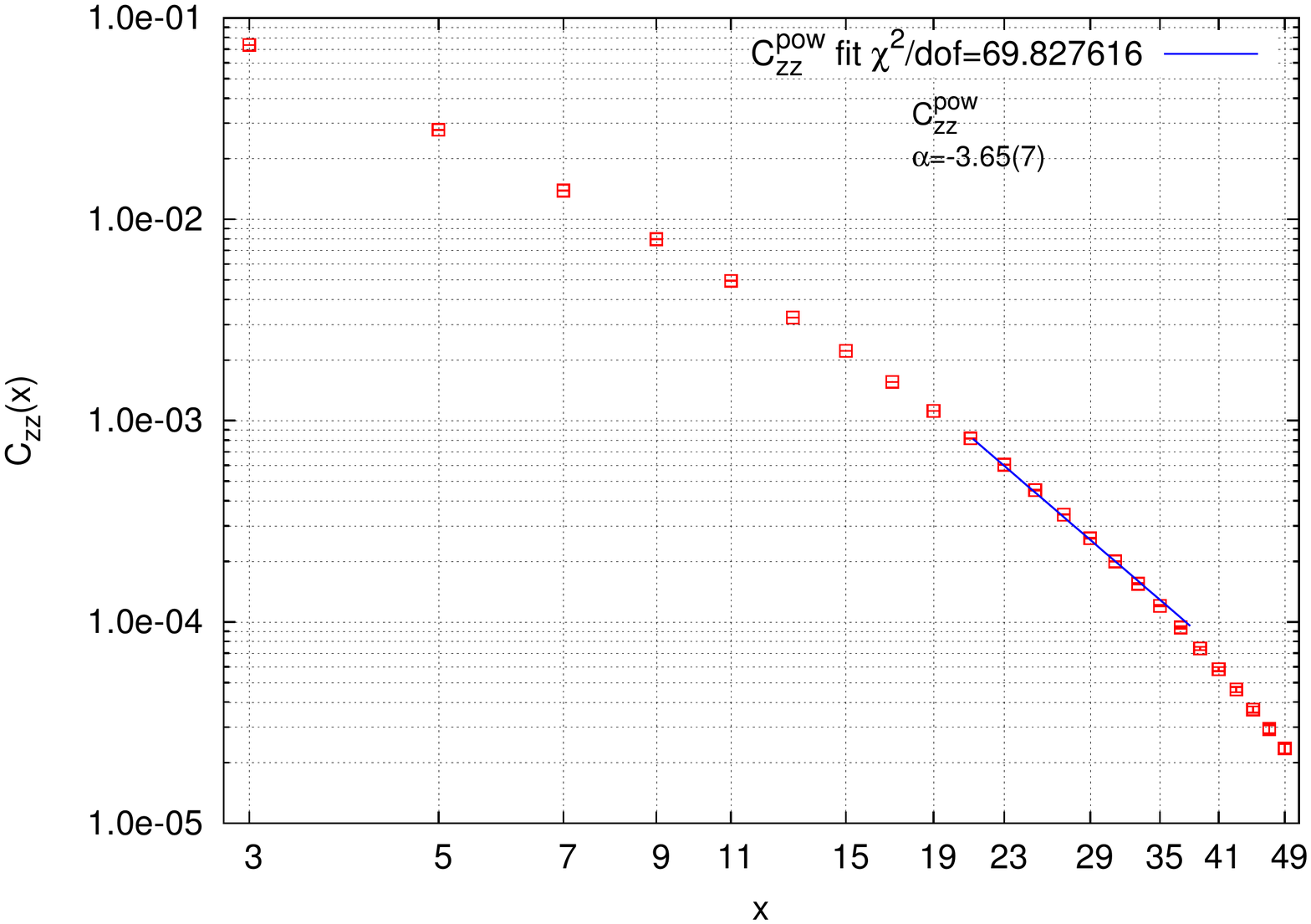}
\hspace{-1.3cm}
\includegraphics[width=9.5cm]{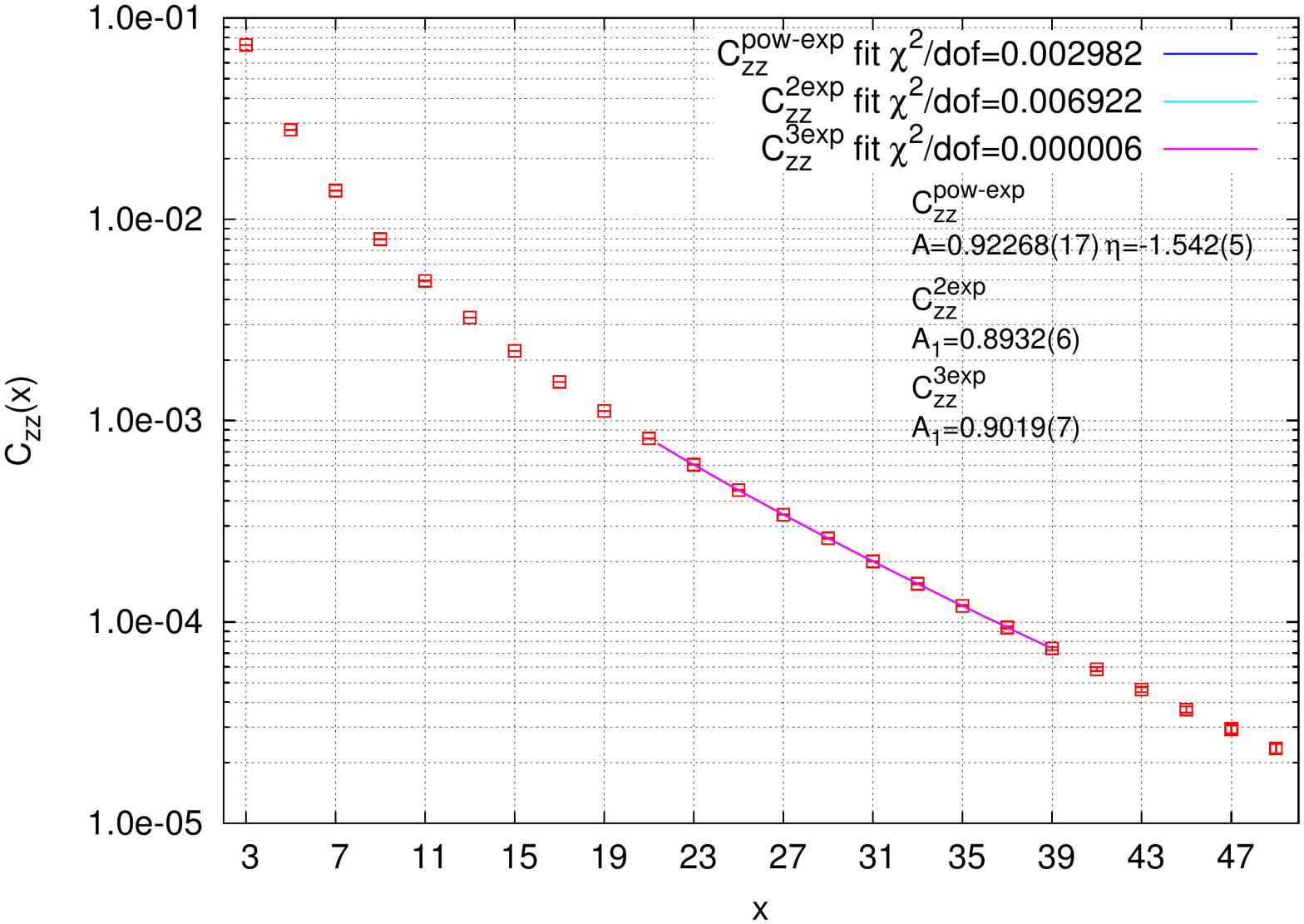}
  \caption{Density-density correlator $C_{\rm zz}(x)$ for $N=1000$, $\tilde{m}_{0}a = 0.02$, $\Delta(g)=0.4$. In the left panel, we show an example fit of the power-law fitting ansatz (\ref{eq:pow}) in the interval $x\in[21,39]$. In the right panel, the fits are for three types of exponential ansatzes (\ref{eq:powexp})-(\ref{eq:3exp}), in the same interval.}
  \label{fig:Czz_g04}
\end{figure}

To illustrate how the gapped phase manifests itself in the considered correlator, we show the fits also for $\Delta(g)=0.4$ (other parameters unchanged), see Fig.\ \ref{fig:Czz_g04}.
Even though the decay looks approximately linear in the log-log plot (left panel), our precision is good enough to conclude that power-law fit does not describe the data well, indicated by the large value of $\chi^2/{\rm dof}$.
The mixed power-exponential ansatz works well, with $A\approx0.923$ and $\eta\approx-1.54$.
Thus, the spectrum of the transfer matrix is gapped and the exponent $\eta$ significantly deviates from $-2$.
We note that this picture is consistent with the one from entanglement entropy, i.e.\ the phase where power-law fits are proper is the one where the central charge is found close to 1, while the region where the exponential correction to the correlator decay is important manifests itself by flat behaviour of the entanglement entropy, i.e.\ the Calabrese-Cardy scaling is not observed.

The correlators are expected to follow a given type of behaviour asymptotically, i.e.\ at large enough distances.
Hence, a proper choice of the fitting interval has to be made.
To analyse the dependence of the fitting parameters on the coupling $\Delta(g)$, we avoid the arbitrary choice of the fitting interval by adopting a systematic procedure, similar to the one used e.g.\ in Ref.\ \cite{Banuls:2013jaa} (see the Appendix of this reference).
We consider all possible fits in the interval $x\in[5,49]$ encompassing a minimum of 10 consecutive distances\footnote{In cases when we observe a systematic tendency in the behaviour of the fitting parameters when increasing the distance, we extend the interval up to $x\in[5,99]$.}.
Each fit is weighted with $\exp(-\chi^2/{\rm dof})$ and we build histograms of the fitting parameters for each fitting ansatz.
The central value for each fitting parameter in a given physical setup (same system size, fermion mass and coupling) is extracted as the median of this distribution and the error as half of the interval in which 68.3\% of the weighted fits around the median are contained (corresponding to a 1-$\sigma$ deviation in the case of an ideally Gaussian distribution).
We note the obtained distributions are approximately Gaussian and the thus extracted systematic error is in most cases a factor 5-10 larger than the error obtained from typical fits.
Finally, we add this systematic error to the one of the fit that best describes the data, defined as the one with the smallest error among the fits with $\chi^2/{\rm dof}\leq1$.

\begin{figure}[!t]
  \centering
\includegraphics[width=9.5cm]{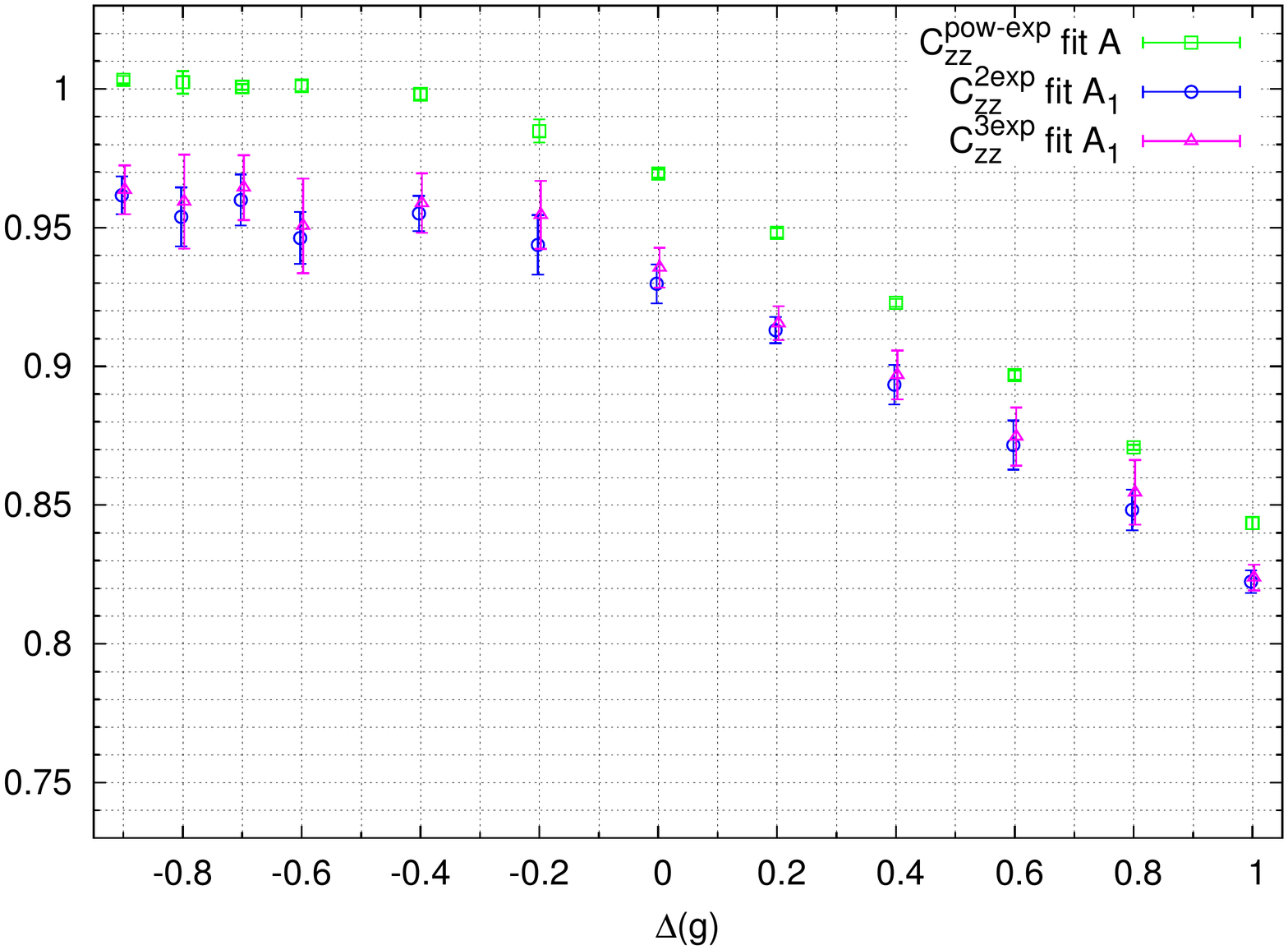}
\hspace{-1.3cm}
\includegraphics[width=9.5cm]{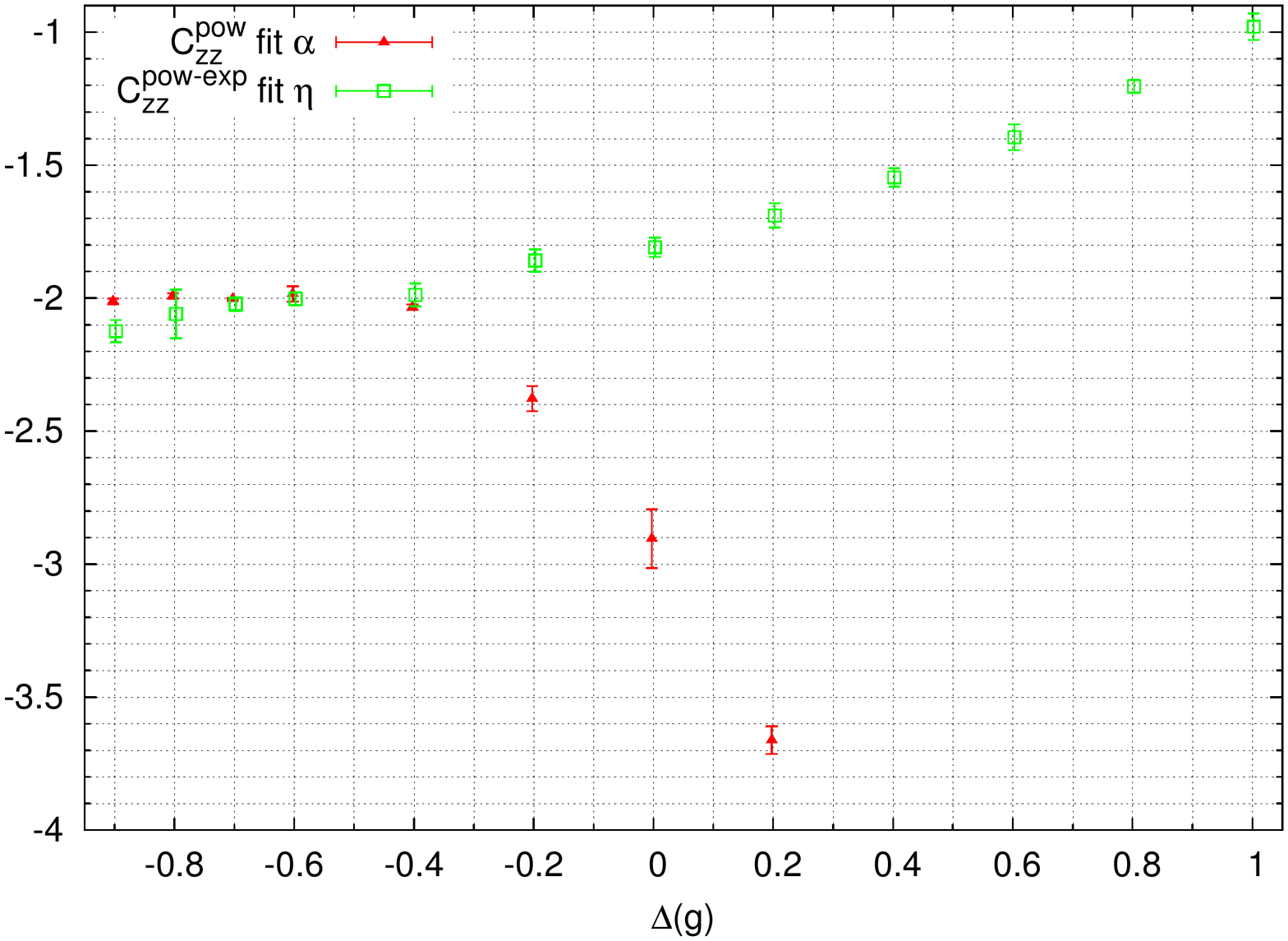}
  \caption{Density-density correlator $C_{\rm zz}(x)$. Left panel: dependence of the parameter $A$ and $A_1$ for three types of exponential ansatzes (\ref{eq:powexp})-(\ref{eq:3exp}) on the coupling $\Delta(g)$. Right panel: dependence of the parameter $\alpha$ and $\eta$ for the power-law and power-exponential fitting ansatzes (\ref{eq:pow})-(\ref{eq:powexp}) on the coupling $\Delta(g)$. Parameters: $N=1000$, $\tilde{m}_{0}a = 0.02$.}
  \label{fig:Czz_syst_m002}
\end{figure}

The result of applying this procedure to the density-density correlator is shown in Fig.\ \ref{fig:Czz_syst_m002}, again for fermion mass $\tilde{m}_{0}a = 0.02$.
In the left panel, we show the $A$ parameters extracted for different couplings.
We note the $A$ parameter of the power-exponential fit becomes compatible with 1 somewhere between $\Delta(g)=-0.4$ and $-0.6$.
The expected location of the BKT crossover is at $\Delta(g)\approx-0.7$, however the smooth transition between the functional forms of the power-law and power-exponential type of behaviour makes it impossible to locate the BKT point at the current level of precision.
We expect that close to the critical point, there is only a small admixture of the exponential factor to the power-law term, impossible to disentangle without much better precision.
Further into the gapped phase, at $\Delta(g)\gtrsim-0.4$, the exponential term becomes clearly visible and $A$ is no longer consistent with 1.
The value of $A$ drops when $\Delta(g)$ is increased and the exponent $\eta$ of the power-law factor in the fitting ansatz increases towards less negative values.
Thus, the exponential decay becomes relatively more important deeper in the gapped phase.
This is also indicated by the smaller difference of the parameter $A$ and the parameter $A_1$ of the 3-exponential fit for positive $\Delta(g)$, which would agree in the limit of purely exponential behaviour.

\begin{figure}[!t]
  \centering
\includegraphics[width=9.5cm]{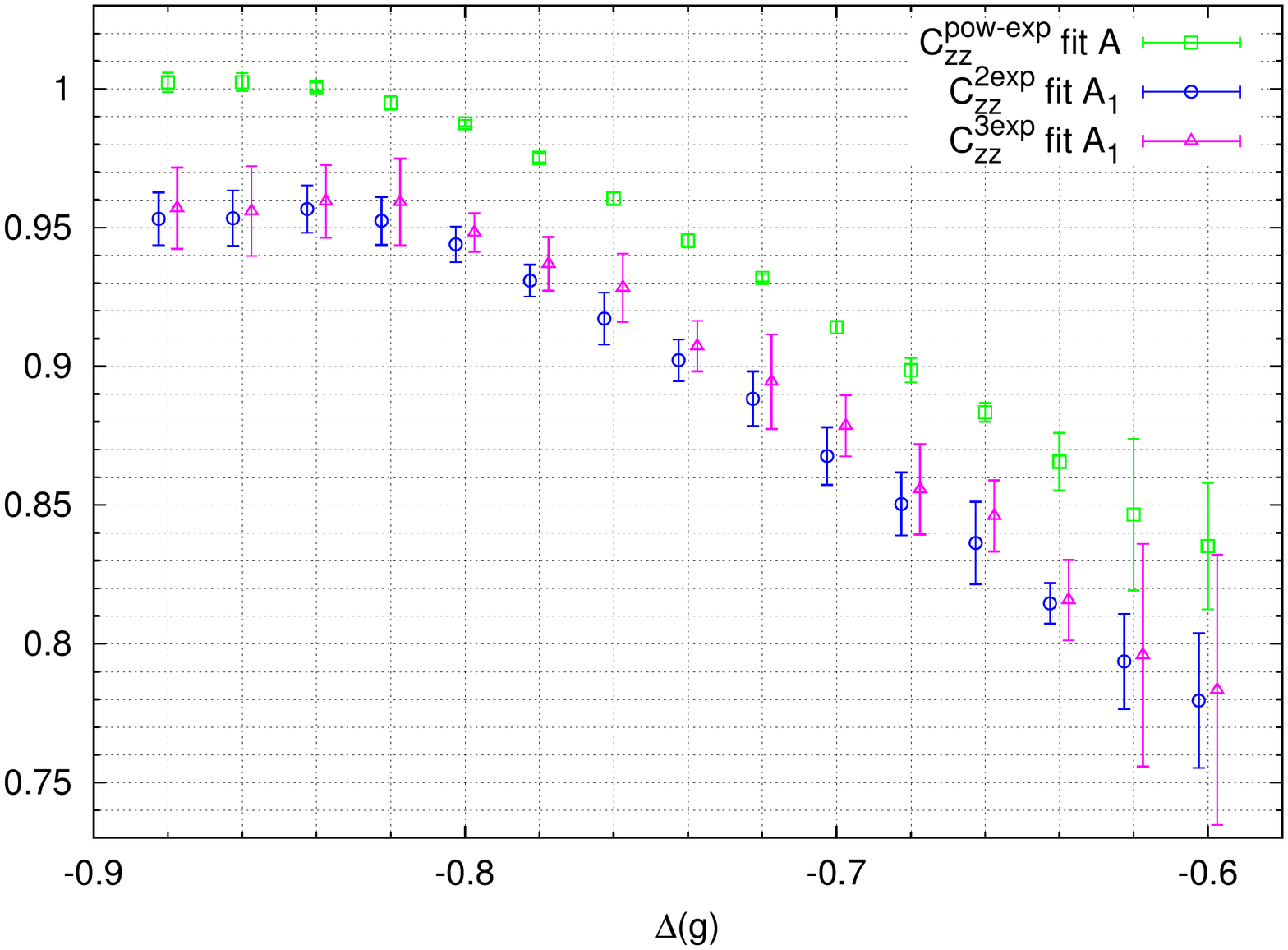}
\hspace{-1.3cm}
\includegraphics[width=9.5cm]{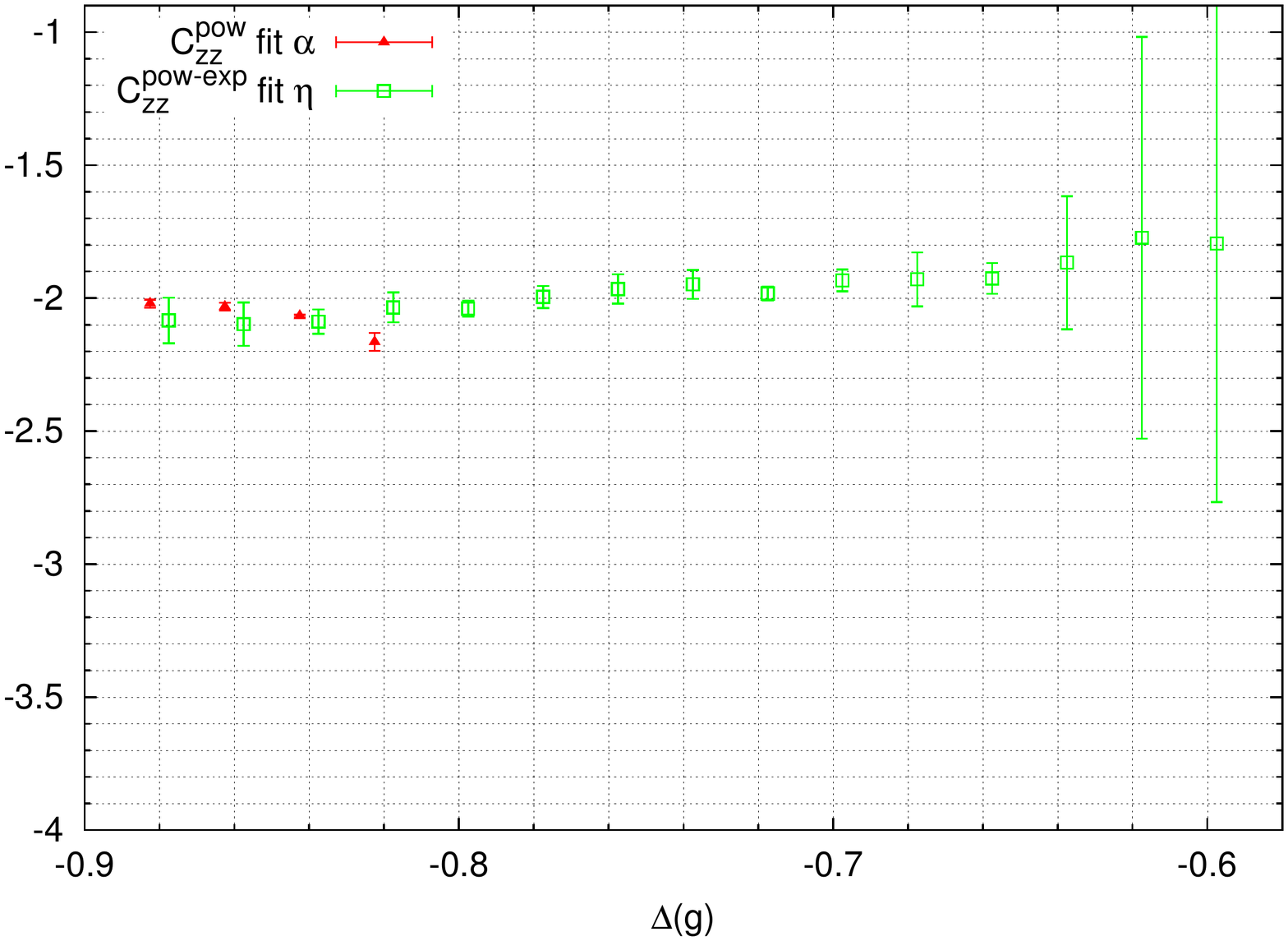}
  \caption{Density-density correlator $C_{\rm zz}(x)$. Left panel: dependence of the parameter $A$ and $A_1$ for three types of exponential ansatzes (\ref{eq:powexp})-(\ref{eq:3exp}) on the coupling $\Delta(g)$. Right panel: dependence of the parameter $\alpha$ and $\eta$ for the power-law and power-exponential fitting ansatzes (\ref{eq:pow})-(\ref{eq:powexp}) on the coupling $\Delta(g)$. Parameters: $N=1000$, $\tilde{m}_{0}a = 0.3$.}
  \label{fig:Czz_syst_m03}
\end{figure}

\begin{figure}[!t]
  \centering
\includegraphics[width=12cm]{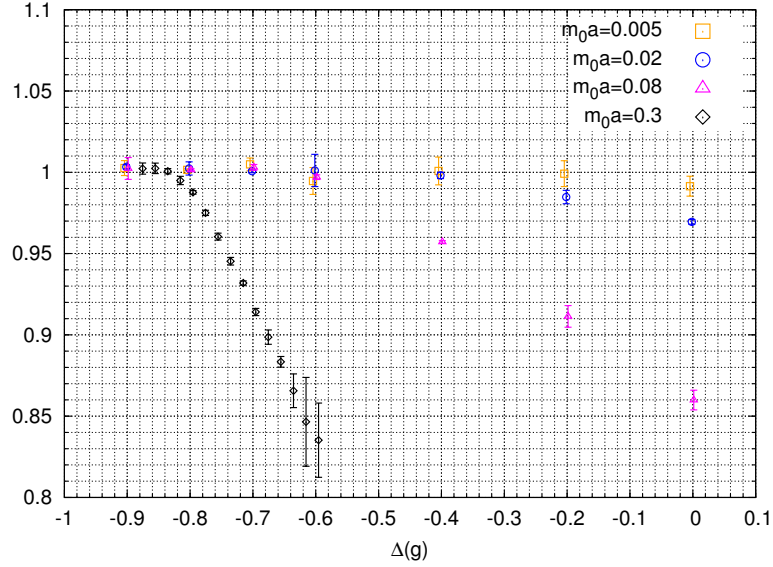}
  \caption{Dependence of the parameter $A$ for the power-exponential ansatz (\ref{eq:powexp}) on the coupling $\Delta(g)$. Shown are four fermion masses, $\tilde{m}_{0}a = 0.005, 0.02,0.08, 0.3$ for a 1000-site system.}
  \label{fig:Czz_syst_m}
\end{figure}

In Fig.\ \ref{fig:Czz_syst_m03}, we show the same kind of plot for a larger fermion mass, $\tilde{m}_{0}a = 0.3$.
In this case, the dependence of the parameter $A$ of the power-exponential fit is much steeper and $A$ becomes compatible with 1 between $\Delta(g)=-0.82$ and $-0.84$.
This signals that the BKT transition moves towards more negative values of the coupling with increasing fermion mass.
For $\Delta(g)>-0.82$, the system is clearly in the gapped phase, which is indicated also by $\chi^2/{\rm dof}\gg1$ for the pure power-law fits.
In contrast, such fits in the small fermion mass case are still reasonable until $\Delta(g)\approx0$, as a consequence of our rather conservative error estimate procedure.
We therefore conclude that the BKT crossover is more pronounced for larger fermion masses.
However, interestingly, the value of the exponent $\eta$ of the power-exponential fit is consistent with $-2$ for all couplings.
A comparison of the coupling dependence of the parameter $A$ for different fermion masses, $\tilde{m}_{0}a = 0.005,\,0.02,\,0.08,\,0.3$ is shown in Fig.\ \ref{fig:Czz_syst_m}.
It offers a rather clear picture -- while for the smallest mass the parameter $A$ is consistent with 1 up to $\Delta(g)\approx0$, increasing the fermion mass confines the location of the BKT transition to smaller and smaller ranges of the coupling.
Nevertheless, only in the largest mass case in this plot, it is possible to determine the transition point with a precision of a few percent.
It is very likely that the critical coupling, expected to be around $g_{\ast}=-\pi/2$ ($\Delta(g_{\ast})\approx-0.7$), moves towards more negative values as the mass is increased, but the numerical evidence for this is convincing only for the largest mass, at our level of precision in this correlator.
For smaller fermion masses, their effect can be interpreted as a perturbation of the massless conformal theory.
The mass perturbation opens the gap at the transition point, but the almost power-law decay of the correlator for small fermion masses is a remnant of the conformal phase that extends considerably into the gapped phase.

\begin{figure}[!t]
  \centering
\includegraphics[width=9.5cm]{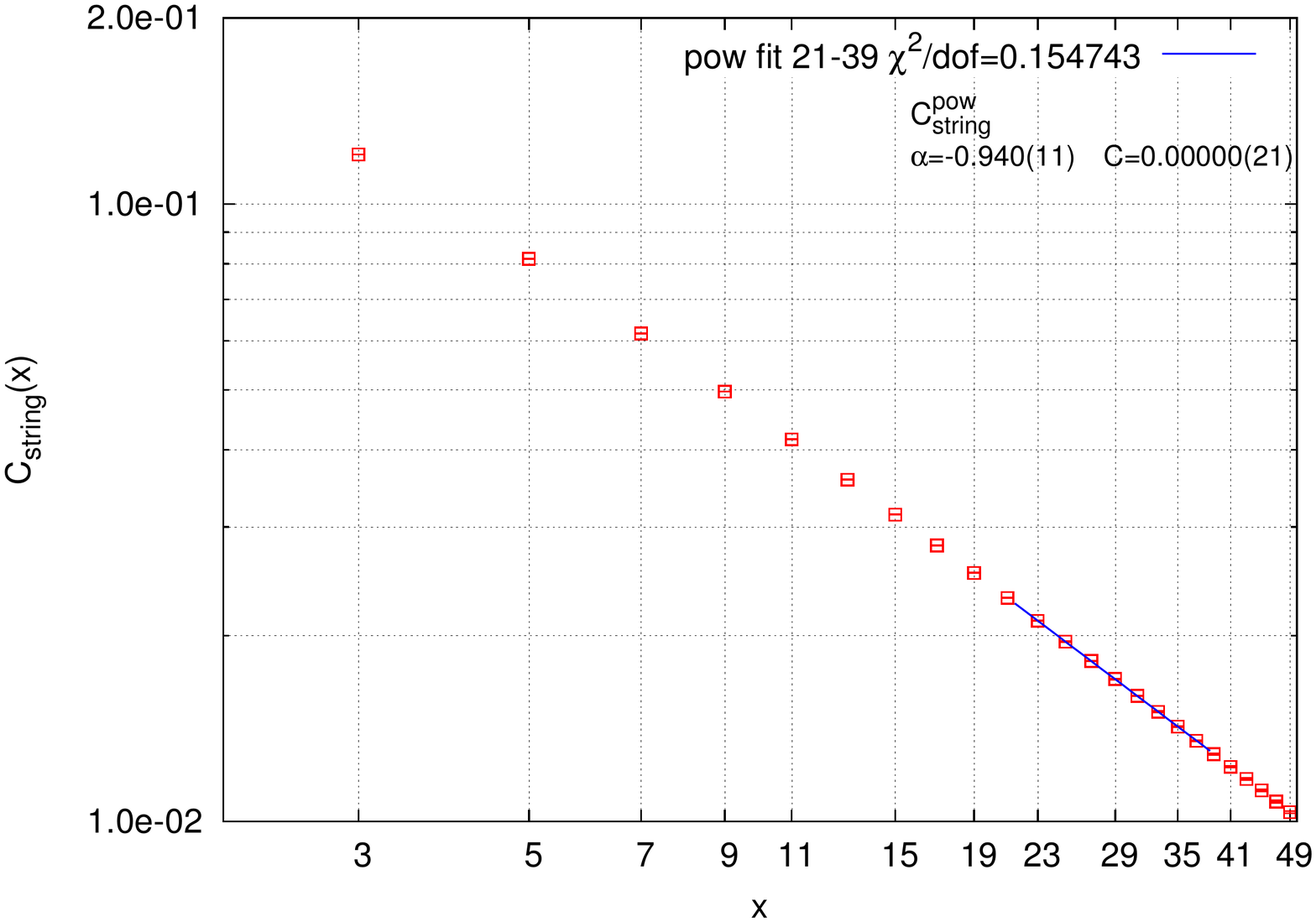}
\hspace{-1.3cm}
\includegraphics[width=9.5cm]{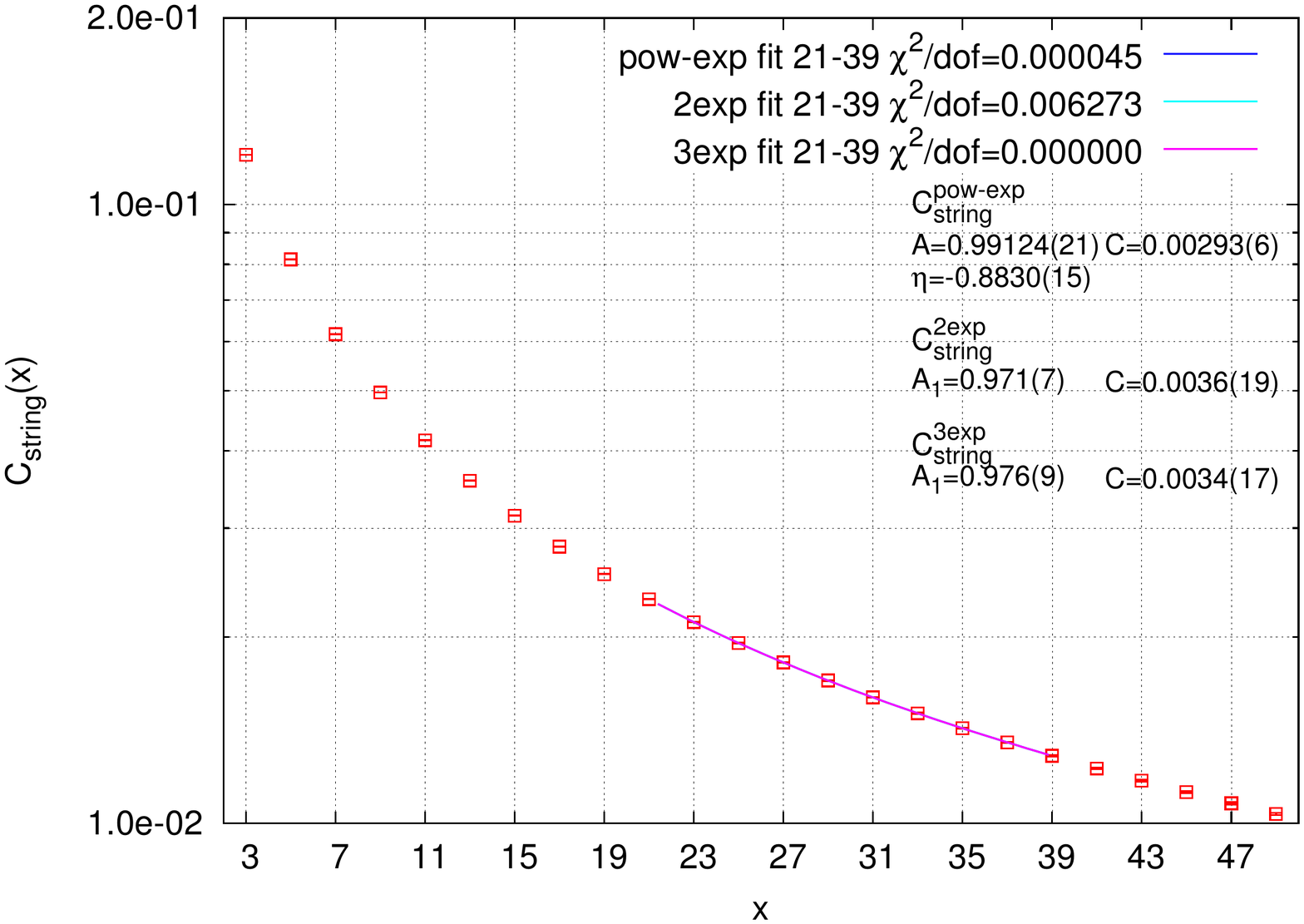}
  \caption{Fermion-antifermion correlator $C_{\rm string}(x)$ for $N=1000$, $\tilde{m}_{0}a = 0.02$, $\Delta(g)=-0.7$. In the left panel, we show an example fit of the power-law fitting ansatz (\ref{eq:pow}) in the interval $x\in[21,39]$. In the right panel, the fits are for three types of exponential ansatzes (\ref{eq:powexp})-(\ref{eq:3exp}), in the same interval.}
  \label{fig:Cstr_g-07}
\end{figure}

\begin{figure}[!t]
  \centering
\includegraphics[width=9.5cm]{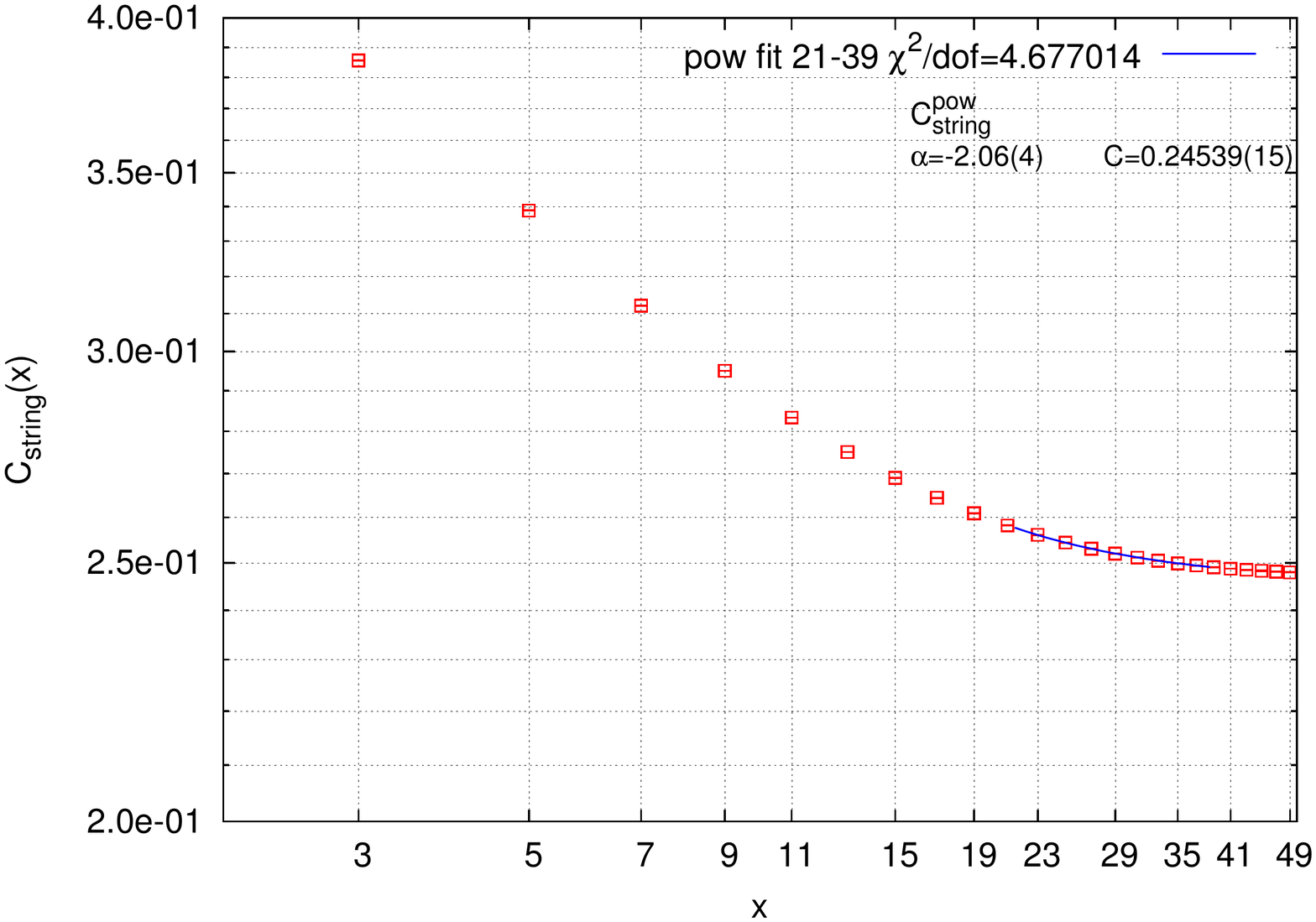}
\hspace{-1.3cm}
\includegraphics[width=9.5cm]{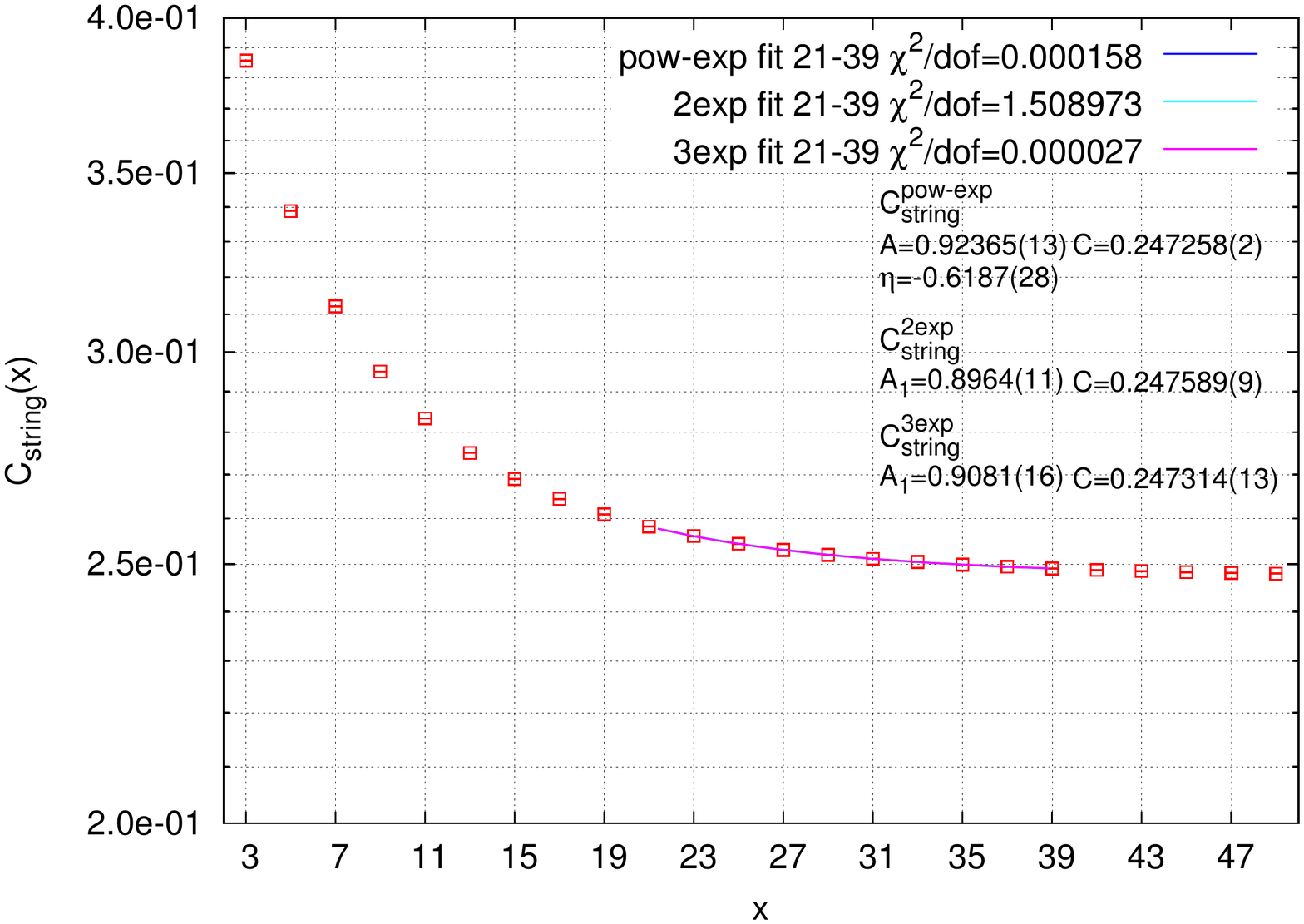}
  \caption{Fermion-antifermion correlator $C_{\rm string}(x)$ for $N=1000$, $\tilde{m}_{0}a = 0.02$, $\Delta(g)=0.4$. In the left panel, we show an example fit of the power-law fitting ansatz (\ref{eq:pow}) in the interval $x\in[21,39]$. In the right panel, the fits are for three types of exponential ansatzes (\ref{eq:powexp})-(\ref{eq:3exp}), in the same interval.}
  \label{fig:Cstr_g04}
\end{figure}

\begin{figure}[!t]
  \centering
\includegraphics[width=9.5cm]{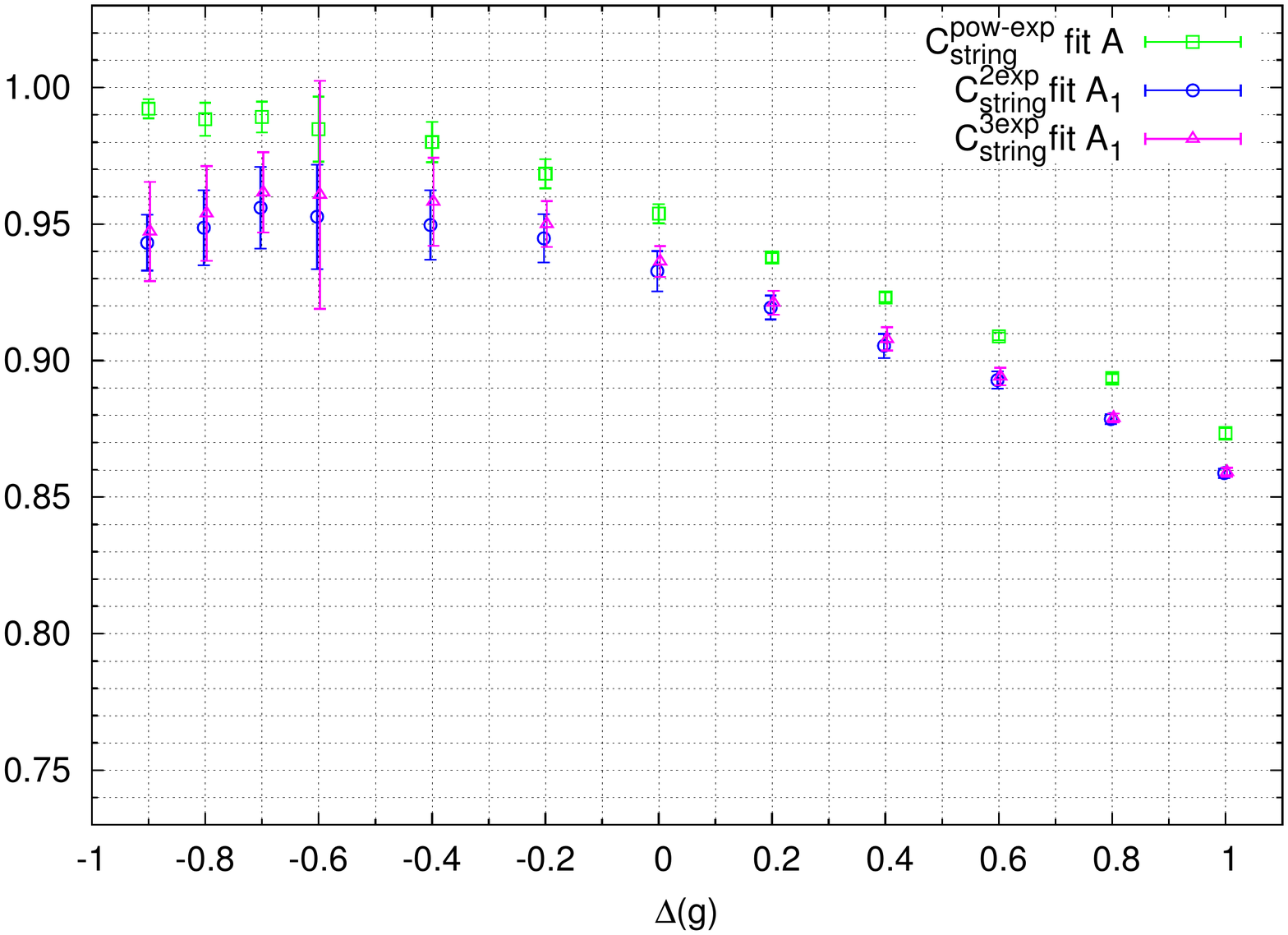}
\hspace{-1.3cm}
\includegraphics[width=9.5cm]{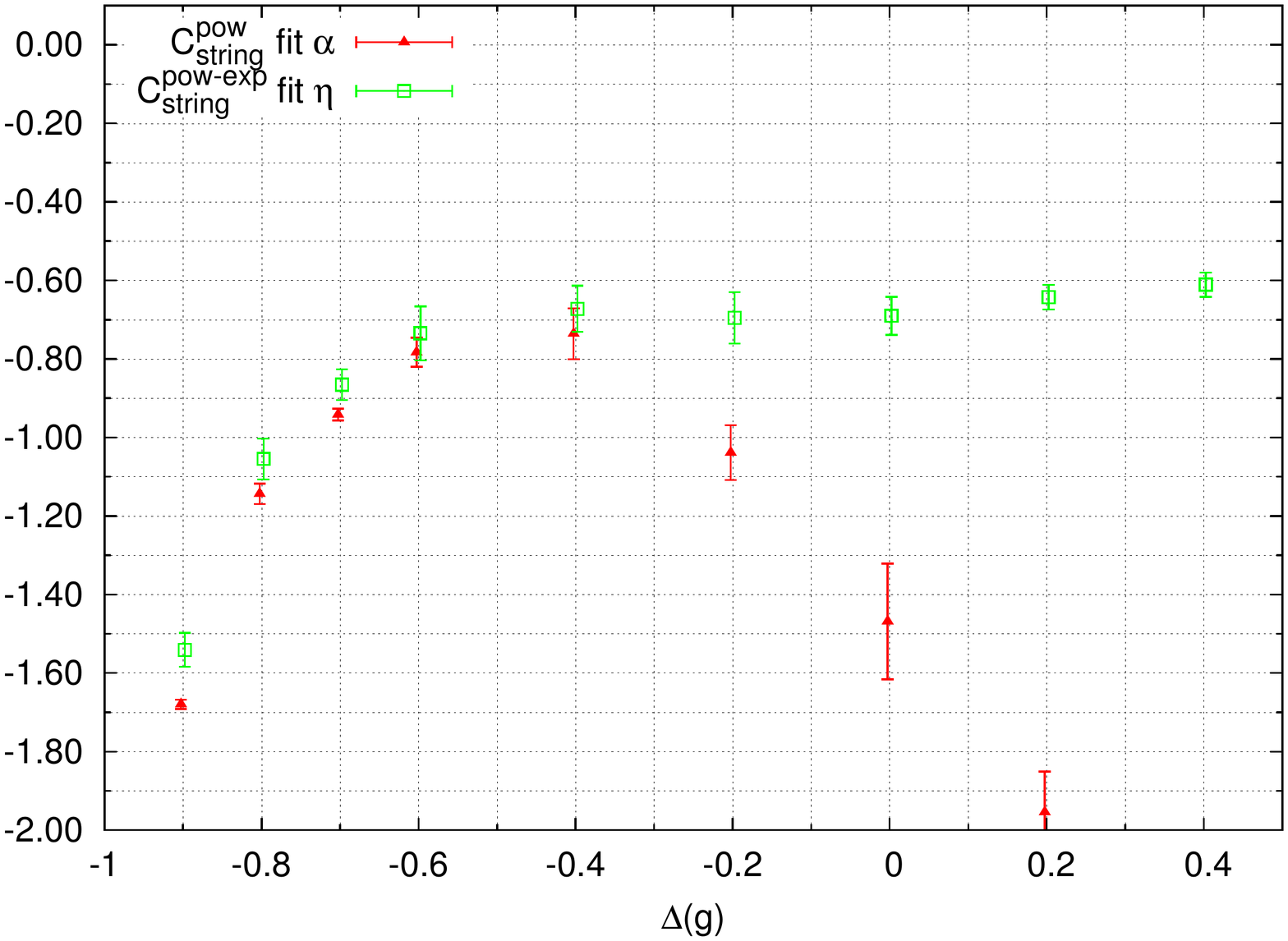}
  \caption{Results from fitting the fermion-antifermion correlator, $C_{\rm string}(x)$.  Left panel: dependence of the parameter $A$ and $A_1$ for three types of exponential ansatzes (\ref{eq:powexp})-(\ref{eq:3exp}) on the coupling $\Delta(g)$. Right panel: dependence of the parameter $\alpha$ and $\eta$ for the power-law and power-exponential fitting ansatzes (\ref{eq:pow})-(\ref{eq:powexp}) on the coupling $\Delta(g)$. Parameters: $N=1000$, $\tilde{m}_{0}a = 0.02$. }
  \label{fig:Cstr_syst_m02}
\end{figure}

\begin{figure}[!t]
  \centering
\includegraphics[width=9.5cm]{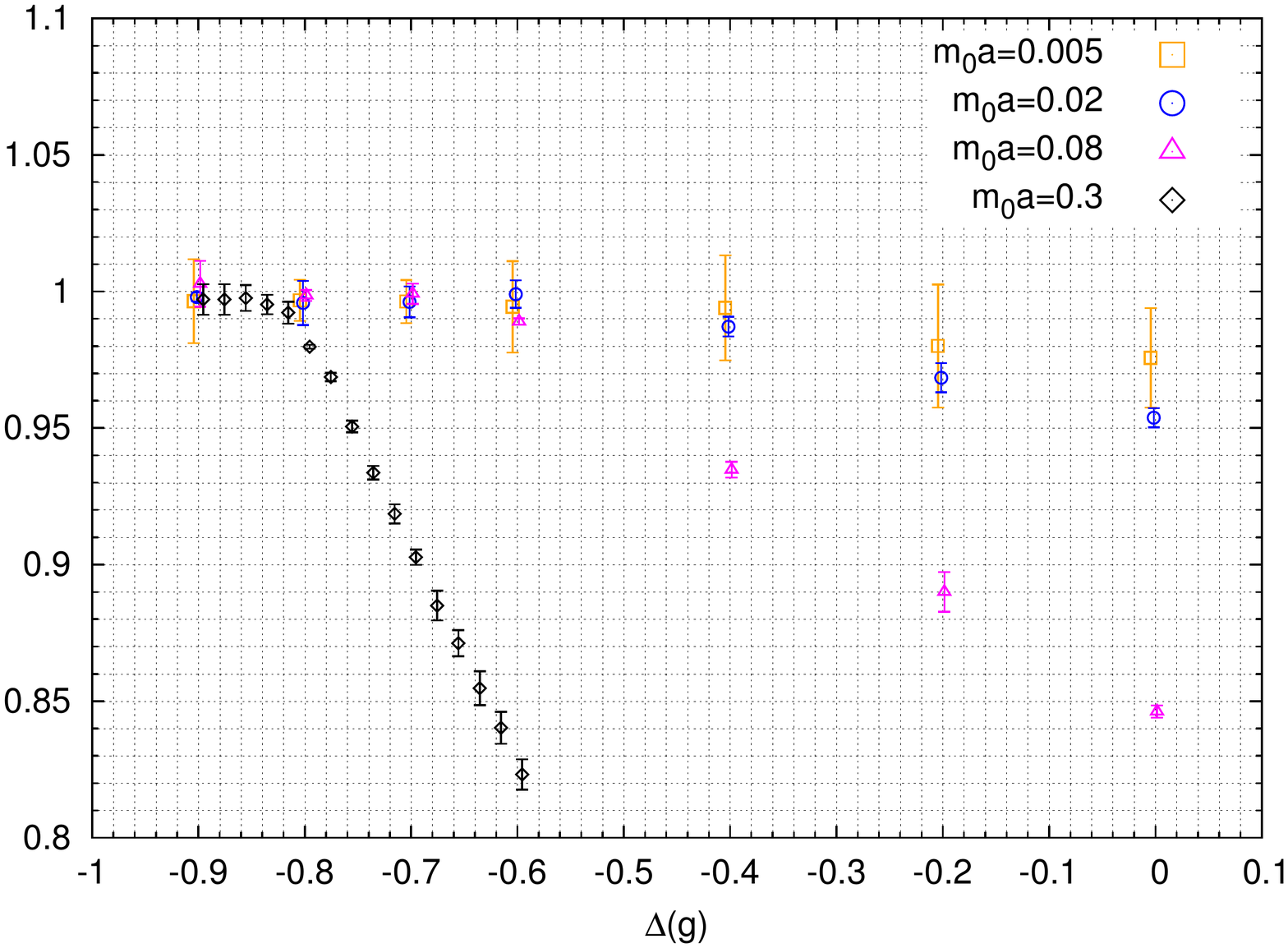}
\hspace{-1.3cm}
\includegraphics[width=9.5cm]{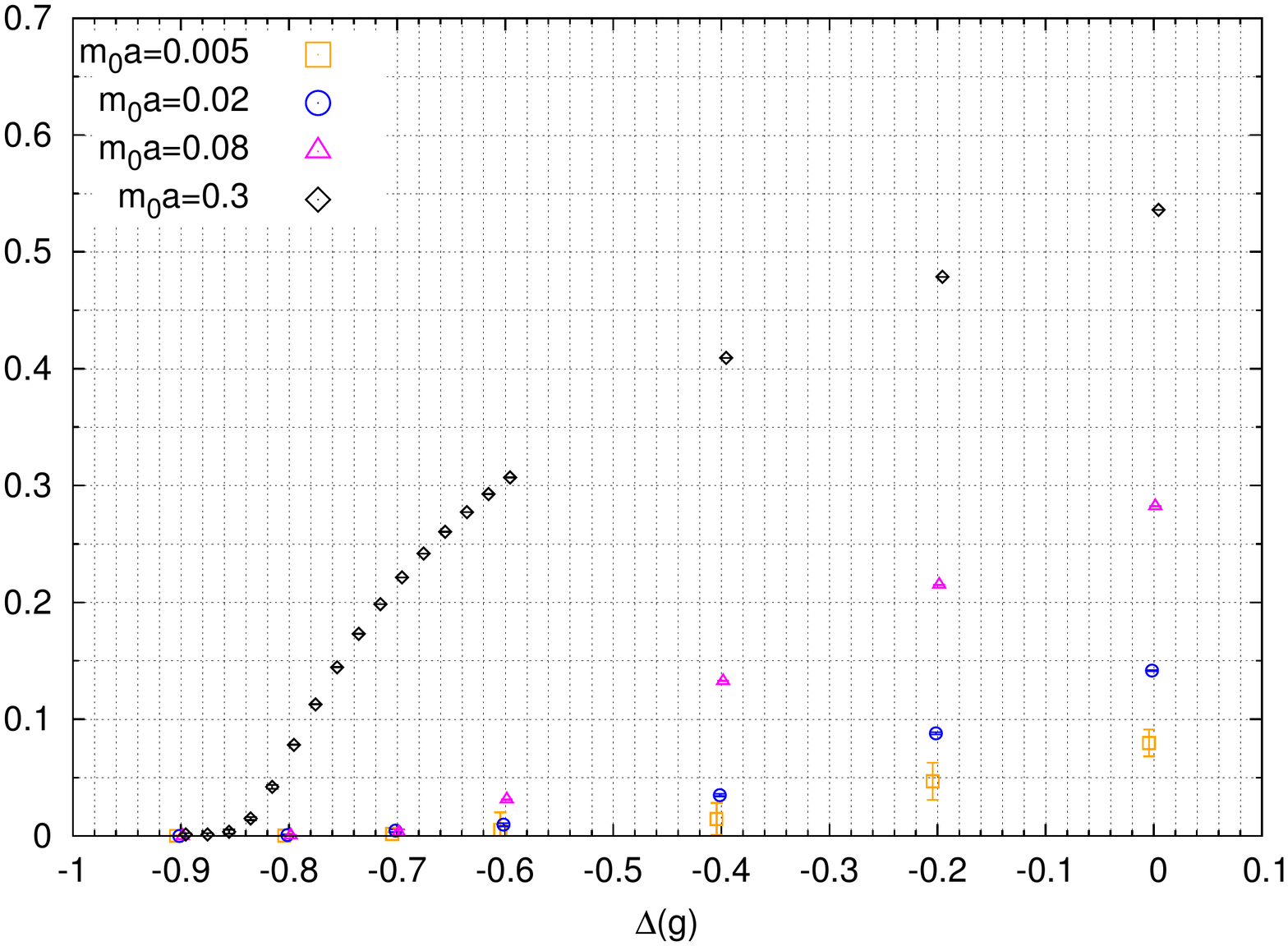}
  \caption{Dependence of the parameter $A$ (left) and $C$ (right) for the power-exponential ansatz (\ref{eq:powexp}) on the coupling $\Delta(g)$, obtained from fitting $C_{\rm string}(x)$. Shown are four fermion masses, $\tilde{m}_{0}a = 0.005, 0.02,0.08, 0.3$ for a 1000-site system.}
  \label{fig:Cstr_syst_m}
\end{figure}

We also consider the fermion-antifermion (string) correlator.
In Figs.\ \ref{fig:Cstr_g-07}, \ref{fig:Cstr_g04}, we show its decay for the same parameters as in Figs.\ \ref{fig:Czz_g-07}, \ref{fig:Czz_g04} ($N=1000$, $\tilde{m}_{0}a = 0.02$ and two couplings, $\Delta(g)=-0.7$ and $\Delta(g)=0.4$).
We illustrate our fits again with examples for the fitting range of $x\in[21,39]$ and we reach conclusions that are consistent with the ones from the density-density correlator.
For coupling close to the BKT transition point, the decay is power-like, indicated by the good $\chi^2/{\rm dof}$ of the power-law and power-exponential fits, the latter having $A$ close to 1 and thus being dominated by the algebraic term.
The constant $C$ of the power-exponential fit is small, but non-zero, indicating that the actual transition may occur for slightly more negative $\Delta(g)$.
As we concluded above for the density-density correlator, it is challenging to locate the transition point more precisely and it would require much more accurate data.
When the coupling is increased and the system is clearly in the gapped phase, the pure power-law fit does not describe well the data and the proper fitting ansatz is the power-exponential one, that can be mimicked by a few exponentials.
A rather large value of the constant $C$ is observed and, similarly as for the density-density correlator, the value of $A$ is much smaller than 1 and decreases towards larger $\Delta(g)$, as can be seen in Fig.\ \ref{fig:Cstr_syst_m02}.
The exponent $\eta$ increases from below $-1$ in the critical phase towards less negative values deep in the gapped phase, indicating that the fits are dominated by the exponential term in the latter.
An important difference with respect to the $C_{zz}$ correlator is that the exponent $\alpha=\eta$ in the critical phase does not have a universal value.
In the density-density correlator, $\alpha=\eta=-2$ for all couplings in the gapless phase, while in the string correlator, there is rather strong dependence of this exponent when moving deep into this phase \cite{Kosterlitz:1973xp}.
For larger fermion masses, the qualitative picture is the same as for the $\tilde{m}_{0}a = 0.02$ case.
We illustrate the $g$-dependence of the fitting parameter $A$ of the power-exponential fit in the left panel of Fig.\ \ref{fig:Cstr_syst_m}.
Comparing with the analogous plot for the density-density correlator, Fig.\ \ref{fig:Czz_syst_m}, we observe that the value of $A$ for a given pair $(\tilde{m}_{0}a,\Delta(g))$ is basically independent of the correlator type.
The right panel of Fig.\ \ref{fig:Cstr_syst_m} shows the coupling dependence of the constant $C$.
Similarly to the case of the parameter $A$, there is a pronounced dependence of $C$ for the largest fermion mass -- its value is zero in the critical phase and steeply increases when moving deeper into the gapped phase.
For smaller masses, this dependence is milder, again as for the $A$ parameter.
However, in general, the relative systematic error of $C$ is smaller than the one of $A$, allowing to pinpoint the transition point a bit more precisely.

\section{Discussion}
\label{sec:discussion}
In this section, we will use the numerical results reported in Sec.~\ref{sec:results} to discuss the phase structure of the massive Thirring model in 1+1 dimensions.   Implication for the scaling behaviour and the continuum limit of the model is also addressed.

\subsection{Summary of the phase structure}
\label{sec:phase_structure_summary}
From the investigation of the entanglement entropy and two types of correlators, as presented in Secs.~\ref{sec:entanglement_entropy} and \ref{sec:correlators}, we observe concrete evidence for the existence of two phases in the massive Thirring model in 1+1 dimensions.  
%
%
In one of these two phases we find power law decaying correlations and critical scaling of the entanglement entropy, indicating a conformal phase with central charge equal to unity.
The other phase is a massive (gapped) phase: it exhibits exponentially decaying correlations and bounded entropy.
Combining this with the results from Sec.~\ref{sec:fermion_bilinear_condensate}, where we showed that the chiral condensate does not vanish in either phase when $m_0 a \neq 0$,
and thus cannot serve as order parameter,  it can be concluded that the relevant phase transition is of the BKT-type. 
Although this phase structure is expected from duality properties of the model, our current work adds important new ingredients to this research direction.    
Our results demonstrate that the MPS methods are still applicable to study the lattice version of a model with this feature, despite the intrinsic difficulty of locating BKT transitions numerically.
This leads to possibilities for further understanding physics of the BKT phase transitions, since the MPS formulation can naturally be implemented for studying real-time dynamics~\cite{Banuls:2020inprep}.

\begin{figure}[!t]
  \centering
\includegraphics[width=14cm, height=10cm]{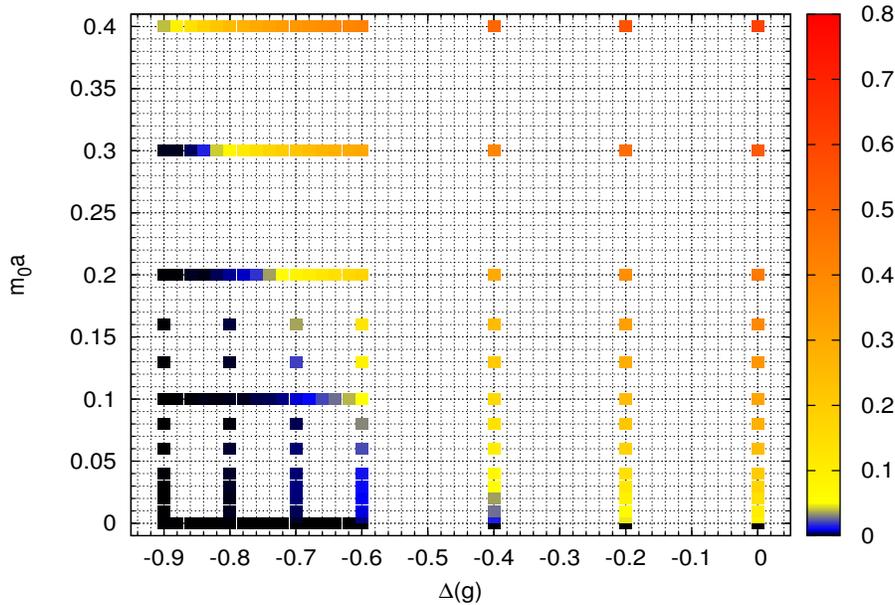}
\caption{The central value of the parameter $C$ for different combinations of the fermion mass $\tilde{m}_{0}a$ and coupling $\Delta(g)$.
}
  \label{fig:Cphase}
\end{figure}

Our numerical calculation confirms that the (1+1)-dimensional Thirring model is a conformal field theory in the massless limit.   At non-vanishing values of the bare mass in lattice units, $\tilde{m}_{0} a$, both the massive and the critical phases are seen.   To depict the phase diagram, we resort to the resultant central value of the constant term, $C$, in fitting the fermion-antifermion correlator defined in Eq.~(\ref{eq:Cstr}) to the ansatz in Eq.~(\ref{eq:powexp}) which describes the power-exponential analysis of the data.   It is reported in Sec.~\ref{sec:correlators} that this ansatz always leads to good reduced $\chi^{2}$ in the fitting procedure.  At criticality, its results also converge well to those given by the power-law analysis using Eq.~(\ref{eq:pow}), with $\eta$ compatible with $\alpha$, and the parameter $A$ in Eq.~(\ref{eq:powexp}) being consistent with one.  As already discussed in Sec.~\ref{sec:correlators}, the parameter $C$ is expected to be non-zero in the gapped phase where the string order in the corresponding spin chain can emerge, while it vanishes in the critical phase.   The reason for choosing $C$ as the main probe of the phase structure is because it is the best determined quantity amongst all the observables computed in this work.  
Figure~\ref{fig:Cphase} shows results for the central value of $C$ at all choices of $[\Delta (g), \tilde{m}_{0}a]$ in our simulations.  
It is clear from this plot that on the $\Delta (g){-}\tilde{m}_{0}a$ plane, there is a region where this parameter is well consistent with zero.  Here we also remind the reader that typical error on $C$ is of percentage or subpercentage level.

We then use the information presented in Fig.~\ref{fig:Cphase} to extract the non-thermal phase structure of the (1+1)-dimensional massive Thirring model.   While $C$ is the most accurately computed quantity in our analysis procedure, it still contains systematic error.   For cases where this parameter is small but non-vanishing in the fermion-antifermion correlator, it is challenging to decide whether the theory is in the conformal or gapped phase.  In fact, we discover that when the central value of $C$ is between 0.001 and 0.01, it is impossible to use our data for a clear judgement on this issue, as all the fit ansatzes in Eqs.~(\ref{eq:pow}), (\ref{eq:powexp}), (\ref{eq:2exp}) and (\ref{eq:3exp}) lead to acceptable reduced $\chi^{2}$.  That is, power-law, power-exponential and pure-exponential functions all describe our data well when $0.001 \leq C \leq 0.01$.  For this reason, we identify the cases in this range as being ``undetermined''.
It may be possible to reduce the size of the undetermined region by means of better (e.g.\ with larger bond dimension) simulations.
Nevertheless, such a high-precision determination of the phase boundary is beyond the scope of this work.

\begin{figure}[!t]
  \centering
\includegraphics[width=11cm]{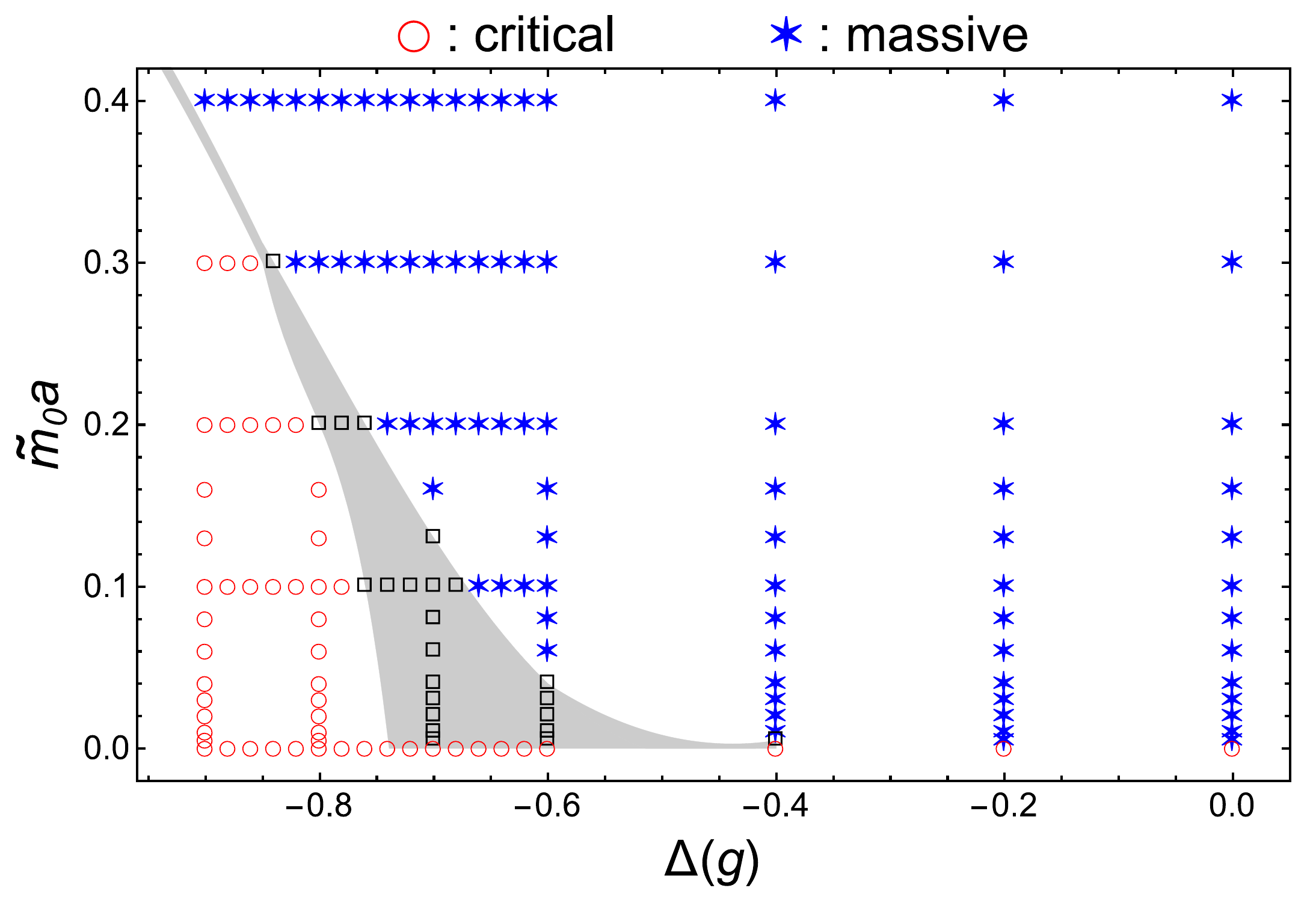}
\caption{Non-thermal phase structure of the massive Thirring model from our numerical investigation.  In addition to the data points that can be identified to be in the gapped phase (blue stars) or at criticality (red circles), there are points (black squares) where our simulations cannot determine which phase the theory is in.  The grey area indicates the regime where we find these ``undetermined'' point.  The BKT phase transition must occur within this grey area.}
  \label{fig:phase_structure}
\end{figure}

From the above discussion, we classify our data points into three groups according the to the central value of the parameter $C$ in fitting the data of fermion-antifermion correlator to Eq.~(\ref{eq:powexp}).
\begin{enumerate}
 \item The critical regime is where $C < 0.001$.
 \item The gapped phase is identified as where $C > 0.01$.
 \item For $0.001 \leq C \leq 0.01$, we label these points as ``undetermined''.
\end{enumerate}
Using this categorization, we plot the non-thermal phase structure of the Thirring model in Fig.~\ref{fig:phase_structure}. 
In this figure, there is a grey area.  This area indicates qualitatively the regime where we find the undetermined points described above.  It is obvious that the BKT transition occurs in this grey region, with the phase boundary described by a function
\beq
\label{eq:phase_boundary}
 \Delta_{\ast} (\tilde{m}_{0}a) = \Delta [ g_{\ast} (\tilde{m}_{0}a) ] \,  ,
\eeq
where $g_{\ast} (\tilde{m}_{0}a)$ is the value of $g$ for the BKT phase transition in the massive Thirring model with bare mass $m_{0}$ and lattice spacing $a$.   At a given value of $\tilde{m}_{0} a$, the theory is conformal when $\Delta$ is less than $\Delta_{\ast} (\tilde{m}_{0}a)$, and becomes gapped otherwise.

With the definition,
\beq
\label{eq:Delta_bar_star_def}
  \bar{\Delta}_{\ast} \equiv \Delta_{\ast} (\bar{g}_{\ast}) = \Delta [ g_{\ast} (\tilde{m}_{0}a) ]|_{\tilde{m}_{0}a = 0} \, ,
\eeq
Figure~\ref{fig:phase_structure} shows that 
\beq
\label{eq:Delta_bar_star_value}
  \bar{\Delta}_{\ast} \sim -0.7 \, .
\eeq
As discussed in Secs.~\ref{sec:intro} and \ref{sec:latt_spin}, we work with the value of the four-fermion coupling, $g$, in the range between $-\pi$ and $\pi$ such that the S-duality is valid.  In this range of the coupling, $\Delta$ is singled-valued and increases monotonically with $g$, as can be seen from Eq.~(\ref{eq:m_tilde_0_and_Delta}).   Therefore, Eq.~(\ref{eq:Delta_bar_star_value}) implies that
\beq
\label{eq:g_bar_star_value_from_data}
 \bar{g}_{\ast} \sim - \frac{\pi}{2} \, ,
\eeq
which is consistent with the result, Eq.~(\ref{eq:g_bar_star}), extracted from the continuum RGE's.   It can also be inferred from Fig.~\ref{fig:phase_structure} that $\Delta_{\ast}[g_{\ast} (\tilde{m}_{0}a)]$ and $g_{\ast} (\tilde{m}_{0}a)$ decrease as the value of the fermion mass grows.  This qualitative feature agrees with the that of the phase boundary presented in Fig.~\ref{fig:RG_flow}.   Here we stress that Eq.~(\ref{eq:g_bar_star}) and Fig.~\ref{fig:RG_flow} are obtained from a perturbative expansion in $m/\Lambda$, while results presented in this section are extracted non-perturbatively.

\subsection{Implication for scaling behaviour and the continuum limit}
\label{sec:scaling_and_continuum_limit}
Numerical results presented in this paper can be employed to infer scaling behaviour of the massive Thirring model in a non-perturbative fashion.  Such investigation is particularly interesting for the critical phase.  As already discussed in Sec.~\ref{sec:phase_structure_summary}, with the phase boundary introduced in Eq.~(\ref{eq:phase_boundary}),  it is found that the theory with non-zero bare mass is conformal in the regime $\Delta (g) < \Delta_{\ast} (\tilde{m}_{0} a)$ at a given value of $\tilde{m}_{0} a$.  Because of chiral symmetry, the bare mass is proportional to the renormalized mass in the Thirring model\footnote{This is the reason why in this section we do not distinguish between the bare and the renormalized fermion masses when it does not affect our conclusion.}.  For a massive theory to be conformal, the only possibility is that the fermion mass is an irrelevant coupling.   This is consistent with the scaling behaviour in the small-mass limit as predicted by Eq.~(\ref{eq:RGE_Thirring_for_m}), and one can use the RG-flow diagram of Fig.~\ref{fig:RG_flow} to illustrate this feature.   In this diagram, there is a stable ``fixed half-line'' at  $(g < \bar{g}_{\ast}, m = 0)$ where $\bar{g}_{\ast} = -\pi/2$.  Let us denote the phase boundary (the dotted blue curve) in Fig.~\ref{fig:RG_flow} by a function $g_{\ast}(m)$.  The whole region spanned by $g < g_{\ast}(m)$ at each value of positive $m$ is then the critical surface of this fixed half-line, where there is no relevant scaling variable.

Corresponding to the above scenario, the $\bar{\psi}\psi$ operator is irrelevant in the critical phase of the massive Thirring model.    
That means, although this operator is of dimension-one at the classical  level, quantum effects from the four-fermion interaction must render it into being irrelevant (with dimension greater than two) in the critical phase.  In other words, a large anomalous dimension has to be generated through interactions.
In Refs.~\cite{Wilson:1970pq,Gomes:1972yb} the scaling dimension of this operator was studied using the operator product expansion (OPE), and was found to increase with decreasing coupling strength, $g$.  It is allowed to be larger than two when $g$ is small enough.  In this regard, our numerical investigation shows the same feature as that from these previous studies.  It would also be possible to have further detailed numerical examination for the scaling properties of the $\bar{\psi}\psi$ operator using the MPS method.  Nevertheless, this calculation requires a dedicated project, and is beyond the scope of our current work.

In the gapped phase, the $\bar{\psi}\psi$ operator is relevant, and a spectrum of excited-state masses can be generated.   We denote the set of these masses by $\{ M_{i} \}$, and assume the hierarchy
\beq
\label{eq:gapped_phase_spectrum}
 M_{1} < M_{2} < M_{3} < \ldots \, .
\eeq
These masses can be computed using variational methods in the framework of MPS~\cite{Banuls:2020alsoinprep}, with the tools developed in Ref.~\cite{Banuls:2013jaa} for the study of the Schwinger model.   With the fermion mass approaching zero, $m \rightarrow 0$, all $M_{i}$ must vanish because the massless Thirring model is scale-invariant.   Nevertheless, we observe that in the regime where the fermion mass $m$ is well below the cut-off scale, the gapped phase can be interpreted as a mass-deformed conformal field theory, and conformality can be restored at any value of the four-fermion coupling on the unstable fixed half-line, $(g > \bar{g}_{\ast}, m=0)$, in Fig.~\ref{fig:RG_flow}.    In this case the scaling behaviour of $M_{i}$ can be studied by employing the hyperscaling techniques detailed in Ref.~\cite{DelDebbio:2010ze}.    To proceed, we first choose a conformal point, $(g = \bar{g}_{+}, m=0)$ where $\bar{g}_{+}$ is greater than $\bar{g}_{\ast} = -\pi/2$.  This conformal field theory is then deformed by introducing a small mass perturbation $(\bar{g}_{+}, m_{{\mathrm{deform}}})$, where $m_{{\mathrm{deform}}}$ in lattice units is much smaller than unity.  In other words, this mass-deformed theory is in the vicinity of the fixed point $(g = \bar{g}_{+}, m=0)$.  With the condition that the four-fermion coupling is not changing under a RG transformation, it is then straightforward to show that~\cite{DelDebbio:2010ze} 
\beq
\label{eq:hyperscaling_M_i}
 a M_{i} = c_{i}(\bar{g}_{+})\, (a m_{{\mathrm{deform}}})^{1/1-\gamma (\bar{g}_{+})} \, , 
\eeq
where $\gamma (\bar{g}_{+})$ is the anomalous dimension of the $\bar{\psi}\psi$ operator evaluated at $\bar{g}_{+}$, and $c_{i}$ is function of $\bar{g}_{+}$.  In other words, when the model is very close  to a particular fixed point, all the excited-state masses will scale to zero with the same exponent.   This interesting behaviour can be tested in future numerical computations for the spectrum of the model ~\cite{Banuls:2020alsoinprep}.  It also offers an approach to determine the anomalous dimension, $\gamma (\bar{g}_{+})$, in the gapped phase near the fixed half-line.  We stress again that Eq.~(\ref{eq:hyperscaling_M_i}) is obtained with the assumption that $g$ takes the value of $\bar{g}_{+}$ and does not change under a RG transformation.

The above scaling properties can serve as a guide to understanding the continuum limit of the massive Thirring model in 1+1 dimensions.
By regarding the field theory as a statistical mechanics system, the latticized model approaches its continuum counterpart at criticality where the correlation, $\xi$, diverges, 
\beq
\label{eq:xi_diverges}
\frac{\xi}{a} \rightarrow \infty \, .
\eeq
One can then associate $\xi$ with typical scale at low energy, e.g., the fermion mass, $m$.  This immediately leads to the conclusion that the theory is in the continuum limit on the line $\tilde{m}_{0} a = 0$ in Fig.~\ref{fig:phase_structure}.  However, because of the BKT phase transition, complication arises in the Thirring model.  This is because the line, $\tilde{m}_{0} a = 0$, is divided into two sectors which are associated with different scaling properties discussed above.  For the critical area with non-vanishing mass in Fig.~\ref{fig:phase_structure}, the model describes the same physics as that on the stable fixed half-line, $(g < \bar{g}_{\ast}, \tilde{m}_{0}a=0)$.  This whole critical regime with $\tilde{m}_{0} a \not= 0$ is then a continuum limit where the theory remains gapless/massless in the IR.   In the gapped phase, approaching the continuum limit, the unstable fixed half-line described by $(g > \bar{g}_{\ast},\tilde{m}_{0}a=0)$, is less straightforward.   Below we address the related issues.

Since the extrapolation to the continuum limit requires the knowledge of how the theory scales when changing the space-time cut-off, the lattice spacing, it is instructive to gain insight from the RGE's\footnote{For a comprehensive treatment of the relation between renormalization and the continuum limit, we refer the reader to Chapter 9 of Ref.~\cite{Creutz:2018dgh}.}.   In addition, we know that the continuum limit of the gapped phase must be on the unstable fixed half-line where the fermion mass vanishes.  Therefore we can resort to the RGE's resulting from a perturbative expansion in $m/\Lambda$, i.e., Eqs.~(\ref{eq:RGE_Thirring_for_g}) and (\ref{eq:RGE_Thirring_for_m}).  Identifying the cut-off as $1/a$, these equations imply
\bea
\label{eq:RGE_Thirring_for_g_in_a}
 \bar{\beta}_{g} &\equiv& a \frac{d g}{d a} = 64 \pi
 \left ( a m \right )^{2} \, , \\
\label{eq:RGE_Thirring_for_m_in_a}
 \bar{\beta}_{m} &\equiv& a \frac{d m}{d a} = m \left [ \frac{ 2 (g +
   \frac{\pi}{2})}{g + \pi} + \frac{256 \pi^{3}}{(g+\pi)^{2}} \left (am \right )^{2} \right ]\, .
\eea
In the limit $a m \ll 1$, they can be approximated as
\bea
\label{eq:RGE_Thirring_for_g_in_a_LO}
 \bar{\beta}_{g} &\equiv& a \frac{d g}{d a} \approx 0 \, , \\
\label{eq:RGE_Thirring_for_m_in_a_LO}
 \bar{\beta}_{m} &\equiv& a \frac{d m}{d a} \approx G_{+} m \, ,
\eea
where 
\beq
\label{eq:G_plus}
 G_{+} = \frac{ 2 (g +
   \frac{\pi}{2})}{g + \pi}  > 0 \, ,
\eeq
since $g > -\pi/2$ near $am = 0$ in the gapped phase.  While Eq.~(\ref{eq:RGE_Thirring_for_m_in_a_LO}) leads to the conclusion that in the continuum limit, the fermion mass is zero, we cannot infer which value of the four-fermion coupling should take from Eq.~(\ref{eq:RGE_Thirring_for_g_in_a_LO}).   This means that any value of $g$ can be a continuum limit in this phase, as long as it is larger than $-\pi/2$.

The above argument establishes the fact that there are infinite number of continuum limits for the massive Thirring model in the gapped phase.   However, it is possible to approach a particular one amongst them in a controlled way, by resorting to information encoded in physical quantities that are computable.  Below we demonstrate how this is implemented using the excited-state masses defined in Eq.~(\ref{eq:gapped_phase_spectrum}).  As described by Eq.~(\ref{eq:hyperscaling_M_i}), these masses all vanish at any point of the fixed half-line, $(g > \bar{g}_{\ast},\tilde{m}_{0}a=0)$.   Meanwhile, the dimensionless ratio
\beq
\label{eq:r_ij}
   r_{ij} = \frac{M_{i}}{M_{j}} \, ,
\eeq
remains finite and positive.  At a particular continuum limit specified by $(g=g_{+}, \tilde{m}_{0}a=0)$, this ratio is
\beq
\label{eq:r_ij_and_c_ij}
    r_{ij} = c_{ij}(g_{+}) \equiv \frac{c_{i}(g_{+})}{c_{j}(g_{+})} \, .
\eeq
That is, one can in principle extract many such ratios by calculating excited-state masses, and
they are uniquely fixed once a value of $g_{+}$ is chosen.  Away from the fixed half-line, numerical results of the masses $\{ M_{i} \}$ contain lattice artefacts, and the hyperscaling behaviour in Eq.~(\ref{eq:hyperscaling_M_i}) receives corrections.    Therefore, $r_{ij}$ depends on the lattice spacing and the couplings, $g$ and $m_{0}$.  
To be able to approach a particular continuum limit, we simply have to hold two such ratios constant.  This is then enough to non-perturbatively determine the two RGE's governing the change in $g$ and $m_{0}$ when the lattice spacing is varied.   For instance, one can perform simulations at several choices of $g$ and $m_{0} a$, and select points with values of $r_{21}$ and $r_{31}$ fixed to be $\bar{r}_{21}$ and $\bar{r}_{31}$, respectively, i.e.,
\bea
\label{eq:fixing_r_ij}
 r_{21} (g, m_{0}, a) &=& \bar{r}_{21} \, , \nonumber\\
 r_{31} (g, m_{0}, a) &=& \bar{r}_{31} \, .
\eea
This specifies a particular continuum limit, and it guides us to this limit by following the curve of the above condition on the $g{-}m_{0} a$ plane to the limit $m_{0} a \rightarrow 0$.  By taking the derivative with respect to $a$, it is also clear that the above procedure is equivalent to that of fixing a renormalization scheme for studying the scaling of $g$ and $m_{0}$.   We stress that this procedure requires information of the excited-state masses, $\{ M_{i}\}$, but they will be computed in our future work~\cite{Banuls:2020alsoinprep}.

\section{Conclusion and outlook}
\label{sec:conclusion}
In this paper, we report our study for the non-thermal phase structure
of the massive Thirring model in 1+1 dimensions employing the MPS
approach.   By restricting the model in the sector of vanishing total fermion
number, it is expected to be S-dual to the sine-Gordon theory where
classical soliton solutions are known.   
Our construction of the Hamiltonian makes use of the 
staggered-fermion discretization in the spatial direction.   Through
the application of the JW transformation, this Hamiltonian
can be shown to be equivalent to the XXZ spin chain coupled to 
uniform and staggered external magnetic fields.  In addition, a ``penalty
term'' is included in the Hamiltonian [Eq.~(\ref{eq-penalty-term})] to lift the energies of other
sectors to the cut-off scale.  This ensures that the variational
optimization of the MPS with the DMRG method will result in the ground
state with zero total fermion number\footnote{It is
    straightforward to perform simulations for other sectors by having appropriate values of
    $S_{{\mathrm{target}}}$ in Eq.~(\ref{eq-penalty-term}).}.   Numerical
simulations are then performed at fourteen values of bare fermion mass,
ranging from 0 to 0.4 in lattice units, and at twenty four choices
of the four-fermion coupling, $g$, straddling the regime where phase
transitions are expected to occur.   Furthermore, we carry out
computations at seven different bond dimensions ($D = 50, 100, 200, 300,
400, 500, 600$), and four system sizes ($N = 400, 600, 800, 1000$).
This facilitates the extrapolation to the limit of infinite $D$, as
well as to the thermodynamic limit.

Results of this work clearly identify the existence of two phases,
one critical and the other gapped, in
the (1+1)-dimensional massive Thirring model.   We also
demonstrate that in the critical phase, the model is equivalent to the
free bosonic field theory with the central charge being unity.  Through the study of
the entanglement entropy, the $\bar{\psi}\psi$ condensate, as well as
the density-density and the fermion-antifermion correlators, we
observe unambiguous numerical evidence that the two phases are
separated by a BKT transition.  The theory in the chiral limit is
found to be in the critical phase.  This is in accordance with the
expectation that the two-dimensional massless Thirring model is a
conformal field theory.  For the case of non-vanishing fermion mass,
$m$, both the critical and the gapped phases appear.  The phase
boundary, as depicted in Fig.~\ref{fig:phase_structure}, can be
specified with a function, $g_{\ast}(m)$, that takes decreasing
value against growing $m$.  The critical phase is in the regime $g <
g_{\ast}(m)$ at a given $m$.  Furthermore, our calculation shows that
\beq
\label{eq:g_bar_star_conclusion}
\bar{g}_{\ast} \equiv \lim_{m\to 0} g_{\ast}(m) \sim -\frac{\pi}{2} \, ,
\eeq
which is consistent with previous studies.

The investigation presented in this article provides information for the scaling
behaviour and the control of the continuum extrapolation in the massive
Thirring model.  One intriguing feature of the theory with non-zero
bare fermion mass is that it can be in the critical phase, as
indicated in Fig.~\ref{fig:phase_structure}, where the $\bar{\psi}\psi$
operator becomes irrelevant although its classical dimension is one.  
This implies the generation of a large anomalous dimension through the
four-fermion interaction.  Detailed numerical examination of the
scaling dimension can be carried out, although it requires a dedicated
implementation and is beyond the scope of current work.  Another interesting
character of the massive Thirring model is that there are two distinct
types of continuum limits.   The first kind is defined on the stable
fixed half-line of $g
< \bar{g}_{\ast}$ on the $m_{0} a=0$ axis of the $g{-}m_{0} a$ plane,
plus the entire region of the critical phase with $m_{0} a \not= 0$.
The second kind of continuum limit is the unstable fixed
half-line, $(g > \bar{g}_{\ast}, m_{0} a = 0)$.   In Sec.~\ref{sec:scaling_and_continuum_limit} of this article, we
argue that this half-line contains an infinite number of possible
continuum limits which can be distinguished by ratios of the
excited-state masses extrapolated to the limit $m_{0} a \to 0$ in the
gapped phase.

Our work described in this paper establishes the possibility of
applying the
MPS method for probing the BKT phase transition in quantum field
theories.  It is also the first step towards further exploration of
the massive Thirring model using this approach.  As mentioned above,
it is interesting to perform detailed studies of the scaling
behaviour and the continuum limit of the model, which requires the computation of
excited-state masses~\cite{Banuls:2020alsoinprep}.   In addition, the
MPS offers another handle of the scaling properties, through
examining how the theory responds to the change of the bond
dimension~\cite{Tagliacozzo:2007rda,pollmann2009theory,Pirvu2012,Vanhecke:2019pez}.   Finally, it
is also essential to understand the dynamical aspects
of the BKT transition in the massive Thirring model, since it can shed
light on the dynamics related to the process of engineering a field
theory into a conformal phase.  In this regard, the MPS technique provides a natural
way to implement real-time evolution with a quench across the phase
boundary, and this is one of our next steps in this research programme~\cite{Banuls:2020inprep}.

Extending the study to higher dimensions is also
  possible in principle. The natural generalization of MPS to two (or
  more) spatial dimensions is the family of projected entangled
  pair states (PEPS), introduced in
  Ref.~\cite{Verstraete:2004cf}. Variational algorithms can also be
  employed to find a PEPS approximation to the ground state of a given
  Hamiltonian (see e.g. recent results in
  Ref.~\cite{corboz2018finite}).  However, the computational cost
  involved is still considerably higher than in the one-dimensional
  case.  Therefore, carrying out such an exhaustive exploration of the phase
  diagram in $2+1$ dimensions is presently a longer-term goal that will require optimizing or adapting the algorithms.

\acknowledgements
We thank Pochung Chen for useful discussions.
The work of MCB is partly supported by the Deutsche Forschungsgemeinschaft (DFG, German Research Foundation) under Germany's Excellence Strategy -- EXC-2111 -- 390814868,
and by the EU-QUANTERA project QTFLAG (BMBF grant No. 13N14780).
Research of KC is funded by National Science Centre (Poland) grant SONATA BIS
2016/22/E/ST2/00013.  The work of YJK is supported in part by
Ministry of Science and Technology (MoST) of Taiwan under Grants No.
105-2112-M-002-023-MY3, 107-2112-M-002 -016 -MY3, and 108-2918-I-002
-032.  CJDL and DTLT acknowledge the Taiwanese MoST
Grant No. 105-2628-M-009-003-MY4.  YPL is sponsored by the Army
Research Office under Grant No. W911NF-17-1-0482. The views and
conclusions contained in this document are those of the authors and
should not be interpreted as representing the official policies,
either expressed or implied, of the Army Research Office or the
U.S. Government. The U.S. Government is authorized to reproduce and
distribute reprints for Government purposes notwithstanding any
copyright notation herein.   Numerical simulations for this work were
performed on the LOEWE-CSC high-performance computer of Johann
Wolfgang Goethe-University Frankfurt am Main, and on the
high-performance computing
facilities at National Chiao-Tung University.



\clearpage

\bibliographystyle{apsrev.bst} 
\bibliography{refs} 
 
\end{document}